\documentclass[a4paper,11pt]{article}

\pdfoutput=1 

\usepackage{jheppub} 

\usepackage{bm}
\usepackage{color}
\usepackage{appendix}
\usepackage{graphicx}
\usepackage{amsmath}
\usepackage{amssymb}
\usepackage{slashed}
\usepackage{tabularx}
\usepackage{wrapfig}
\usepackage{hyperref}
\usepackage{mathabx}

\usepackage[utf8]{inputenc}

\newcommand{\be}{\begin{equation}} \newcommand{\ee}{\end{equation}}
\newcommand{\ba}{\begin{array}{c}} \newcommand{\ea}{\end{array}}
\newcommand{\bea}{\begin{eqnarray}} \newcommand{\eea}{\end{eqnarray}}

\newcommand\tstrut{\rule{0pt}{2.9ex}}       

\title{\Large New physics and the tau polarization vector in $b\to c \tau \bar\nu_\tau$
decays}

\author[a]{Neus Penalva,}
\author[b]{Eliecer Hern\'andez}
\author[a,1]{and Juan Nieves%
\note{Corresponding author.}}

\affiliation[a]{Instituto de F\'{\i}sica Corpuscular (centro mixto CSIC-UV), Institutos de Investigaci\'on de Paterna,
Apartado 22085, 46071, Valencia, Spain}

\affiliation[b]{Departamento de F\'\i sica Fundamental 
  e IUFFyM,\\ Universidad de Salamanca, E-37008 Salamanca, Spain}
\emailAdd{Neus.Penalva@ific.uv.es}
\emailAdd{gajatee@usal.es}
\emailAdd{jmnieves@ific.uv.es}

\date{\today}
\abstract{
For a general  $H_b\to H_c\tau\bar\nu_\tau$  decay we 
analyze the role of the $\tau$ polarization vector ${\cal P}^\mu$ in the
 context of lepton flavor universality violation studies. We use
a general phenomenological approach that includes, in addition to the 
Standard Model (SM) contribution,  vector, axial, scalar, pseudoscalar
 and tensor new physics (NP) terms which strength is governed by, 
 complex in general, Wilson coefficients. We show that both in the
laboratory frame, where the initial hadron is at rest, and in the
center of mass of the two final leptons, a $\vec {\cal P}$ component
perpendicular to the plane defined by the three-momenta of the
final hadron and the $\tau$ lepton is only possible for complex Wilson
coefficients, being  a clear signal for physics beyond the SM as well
as time reversal (or CP-symmetry) violation.  We make specific evaluations of the different polarization vector components  
for the $\Lambda_b\to\Lambda_c$, $\bar B_c\to\eta_c,J/\psi$ 
and $\bar B\to D^{(*)}$ semileptonic decays, and describe NP effects 
in the complete two-dimensional space associated with the independent 
kinematic variables on which the polarization vector depends. 
   We find that the detailed 
study of ${\cal P}^\mu$  has great potential to discriminate between 
different NP scenarios for $0^-\to 0^-$ decays,  but also for 
$\Lambda_b \to \Lambda_c$ transitions. For this latter reaction, 
we pay special attention to corrections to the SM predictions  derived 
from  complex  Wilson coefficients contributions.  }

\begin{document}
\maketitle

\section{Introduction}
\vspace{1cm}
The tension between the Standard Model (SM) predictions and experimental data 
in semileptonic decays involving the third quark and lepton generations points to the possible existence of new physics (NP) affecting those decays. The strongest evidence for this lepton flavor universality violation (LFUV) is in the ratios ($\ell=e,\mu$)
\bea
{\cal R}_{D}
=\frac{\Gamma(\bar B\to D\tau\bar\nu_\tau)}
{\tstrut\Gamma(\bar B\to D \ell\bar\nu_\ell)}=0.340\pm0.027\pm0.013,\nonumber\\
{\cal R}_{D^*}
=\frac{\Gamma(\bar B\to D^*\tau\bar\nu_\tau)}
{\Gamma(\bar B\to D^*\ell\bar\nu_\ell)}=0.295\pm0.011\pm0.008,\nonumber\\
{\cal R}_{J/\psi}=\frac{\Gamma(\bar B_c\to J/\psi\tau\bar\nu_\tau)}
{\Gamma(\bar B_c\to J/\psi\mu\bar\nu_\mu)}=0.71\pm0.17\pm0.18. \label{eq:ratios}
\eea
The ${\cal R}_{D^{(*)}}$ values have been obtained by the Heavy Flavor Averaging Group 
 (HFLAV)~\cite{Amhis:2019ckw},  combining
   different
experimental data  by the 
BaBar~\cite{Lees:2012xj,Lees:2013uzd}, 
Belle~\cite{Huschle:2015rga,Sato:2016svk,Hirose:2016wfn,Belle:2019rba} 
and LHCb~\cite{Aaij:2015yra,Aaij:2017uff} collaborations. The corresponding 
SM  results given in Ref.~\cite{Amhis:2019ckw},
 ${\cal R}_{D}=0.299\pm 0.003$ and ${\cal R}_{D^*}=0.258\pm0.05$, are
obtained from the SM
predictions in Refs.~\cite{Aoki:2016frl, Bigi:2016mdz, Bigi:2017jbd,
Jaiswal:2017rve, Bernlochner:2017jka}. Similar results are obtained in 
Ref.~\cite{Iguro:2020cpg} using the heavy quark effective theory
parameterization of the form factors with up to ${\cal O}(1/m^2_c)$ corrections. The
 tension  with the SM is at the level of  $3.1\,\sigma$, although it will reduce 
 to just $0.8\,\sigma$ if only the latest Belle results from Ref.~\cite{Belle:2019rba}
  were considered. In this respect, in Refs.~\cite{Alok:2019uqc, Kumbhakar:2020jdz} it is argued that
  the inclusion of the new Belle data heavily restricts the number of allowed NP 
  solutions,  claiming that a precise measurement of the $\bar B_c\to \tau\bar\nu_\tau$ branching
  ratio can distinguish among them. The important constraints, on new-physics interpretations of the anomalies observed in $\bar B\to D^{(*)}\tau\bar\nu_\tau$ 
  decays, derived from the lifetime of the $\bar B_c$ meson were firstly pointed out in  \cite{Alonso:2016oyd}, and they have commonly be considered in all subsequent analyses. 
  
The ratio ${\cal R}_{J/\psi}$ has been recently measured by the LHCb Collaboration~\cite{Aaij:2017tyk} and it shows 
a  $1.8\,\sigma$ discrepancy with SM results, which are in the range
$R^{\rm SM}_{J/\psi} \sim 0.25-0.28$~\cite{Anisimov:1998uk,Ivanov:2006ni,
Hernandez:2006gt,Huang:2007kb,Wang:2008xt,Wen-Fei:2013uea, Watanabe:2017mip, Issadykov:2018myx,Tran:2018kuv,
Hu:2019qcn,Leljak:2019eyw,Azizi:2019aaf,Wang:2018duy}. $\bar B_c$ decays 
induced by the $c\to  s, d$ transition at the quark level are also being 
investigated as a possible source of information on NP~\cite{Colangelo:2021dnv}, taking advantage of the recent results of Ref.~\cite{Becirevic:2020rzi}. In this latter work, the possibilities of extracting constraints on NP by using the current data on the leptonic and semileptonic decays of pseudoscalar mesons, not only driven by the $b\to c$ transition,  have been exhaustively discussed.

 NP effects on ${\cal R}_{D^{(*)}}$ and ${\cal R}_{J/\psi}$  are studied in a phenomenological way 
 including   scalar, pseudoscalar and tensor $b\to c\tau\bar\nu_\tau$ effective
  operators, as well as NP corrections to the SM vector and axial ones. NP 
  terms are governed by Wilson coefficients which are complex in general and 
  should be fitted to data. As a result of this fitting procedure, different 
  NP scenarios  actually lead to an equally good reproduction of the
 above ratios in Eq.~\eqref{eq:ratios} (see for instance Refs.~\cite{Bhattacharya:2018kig, Murgui:2019czp, Shi:2019gxi}\footnote{The latest measurements of  ${\cal R}_{D^{(*)}}$ reported by Belle~\cite{Belle:2019rba} have a strong influence in the admissible extensions of the SM~\cite{Shi:2019gxi}, strongly disfavoring, for instance, large pure tensor NP scenarios which were possible~\cite{Bhattacharya:2018kig} with the 2018 HFLAV averages~\cite{Amhis:2016xyh}.}). Then, other observables are needed to 
 constrain and determine the
 most plausible NP extension of the SM. Typically, the $\tau$-forward-backward 
 (${\cal A}_{FB}$) and $\tau$-polarization (${\cal A}_{\lambda_\tau}$) 
 asymmetries have  also been considered. A greater discriminating power can 
 be reached by analyzing the four-body
$\bar B\to D^*(D\pi, D\gamma)\tau\bar\nu_\tau$
~\cite{Duraisamy:2013pia,Duraisamy:2014sna,Becirevic:2016hea,Colangelo:2018cnj} and the full five-body $\bar B\to D^*(D Y)\tau(X 
\nu_\tau)\bar\nu_\tau$~\cite{Ligeti:2016npd,Bhattacharya:2020lfm}  angular distributions.

Another test of this non-universality can
be obtained from the analog semileptonic ${\cal R}_{\Lambda_c}$ ratio, which has been predicted within the SM
in several works~\cite{Gutsche:2015mxa,Azizi:2018axf, Bernlochner:2018kxh}. In Ref.~\cite{Bernlochner:2018kxh}, the result from a solid calculation including leading and sub-leading heavy quark spin symmetry (HQSS) Isgur-Wise (IW) functions, which were simultaneously fitted to LQCD  results and LHCb data, was provided.  The effects of different NP scenarios have been also examined in Refs.~\cite{Shivashankara:2015cta,Ray:2018hrx, Li:2016pdv,Datta:2017aue,Blanke:2018yud,Bernlochner:2018bfn,DiSalvo:2018ngq,Blanke:2019qrx,Boer:2019zmp,Murgui:2019czp,Ferrillo:2019owd}.  We note that the case of a polarized  decaying $\Lambda_b$
baryon has also been addressed in Ref.~\cite{Colangelo:2020vhu}.

 In Refs.~\cite{Penalva:2019rgt,Penalva:2020xup,Penalva:2020ftd}, 
 we have analyzed the   relevant role   that  different contributions to the   differential decay widths $d^2\Gamma/(d\omega d\cos\theta_\tau)$ and $d^2\Gamma/(d\omega dE_\tau)$ could play to the NP search, both for  unpolarized and helicity-polarized final $\tau$-lepton. Here, $\omega$ is the product of
the two hadron four-velocities, $\theta_\tau$ is the angle made by the tau lepton
and final hadron three-momenta in the center of mass of the final 
two-lepton pair (CM), and $E_\tau$ is the final tau energy in the laboratory frame 
(LAB), where the initial hadron is at rest. In Refs.~\cite{Penalva:2019rgt,
Penalva:2020xup}, we give a  general description of our formalism,  based on the
 use of general hadron tensors parameterized in terms of Lorentz scalar functions. 
 It is an alternative to the helicity amplitude scheme, and becomes very useful to 
 describe processes where all hadron polarizations are summed up and/or averaged.   
 In these two works,  we presented results for the $\Lambda_b\to\Lambda_c\tau\bar
 \nu_\tau$ decay and  showed that the helicity-polarized distributions in the LAB 
 frame provide additional information about the NP contributions, which cannot be 
 accessed only by analyzing the CM differential decay widths, as is commonly proposed in 
 the literature.
In Ref.~\cite{Penalva:2020ftd} we extended  the study to $\bar B_c\to\eta_c\tau\bar
\nu_\tau$, $\bar B_c\to J/\psi\tau\bar\nu_\tau$ as well as the $\bar B\to D^{(*)}
\tau\bar\nu_\tau$ decays.  What we have found is  that the discriminating power between 
different NP scenarios was better for $0^-\to 0^-$ and $1/2^+\to 1/2^+$ decays than 
for   $0^-\to 1^-$ reactions.

In this work, we present our results in a different way by looking at the $\tau$ polarization vector ${\cal P}^\mu$. Furthermore, the transverse (referred to  the direction of the $\tau$) components of ${\cal P}^\mu$ allows us to evaluate new 
observables, which do not appear in  the study of LAB and CM helicity-polarized decays. 
The possibility of searching for NP signatures in different $\tau-$polarization related contributions was suggested already 
twenty five years ago in Ref.~\cite{Tanaka:1994ay} for $\bar B \to D^{(*)}$-decays, in the context of SM extensions with charged Higgs bosons. The  
idea has been further developed in more recent works~\cite{Nierste:2008qe, Tanaka:2012nw, Ivanov:2017mrj, Alonso:2017ktd,Blanke:2018yud,  Asadi:2020fdo}, and in particular, a complete framework  to obtain the maximum information with polarized $\tau$ leptons and unpolarized $D^{(*)}$ mesons is discussed in Ref.~\cite{Asadi:2020fdo}, where the full decay chain down to the detectable particles stemming from the
$\tau$ is considered. As mentioned above, we use here a technique different to the usual helicity-amplitude method, and we also show results for the $\Lambda_b$ and $\bar B_c$ decays, for which  such exhaustive analyses are not available yet.

We provide an overview of the spin density matrix formalism for semileptonic decay 
reactions, including NP operators, and discuss how $ {\cal P}^\mu $ is defined in 
that context. As we shall show, for a given configuration of the momenta of the 
involved particles, the  polarization vector  components (projections of ${\cal P}^\mu$
 onto some spatial-like unit four-vectors)
 depend on two variables $(\omega, \cos\theta_\tau)$ or equivalently $(\omega, E_\tau)$,
  and they can be used as extra observables in the search for NP. To our knowledge, 
  this is the first time that such a study has been performed in the context of LFU 
  anomalies.  For fixed $\omega$, the dependence on $\cos\theta_\tau$ (or $E_\tau$) 
  of these observables could be inferred from the general results of Ref.~\cite{Penalva:2020xup}, since the polarization components 
  turn out to be ratios of linear or quadratic functions of the product of  the initial hadron  and final $\bar\nu_\tau$ (or $\tau$)  four-momenta\footnote{For $\bar B$ meson semileptonic decays, the dependence on the CM variable $\cos\theta_\tau$ should 
  be also deduced from the partial wave expansion of the leptonic amplitude within 
  the helicity formalism~\cite{Korner:1989qb,Tanaka:2012nw,Zhang:2020dla}.}. The 
  denominators of these aforementioned ratios are determined by the unpolarized 
  differential decay widths, which can be straightforward seen  in our previous works 
  of Refs.~\cite{Penalva:2020xup, Penalva:2020ftd}. Thus, we will show here results 
  for the coefficients of the polynomials that appear in the numerators of these ratios 
  for  $\Lambda_b\to\Lambda_c\tau\bar
 \nu_\tau$ and  $\bar B\to D^{(*)} \tau\bar\nu_\tau$ decays.
  
  Certain CM angular averages of these components\footnote{We refer to observables additional to  the CM longitudinal $\tau$-polarization asymmetry, ${\cal A}_{\lambda_\tau}$, which is often presented in the literature.}, also addressed in this 
  work, and that might be experimentally accessed through measurements of subsequent 
  hadronic $\tau-$decays, have  already been discussed in Refs.~\cite{Alonso:2017ktd,
  Ivanov:2017mrj,Asadi:2020fdo, Tanaka:2012nw}, \cite{Tran:2018kuv} and \cite{Ray:2018hrx,Li:2016pdv} for the $\bar B\to D^{(*)}\tau\bar\nu_\tau$, $\bar B_c\to J/\psi\tau\bar\nu_\tau$ and  $\Lambda_b\to\Lambda_c\tau\bar
 \nu_\tau$ decays, respectively. 
 
  
     Here we will present results for all the semileptonic decays mentioned above, 
     keeping in mind that a combined analysis of all them can better restrict the 
     possible extensions of NP. We will pay special attention to the 
$\Lambda_b\to\Lambda_c\tau\bar\nu_\tau$ reaction,  as there are good prospects that 
LHCb can measure it in the near future, given the large number 
of $\Lambda_b$ baryons which are produced at the LHC. Indeed, the shape of the 
$\Lambda_b\to\Lambda_c\mu\bar\nu_\mu$ differential decay rate  was already 
reported by LHCb in 2017~\cite{Aaij:2017svr}. Any measurement for the tau mode 
will be extremely valuable, since the evidences for SM anomalies in $b\to c $ 
semileptonic decays  are currently restricted  to the meson sector, and the 
sensitivity of $\Lambda_b$-decay observables to NP operators would likely be  
different. 

The work is organized as follows. In Sec.~\ref{sec:sdm} we introduce the general 
theory on the spin-density matrix and the polarization vector ${\cal P}^\mu$ for a 
$H_b\to H_c\tau\bar\nu_\tau$ decay. Analytical expressions for 
${\cal P}^\mu$  including NP terms are then given in
Sec.~\ref{sec:eh}, and a detailed analysis of parity and time-reversal violation in 
the decay is presented in Sec.~\ref{sec:ptv}. The results are presented and discussed 
in Sec.~\ref{sec:results}. In Appendix~\ref{app:cmlabkin} we give useful information 
on the kinematics in the CM and LAB frames and in Appendix~\ref{app:pltttavg} we give 
some angular averages of the ${\cal P}^\mu$-components in the CM and LAB frames.

\section{Spin-density matrix and polarization vector in semileptonic decays}
\label{sec:sdm}
We obtain in this section general results valid for any baryon/meson semileptonic
 decay for unpolarized hadrons, though we refer explicitly to those
induced by the $b\to c$ transition. 

\subsection{Spin-density operator}

Let us consider a $H_b\to H_c\tau\bar\nu_\tau$ semileptonic decay of a bottomed hadron
($H_b$) of mass $M$ into a charmed one ($H_c$) with mass $M'$. For a given
momentum configuration of all the particles involved, and when the polarizations of
all particles except the $\tau$ lepton are being summed up (averaged or sum over 
polarizations of the initial or final particles, respectively),\footnote{This is equivalent to say that we only measure the  spin state of the $\tau-$lepton  .} the modulus squared of
the invariant amplitude for the production of a final $\tau-$lepton in a $u(k')$ state\footnote{We use Dirac spinors with square root mass dimensions.}
 can always be written as
\be
{\overline{\sum_{rr'}}\, |{\cal M}|^2 } = \bar u(k'){\cal O}u(k'),
\label{eq:Ooper}
\ee
with $k^{\prime} $ the  four-momentum of the final $\tau-$lepton and $r,r'$ 
 hadron polarization indexes. The differential decay rate  is given by~\cite{Zyla:2020zbs} 
\begin{equation}
  \frac{d^2\Gamma}{ds_{23} ds_{13}} = \frac{G^2_F|V_{cb}|^2
    M'}{16\pi^3 M^2}\, \overline{\sum_{rr'}} |{\cal M}|^2, \label{eq:defsec}
  \end{equation}
where $G_F=1.166\times 10^{-5}$~GeV$^{-2}$  is the Fermi coupling
constant and $s_{23}$ ($s_{13}$) is the invariant mass squared of the outgoing $\tau\bar\nu_\tau$ ($H_c\tau$) pair.

 The operator ${\cal O}$, which depends on the momenta of all particles,
is determined by the physics that governs the $H_b\to H_c\tau\bar\nu_\tau$ transition and satisfies 
\be
{\cal O}^\dagger=\gamma^0 {\cal O}\gamma^0. \label{eq:dynamics}
\ee
Note that
\bea
\bar \rho=\frac{(\slashed{k'}+m_\tau){\cal O}(\slashed{k'}+m_\tau)}{{\rm Tr}\,[(\slashed{k'}+m_\tau){\cal O}(\slashed{k'}+m_\tau)]}\label{eq:rhobardef}
\eea
defines a trace-one hermitian operator ($\bar\rho^\dagger = \bar\rho$) in the two-dimensional Hilbert space spanned by the spin states of the $\tau$ particle\footnote{The formalism for antiparticles runs in parallel to the one that will be discussed below, with the obvious replacements of $(\slashed{k'}+m_\tau)$ by $(m_\tau-\slashed{k'})$ and of Dirac $u-$spinors by $v-$spinors. Besides, ${\cal O}$ will also change. }. 
A general polarization basis (covariant spin) for the $\tau$ states with
four-momentum $k'$, can be constructed as follows. For the $\tau$ at rest, we take the two states $u^{\vec n}_{\pm1}(m_\tau,\vec 0\,)$
 corresponding to spin $\pm 1/2$ along the direction defined by a normalized 
 three vector $\vec n$, then apply to these states  a boost of velocity 
 $\vec k\,'/k^{\prime 0}$.  The resulting $u_{\pm 1}^N(k')$ spinors are 
 eigenstates, with  corresponding   eigenvalues $\pm 1$,  of the  $\gamma_5\slashed{N}$ operator, where $N^\mu$ is the transformed  of the 
  four-vector $(0,\vec n)$ by the boost~\cite{Mandl:1985bg}. The  projectors onto the  $u_{\pm1}^N(k')$ states are given by $P^N_{\pm 1}=
\frac12(1\pm\gamma_5\slashed{N})$. Notice that $N^2=-\vec n\,^2=-1$ and  that $N\cdot k'=0$. Helicity is a particular case of  covariant spin where 
$\vec n= \hat k'= \vec k\,'/|\vec k\,'|$ and $N^\mu\equiv \widetilde s^\mu  = (|\vec k\,'|, k^{\prime 0}\hat k')/ m_\tau$.

For the given configuration of momenta, the spin-density operator $\bar\rho$ encodes all information that can be obtained 
on the spin of the $\tau$ leptons produced in the $H_b\to H_c\tau\bar\nu_\tau$ decay when no other particle spin state is measured.  Actually, the matrix elements 
of  $\bar\rho$ read
\bea
\bar\rho^{S}_{\pm 1}&=& \frac1{2m_\tau}\bar u^{S}_{\pm1}(k')\bar\rho\, u^{S}_{\pm 1}(k')
=  \frac{\bar u^{S}_{\pm1}(k'){\cal O}u^{S}_{\pm1}(k')}{{\rm Tr}[(\slashed{k'}+m_\tau){\cal O}]}=\frac{\bar u^{S}_{\pm1}(k'){\cal O}u^{S}_{\pm1}(k')}{\sum_{h=\pm1}\bar u^{S}_{h}(k'){\cal O}u^{S}_{h}(k')} \nonumber \\
&=& P[u^{S}_{\pm1}(k')] \label{eq:prb}
\eea
and give the probability that  in an actual measurement the $\tau$ is found in the
$u^{S}_{\pm1}(k')$  state, as follows  from Eq.~\eqref{eq:Ooper}.  

\subsection{Polarization vector: definition and properties}

Since $\bar\rho$ is hermitian, it can be diagonalized, and there exists  a polarization basis
$u^{N'}_{\pm1}(k')$ for which the corresponding matrix elements satisfy
\be
\bar\rho^{N'}_{hh'}=\frac1{2m_\tau}\bar u^{N'}_{h'}(k')\bar\rho\, u^{N'}_{h}(k')=\bar\rho'_h \delta_{hh'}, 
\ee
where the eigenvalues, $\bar\rho'_h$, are positive real numbers,  as they are just the probabilities of finding  the $\tau$ in the $u^{N'}_{\pm1}(k')$ states.
In this basis of eigenstates, the spin-density matrix can be written as
\begin{eqnarray}
\bar \rho&=&\frac{1}{2m_\tau}\left[  \bar\rho'_{+1}\, u^{N'}_{+1}(k')
\bar u^{N'}_{+1}(k')
+\bar\rho'_{-1}\, u^{N'}_{-1}(k')\bar u^{N'}_{-1}(k') \right]
\nonumber\\
&=&\frac{1}{2m_\tau}\bigg[  \bar\rho'_{+1}\, \sum_{r=\pm1}u^{N'}_{r}(k')
\bar u^{N'}_{r}(k')P^{N'}_{+1}
+\bar\rho'_{-1}\, \sum_{r=\pm1}u^{N'}_{r}(k')\bar u^{N'}_{r}(k')P^{N'}_{-1} \bigg]
\nonumber\\
&=&\frac{\slashed{k'} +m_\tau}{2m_\tau}
\left(\bar\rho'_{+1}{ P}^{N'}_{+1}+\bar\rho'_{-1}{ P}^{N'}_{-1}\right)= \frac{\slashed{k'} +m_\tau}{4m_\tau}
\left[I-\gamma_5\,\left(\bar\rho'_{-1}-\bar\rho'_{+1}\right)\slashed{N'}\right]\nonumber\\
&=&\frac{\slashed{k'} +m_\tau}{4m_\tau}
\left[I-\gamma_5\,\slashed{\cal P}\right],
\label{eq:rhop}
\end{eqnarray}
where we have defined the  polarization vector ${\cal P}^\mu$ as
\be 
{\cal P}^\mu=\left(\bar\rho'_{-1}-\bar\rho'_{+1}\right)N^{\prime\mu}.
\ee
The four-vector ${\cal P}^\mu$ depends  on the dynamics that governs the  
$H_b\to H_c\tau\bar\nu_\tau$ decay, through the operator ${\cal O}$, and it trivially satisfies 
\be 
{\cal P}^{\mu*}={\cal P}^\mu\ ,\ k'\cdot {\cal P}=0\ ,\   {\cal P}^\mu={\rm Tr}[\bar\rho\gamma_5\gamma^\mu]=\frac{{\rm Tr}[(\slashed{k'}+m_\tau){\cal O}(\slashed{k'}+m_\tau)\gamma_5\gamma^\mu]}{{\rm Tr}[(\slashed{k'}+m_\tau){\cal O}(\slashed{k'}+m_\tau)]}
\label{eq:pproperties}
\ee
Note that, for a given momentum configuration of all the particles involved,  
${\cal P}^\mu$ depends only on three independent quantities\footnote{This 
follows trivially considering that  $\bar\rho$ is a hermitian operator 
with trace one in a two-dimensional Hilbert space.}. In the present  
context, it seems natural to take those quantities as one of the two 
eigenvalues   of  $\bar\rho$  and the two angles that fix the privileged 
direction $\vec{n}^{\,\prime}$ in the $\tau$ rest frame, which gives rise to the polarization 
eigenbasis $u^{N'}_{\pm1}(k')$. All 
three are determined by the dynamics of the transition, which enters through 
the operator ${\cal O}$ introduced in Eq~\eqref{eq:Ooper}.

The information on the  spin  of the produced $\tau$ is  solely contained in the polarization vector ${\cal P}^\mu$. 
Thus, the probability of measuring a $\tau$ in a state $u^{S}_h(k')$,
 with $h=\pm1$,  is given by
 \bea
P[u^{S}_h(k')]&=&\frac1{2m_\tau}\bar u^{S}_{h}(k')\bar\rho\, u^{S}_{h}(k')= \frac1{4m_\tau}\bar u^{S}_h(k')(I-\gamma_5\slashed{\cal P})u^{S}_h(k')\nonumber \\
&=&\frac12 \left[1
-\frac1{2m_\tau}\bar u^{S}_h(k')\gamma_5\slashed{\cal P}u^{S}_h(k')\right]=\frac12(1+h\,{\cal P}\cdot S).
\label{eq:probS}
\eea
where we have used that $\bar u^{S}_h(k')\gamma_5\slashed{\cal P}u^{S}_h(k')=
-2m_\tau h\, {\cal P}\cdot S$.\footnote{ It is obtained by replacing  $\bar u^{S}_h (k')$ and $u^{S}_h (k')$ by  $h\,\bar u^{S}_h (k')\gamma_5 \slashed{S}$ and $h\, \gamma_5 \slashed{S} u^{S}_h(k')$ respectively.} The same result also leads  to
\begin{eqnarray}
 {\overline{\sum_{rr'}}\, |{\cal M}|^2 } &=&  \bar u^S_h(k') {\cal O} u^S_h(k') = \bar u^S_h(k') {\cal O} \frac{\slashed{k'}+m_\tau}{2m_\tau}u^S_h(k')
 \nonumber\\
 &=& \frac1{2m_\tau}
\sum_{h'=\pm1} \bar u^S_{h'}(k') {\cal O}(\slashed{k'}+m_\tau) \Big(\frac{1+h\gamma_5\slashed{S}}{2}\Big)u^S_{h'}(k')\nonumber\\
 &=& \frac{1}{2m_\tau}{ \rm Tr}\Big[(\slashed{k'}+m_\tau){\cal O}(\slashed{k'}+m_\tau)\Big(\frac{1+h\gamma_5\slashed{S}}{2}\Big)\Big] \nonumber \\
 &=&\frac12 { \rm Tr}\left[(\slashed{k'}+m_\tau){\cal O}\right]\left(1+h{\rm Tr}[\bar\rho\gamma_5\slashed{S}]\right) \nonumber \\
 &=&\frac12 {\rm Tr}\left[(\slashed{k'}+m_\tau){\cal O}\right]\left( 1+h\,{\cal P}\cdot S\right).\label{eq:GammafromO}
\end{eqnarray}
Moreover, since $\bar\rho^{\,\prime}_{+1},\bar\rho^{\,\prime}_{-1}\ge 0$ and   ${\rm Tr}[\bar \rho]=(\bar\rho^{\,\prime}_{+1}+\bar\rho^{\,\prime}_{-1}) =1$, we have that  ${\cal P}^2$ is then limited to the interval
\be
-1\le {\cal P}^2=-(\bar\rho^{\,\prime}_{-1}-\bar\rho^{\,\prime}_{+1})^2\le 0.
\ee
The case ${\cal P}^2=0$ implies ${\cal P}^\mu=0$ and it corresponds to the physical situation in which the 
emitted $\tau$ is unpolarized, i.e., the probability of 
measuring any polarization state is the same and equal to $\frac12$. 
The case ${\cal P}^2=-1$ corresponds to a fully polarized $\tau$, and either 
$\bar\rho^{\,\prime}_{-1}=0$ or $\bar\rho^{\,\prime}_{+1}=0$,  and  the $\tau$ is  
produced in the $u^{N'}_{+1}(k')$  or  the $u^{N'}_{-1}(k')$ eigenstates, respectively. 
The case with $-1<{\cal P}^2<0$ corresponds to a partial polarization scenario, in which the $\tau$ is produced in an admixture of the $u^{N'}_{+1}(k')$ and $u^{N'}_{-1}(k')$ states, with probabilities given by $\bar\rho^{\,\prime}_{+1}$ and $\bar\rho^{\,\prime}_{-1}$ respectively. This latter interpretation is substantiated by the following result
\bea
P[u(k')]=\frac1{2m_\tau}\bar u(k')\,\bar\rho\, u(k')=\bar\rho^{\,\prime}_{+1}\bigg|
\frac{\bar u(k')u^{N'}_{+1}(k')}{2m_\tau}\bigg|^2+\bar\rho^{\,\prime}_{-1}\bigg|
\frac{\bar u(k')u^{N'}_{-1}(k')}{2m_\tau}\bigg|^2,
\eea
that gives the probability of finding the $\tau$ in a  $u(k')$ state as a sum over the probabilities that the $\tau$ is produced in the $u^{N'}_{\pm1}(k')$ states times the probabilities that, upon measurement, the latter are found in the $u(k')$ state.

\section{Tau polarization vector for $H_b\to H_c \tau^-\bar\nu_\tau$ decays in the presence of NP}
\label{sec:eh}

We shall consider  the  general effective  Hamiltonian 
\bea
H_{\rm eff}&=&\frac{4G_FV_{cb}}{\sqrt2}[(1+C_{V_L}){\cal O}_{V_L}+
C_{V_R}{\cal O}_{V_R}+C_{S_L}{\cal O}_{S_L}+C_{S_R}{\cal O}_{S_R}
+C_{T}{\cal O}_{T}]
\label{eq:hnp}
\eea
that is discussed in detail for instance in Ref.~\cite{Murgui:2019czp}.  The fermionic 
operators involve only neutrino left-handed currents, while the, complex 
in general, Wilson coefficients $C_i$ parameterize possible deviations from 
the SM, the latter given by  the $4G_F V_{cb}{\cal O}_{V_L}/\sqrt2 $ term. 
The Wilson coefficients could be lepton and flavor dependent, though normally they are 
assumed to be present only for the third quark and lepton generations, where 
anomalies have been seen.

In terms of the above effective Hamiltonian the invariant amplitude for the
$H_b\to H_c\tau\bar\nu_\tau$ process is written as~\cite{Penalva:2020xup}
\bea
{\cal M} = J_{H}^\alpha J^{L}_\alpha+ J_{H} J^{L}+ 
J_{H}^{\alpha\beta} J^{L}_{\alpha\beta}
\eea
The  lepton currents are given by 
\be
J^{L}_{(\alpha\beta)}(k,k';h) = \frac{1}{\sqrt{8}} 
\bar u_h^S (k') \Gamma_{(\alpha\beta)} 
 (1-\gamma_5) v_{\nu_\tau}(k)\,, 
\quad  \Gamma_{(\alpha\beta)}= 1,\gamma_\alpha,\sigma_{\alpha\beta}\, \label{eq:lepton-current}
 \ee
with $k$ the final antineutrino four-momentum and where $h=\pm 1$ stands for the two possible $\tau$ lepton polarizations (covariant spin)
along a certain four vector $S^\mu$ that we choose to measure in the experiment. The dimensionless hadron currents read (here $c(x)$ and $b(x)$ are Dirac  fields in coordinate space),
\bea
J_{Hrr'}^{(\alpha\beta)}(p,p') &=&  \langle H_c; p',r'| \bar c(0) 
O_H^{(\alpha\beta)}b(0) | H_b; p, r\rangle, \nonumber  \\
 O_H &=& C_S- C_P \gamma_5,\, O_H^\alpha=\gamma^\alpha(C_V- C_A \gamma_5),
 \, O_H^{\alpha\beta}= C_T\sigma_{\alpha\beta} (1-\gamma_5), \label{eq:Jh}
\eea
with $C_{V,A}=(1+C_{V_L}\pm C_{V_R})$,  $C_{S,P}=(C_{S_L}\pm C_{S_R})$ and hadron states  normalized as $\langle \vec{p}\,', r'| \vec{p},
r\rangle= (2\pi)^3(E/M)\delta^3(\vec{p}-\vec{p}\,')\delta_{rr'}$, with $r,r'$ 
 polarization indexes. In addition,  $p$ and $p'$ are the four-momenta of the initial and final hadrons, respectively. 

Summing/averaging over the final/initial hadron polarizations one can 
identify the ${\cal O}$ operator in Eq.~(\ref{eq:Ooper}) to be
\bea
{\cal O}=\frac14 \sum_{(\alpha\beta)}\sum_{(\rho\lambda)} \Gamma_{(\alpha\beta)}(1-\gamma_5)\slashed{k}\gamma^0\Gamma^\dagger_{(\rho\lambda)}\gamma^0\Big[\,
\overline{\sum_{rr'}}J^{(\alpha\beta)}_{H rr'}(p,p')
J^{(\rho\lambda)\dagger}_{H rr'}(p,p')\,\Big].
\eea
While this can be used to obtain the $\tau$ polarization vector ${\cal P}^\mu$ through  Eq.~(\ref{eq:rhobardef}) and the last of Eq.~(\ref{eq:pproperties}), 
in fact this work was already done  in Ref.~\cite{Penalva:2020xup}, 
where it was found that for a final  $\tau$ with well defined helicity 
$h$ one has\footnote{We use the notation $\epsilon^{\tilde sk'qp}=
\epsilon^{\mu\nu\alpha\beta}\tilde s_\mu k'_\nu q_\alpha p_\beta$, and take $\epsilon_{0123}=+1$.}
\begin{eqnarray}
\frac{2}{M^2} \overline{\sum_{rr'}} |{\cal M}|^2 &= &
{\cal N}(\omega, p\cdot k) + h\bigg\{ \frac{(p\cdot \tilde s)}{M}\,
{\cal N_{H_{\rm 1}}}(\omega, p\cdot k) +\frac{(q\cdot \tilde s)}{M}\,
{\cal N_{H_{\rm 2}}}(\omega, p\cdot k)\nonumber\\
&&+\frac{\epsilon^{\tilde s k' qp}}{M^3}\,{\cal N_{H_{\rm 3}}}(\omega, p\cdot k)
 \ \bigg\},\label{eq:pol}
\end{eqnarray}
where $q=p-p'=k+k'$ is the four-momentum transferred and  $\omega$ is the 
product  of the initial and final hadron four-velocities (related to the 
invariant mass squared of the outgoing $\tau\bar\nu_\tau$  pair via 
$q^2=M^2+M^{\prime2}
-2MM'\omega)$, which varies from 1 to $\omega_{\rm max}= (M^2+M'^2-m^2_{\tau})
/(2MM')$. We note that  the term in ${\cal  N_{H_{\rm 3}}}(\omega, p\cdot k)$
   was not explicitly shown in  Ref.~\cite{Penalva:2020xup} 
since for  the CM and LAB frames considered in that work 
one has  $\epsilon^{\tilde sk'qp}=0$ for  $\widetilde s^\mu  = (|\vec k\,'|, k^{\prime 0}\hat k')/ m_\tau$. The ${\cal N}$ and  $\cal N_{H_{\rm 123}}$ scalar functions are given by
 \bea
 {\cal N}(\omega, k\cdot p)&=& \frac1{M^2}{\rm Tr}\left[(\slashed{k'}+m_\tau){\cal O}\right]= \frac12\Big[{\cal A}(\omega)
+{\cal B}(\omega) \frac{(k\cdot p)}{M^2}+ {\cal C}(\omega) 
\frac{(k\cdot p)^2}{M^4}\Big],\nonumber\\
{\cal N_{H_{\rm 1}}}(\omega, k\cdot p)&=&{\cal A_H}(\omega)
+ {\cal C_H}(\omega)
 \frac{(k\cdot p)}{M^2},\nonumber\\
{\cal N_{H_{\rm 2}}}(\omega, k\cdot p)&=& {\cal B_H}(\omega)
  + {\cal D_H}(\omega) \frac{(k\cdot p)}{M^2}+ {\cal E_H}(\omega) 
  \frac{(k\cdot p)^2}
  {M^4},\nonumber\\ 
 {\cal  N_{H_{\rm 3}}}(\omega, k\cdot p)&=&{\cal F_H}(\omega)+ 
 {\cal G_H}(\omega)\frac{(k\cdot p)}{M^2}.\label{eq:pol2}
  \eea
 The ten functions,  ${\cal A}, {\cal B}$, ${\cal C}$, ${\cal A_H}, {\cal B_H}, 
{\cal C_H}, {\cal D_H}$, ${\cal E_H}$, ${\cal F_H}$ and  ${\cal G_H}$, above are linear combinations  of the 16 Lorentz scalar structure functions (SFs)   introduced in  
Ref.~\cite{Penalva:2020xup}, and denoted as $\widetilde W's$ in that work. 
These $\widetilde W's$ SFs describe the hadron input to the decay, and they are constructed out of the NP complex Wilson coefficients 
 ($C's$) and the  genuine hadronic responses ($W's$). The latter are expressed in terms of 
 the form-factors  used to parameterize the matrix elements of  the hadron operators. Symbolically, we have $\widetilde W = C W $.  The functions ${\cal A},{\cal B},{\cal C}$ and 
 ${\cal A_H}, {\cal B_H}, {\cal C_H}, {\cal D_H}, {\cal E_H}$ in Eq.~\eqref{eq:pol2} are given in  Appendix D of Ref.~\cite{Penalva:2020xup}. As for ${\cal F_H}$  and ${\cal G_H}$ they read
\begin{eqnarray}
 {\cal F_H}(\omega)&=&4\,{\rm Im\,}\bigg[
 \frac{\widetilde W_{I1}}{4} +\frac{m_\tau}{M}\widetilde W_{I3}+\frac{p\cdot q}{M^2}\widetilde W_{I4}
 +\frac{m^2_\tau}{M^2}\widetilde W_{I5}- \widetilde W_{I6} \bigg], \nonumber \\  
 {\cal G_H}(\omega)&=&-8\,{\rm Im\,}\big[\widetilde W_{I4}\big] \label{eq:FyG}
\end{eqnarray}
where the  involved $\widetilde W_{Ii}$ SFs  are also defined in 
Ref.~\cite{Penalva:2020xup}.  Now, from Eqs.~(\ref{eq:pol}) and \eqref{eq:GammafromO} 
(or equivalently  Eq.~(\ref{eq:probS})), the latter particularized for $S=\tilde s$, 
one  immediately  gets  
\bea
{\cal P}^\mu=\frac1{{\cal N}(\omega, k\cdot p)}\bigg[\ \frac{p^\mu_\perp}{M}
{\cal  N_{H_{\rm 1}}}(\omega, k\cdot p)+\frac{q^\mu_\perp}
 {M}{\cal  N_{H_{\rm 2}}}(\omega, k\cdot p) +\frac{\epsilon^{\mu k'qp}}{M^3}
 {\cal  N_{H_{\rm 3}}}(\omega, k\cdot p)
 \bigg],
\label{eq:PNP}
\eea
with $\ell_\perp= [\ell-(\ell\cdot k'/m_\tau^2) k^{\prime}]$ ($\ell = p,q$), which appears because we have removed the projection of $p$ and $q$ along $k'$ since 
${\cal P}^\mu$ is orthogonal to $k^{\prime\mu}$.

As can be seen from the general results of  Ref.~\cite{Penalva:2020xup}, 
the $\widetilde W$ SFs present in ${\cal  N_{H_{\rm 3}}}$ 
are generated from the interference of vector-axial with scalar-pseudoscalar terms ($\widetilde W_{I1}$), scalar-pseudoscalar with tensor terms ($\widetilde W_{I3}$), and vector-axial with tensor terms ($\widetilde W_{I4,I5,I6}$). Since the vector-axial terms are already present in the SM,  at least one of 
the $C_S,C_P,C_T$ Wilson coefficients must be nonzero for ${\cal  N_{H_{\rm 3}}}$ to be nonzero. Besides, 
${\cal N_{H_{\rm 3}}}$ is proportional to the imaginary part of SFs, which 
requires complex   Wilson coefficients, thus incorporating
violation of the CP symmetry in the NP effective Hamiltonian. This feature makes the study of such contribution to the polarization vector of special relevance and it has been discussed before in the context of $\bar B \to D^{(*)}$ decays~\cite{Tanaka:1994ay,Ivanov:2017mrj}. Moreover for $\bar B \to D^*$, some CP-odd observables, defined using angular distributions involving the kinematics of the products of the $D^*$ decay,  have been also presented~\cite{Duraisamy:2013pia, Duraisamy:2014sna, Ligeti:2016npd, Bhattacharya:2020lfm}. These are known as the CP violating triple product asymmetries, which should be sensitive to the relative phases of the Wilson coefficients, as the ${\cal F_H}$ and ${\cal G_H}$ scalar functions are.

We note that the knowledge of the ten functions  ${\cal A}, {\cal B}$, 
${\cal C}$, ${\cal A_H}, {\cal B_H}, 
{\cal C_H}, {\cal D_H}$, ${\cal E_H}$, ${\cal F_H}$ and  ${\cal G_H}$ 
fully determines $\overline{\sum}_{rr'}|{\cal M}|^2$, obtained after summing/averaging  all spin third components of all particles except the $\tau$ lepton. These functions contain then the 
maximum information on NP that can be inferred by analyzing  the $H_b\to H_c\tau\bar\nu_\tau$ decay. 
As discussed in Ref.~\cite{Penalva:2020xup}, for a fixed value of $\omega$,  ${\cal A}(\omega), {\cal B}(\omega)$ and ${\cal C}(\omega)$  can be 
indistinctly obtained by looking
at the dependence on $\cos\theta_\ell$ or on $E_\ell$ of the CM $d^2\Gamma/(d\omega d\cos\theta_\ell)$ or the LAB 
$d^2\Gamma(d\omega dE_\ell)$ unpolarized differential decay widths, respectively. To obtain all the rest of CP-conserving ${\cal A_H}, {\cal B_H}, {\cal C_H}, {\cal D_H}$ and ${\cal E_H}$ functions, it is however necessary to simultaneously use the $\cos\theta_\ell$  and $E_\ell$
  dependencies of the $ \tau$-helicity polarized CM and LAB distributions, which provide complementary information.  Since those two
distributions do not depend on ${\cal F_H}(\omega)$ and  ${\cal G_H}(\omega)$, further measurements 
are needed to obtain these two latter CP odd quantities.

\subsection{Parity and time-reversal violations in the decay width}
\label{sec:ptv}

Note that, in the most general case reflected in Eq.(\ref{eq:PNP}), ${\cal P}^\mu$ 
contains both vectors and pseudovectors and then it does 
not have  well defined properties under parity and time reversal 
transformations\footnote{The different terms of ${\cal P}^\mu$ in 
Eq.~\eqref{eq:PNP} behave under these symmetries as deduced from their 
momentum content and taking into account that for both type of transformations $\ell^\mu \to \ell_\mu$,
 with $\ell=p,q$ or $k'$.}.  This will give rise to parity and time-reversal 
 violating contributions to the probability $P[u^S_h(k')] \propto (1+h\,
 {\cal P}\cdot S)$ or equivalently in the decay width. To see that  we also  
 need  to know how  $hS^\mu$ transforms under parity ([$hS]^{P\mu}$) and time 
 reversal ([$hS]^{T\mu}$). By using $\gamma_5(h\slashed{S})u^S_h(k')
 =u^S_h(k')$, we find~\cite{Itzykson:1980rh}
\bea
[u^S_h(k')]^P&= &\gamma^0u^S_h(k')=\gamma^0\gamma_5\gamma_\mu(hS^\mu)u^S_h(k')=
\gamma_5\gamma^\mu(-hS^\mu)
\gamma^0u^S_h(k')\nonumber \\
&=&\gamma_5[h\slashed{S}]^P\, [u^S_h(k')]^P
,\\
\protect{[}u^S_h(k')]^T&=&\tau[u^S_h(k')]^*=\tau\gamma_5^*\gamma_\mu^*(hS^\mu)[u^S_h(k')]^*=
\gamma_5\gamma^\mu(hS^\mu)
\tau[u^S_h(k')]^*\nonumber \\
&=&\gamma_5(h\slashed{S})^T
[u^S_h(k')]^T\, , \qquad \tau= i \gamma_5 C =\gamma_5\gamma^0\gamma^2.
\eea
where we have ignored  possible overall phases, that do not affect the  
transformation properties of $hS^\mu$, and we have
 used that $\gamma^0 \gamma_\mu \gamma^0 = \gamma^\mu$ 
 and $\tau \gamma_5^*\gamma_\mu^* \tau^{-1}= \gamma_5\gamma^\mu$. 
 Finally, we deduce
\bea
[hS]^{P\mu}=-hS_\mu \, , \quad
[hS]^{T\mu}=hS_\mu.
\eea
We conclude that the quantity $(1+h\,{\cal P}\cdot S)$, and hence the polarized differential decay width, is not invariant under parity
 due to the presence of the 
$p^\mu_\perp$ and 
$q^\mu_\perp$ terms in ${\cal P}^\mu$. 
Similarly, $(1+h\,{\cal P}\cdot S)$ is not invariant under time reversal  
due to the presence of the  $\epsilon^{\mu k' qp}$ 
contribution in ${\cal P}^\mu$.  This latter result is  expected since, as noted above, the 
very existence  of the $\epsilon^{\mu k'qp}$ 
term in ${\cal P}^\mu$ relies on some of the Wilson coefficients not being real\footnote{
Strictly speaking, what one needs is that not all of them are relatively real.}.
\subsection{Different components of the  polarization vector}
\label{sec:cpv}
In this section we are interested in giving a decomposition of the 
polarization vector in the  CM and LAB reference systems in which either the final pair of two leptons (CM) or the initial hadron (LAB) are at rest. For both frames, we choose as an orthogonal basis of the four-vector Minkowski space 
\bea
N_0^\mu&=&\frac{k^{\prime\mu}}{m_\tau}\,,\qquad N_L^\mu=\tilde s^\mu=\Big(\frac{|\vec k\,'|}{m_\tau},\frac{k^{\prime0}\vec k\,'} 
{m_\tau|\vec k\,'|}\Big) ,\nonumber \\
N_{T}^\mu&=&\Big(0,\frac{(\vec k\,'\times\vec p\,')\times\vec k\,'}
{|(\vec k\,'\times\vec p\,')\times\vec k\,'|}\Big)\,,\qquad
N_{TT}^\mu=\Big(0,\frac{\vec k\,'\times\vec p\,'}
{|\vec k\,'\times\vec p\,'|}\Big)\,,
\eea
where the vectors used in their construction are understood to be measured in 
the corresponding frame. Note that $N_L^\mu,\,N_T^\mu$ and $N_{TT}^\mu$ define polarization states corresponding to 
$\vec n_L=\vec k\,'/|\vec k\,'|$, $\vec n_T=
[(\vec k\,'\times\vec p\,')\times\vec k\,']/
|(\vec k\,'\times\vec p\,')\times\vec k\,'|$  and 
$\vec n_{TT}= (\vec k\,'\times\vec p\,')/
|\vec k\,'\times\vec p\,'|$, respectively. Since ${\cal P}\cdot k'=0$, we will have that in a given reference system
\bea
{\cal P}^\mu={\cal P}_L\, N_L^\mu+{\cal P}_T\, N_T^\mu+{\cal P}_{TT}\, N_{TT}^\mu\,, \qquad {\cal P}_a\ = -({\cal P} \cdot N_a),\,a=L,T,TT
\eea
Note that the quantity
\bea
{\cal P}^ 2=-({\cal P}^ 2_{T}
+{\cal P}^ 2_{TT}+{\cal P}^ 2_{L}),
\eea
which gives the degree of polarization of the $\tau$, is a true scalar under  
Lorentz transformations as can be inferred from Eq.~(\ref{eq:PNP}).
However, the ${\cal P}_L$ and ${\cal P}_{T}$ components 
are different in the two frames. This derives from the fact that 
$N^{\rm CM\, \mu}_{L,T} \ne  \Lambda^\mu_{\  \nu} N^{\rm LAB\, \nu}_{L,T}$, 
with $\Lambda$ the boost which takes four-momenta from the LAB system 
to the CM one. This is so because the corresponding auxiliary three  
vectors $\vec n^{\, \rm LAB,CM}_{L,T}$ depend on the reference frame. On the other hand,
${\cal P}_{TT}$ is  the same in the two systems since it is a component
perpendicular to the velocity $\vec p_{\rm LAB}^{\,\prime}/(M-M'\omega)$ defining the 
LAB-to-CM boost. Indeed, in this case $\vec n^{\,\rm CM}_{TT}=\vec n^{\,\rm LAB}_{TT}$, 
because $\vec p_{\rm CM}^{\,\prime}= M \vec p_{\rm LAB}^{\,\prime}/\sqrt{q^2}$ and the 
components of $\vec{k}^{\,\prime}$ orthogonal to the direction $\vec p_{\rm CM}^{\,\prime} / |\vec p_{\rm CM}^{\,\prime}\,| = \vec p_{\rm LAB}^{\,\prime}/ |\vec p_{\rm LAB}^{\,\prime}\,|$ 
are unaltered by the boost\footnote{I.e.,
$
(\vec{k}^{\,\prime}_{\rm LAB}\times\vec p_{\rm LAB}^{\,\prime})/|\vec p_{\rm LAB}^{\,\prime}|=(\vec{k}^{\,\prime}_{\rm CM}\times\vec p_{\rm CM}^{\,\prime})/|\vec p_{\rm CM}^{\,\prime}|.$}. 

What is  true is that 
\bea
{\cal P}^{\rm CM}_a=-{\cal P}^{\rm LAB}_L\,\left[(\Lambda N^{\rm LAB}_{L})\cdot 
N^{\rm CM}_{a}\right]+ {\cal P}^{\rm LAB}_T\,\left[(\Lambda N^{\rm LAB}_{T})\cdot 
N^{\rm CM}_{a}\right],\quad a=L,T,
\label{eq:cmlabrel}
\eea
which trivially follows from 
\be
{\cal P}^{{\rm CM}\, \mu}= \Lambda^\mu_{\cdot\, \nu} {\cal P}^{{\rm LAB}\,\nu} = \Lambda^\mu_{\cdot\, \nu}\left[ {\cal P}_L\, N_L^\nu+{\cal P}_T\, N_T^\nu+{\cal P}_{TT}\, N_{TT}^\nu\right]^{\rm LAB}
\ee
As a consequence,
for a given tau kinematics determined by a pair $(\omega,E_\tau)$ or $(\omega,\cos\theta_\tau)$, one can express the ${\cal P}^{\rm LAB}_{L,T}(\omega,E_\tau)$ as linear combinations of ${\cal P}^{\rm CM}_{L}(\omega,\cos\theta_\tau)$ and ${\cal P}^{\rm CM}_{T}(\omega,\cos\theta_\tau)$\footnote{
Note that the  $(\Lambda N^{\rm LAB}_{b})\cdot 
N^{\rm CM}_{a}$ products are fully determined by the pair of variables $(\omega,E_\tau)$ 
or equivalently by $(\omega,\cos\theta_\tau)$ with  $E_\tau$ and $\cos\theta_\tau$ 
 related via 
\bea
M\left(M_\omega-E_\tau\right)=k\cdot p=
\frac{M}2\Big(1-\frac{m_\tau^2}{q^2}\Big)\left(M_\omega+M'\sqrt{\omega^2-1}
\cos\theta_\tau\right), \label{eq:relacEcso}
\eea
with $M_\omega=M-M'\omega$.}, and   
thus the  LAB and CM ${\cal P}_{L,T,TT}$ components carry the same information. 
Note however that this equivalence is lost for the averages 
$\langle{\cal P}_a^{\rm LAB,\,CM}\rangle(\omega)$ that we discuss below.

In any of the CM or LAB frames, 
${\cal P}_L^{\rm  CM\,, LAB }$ is given by
\bea
{\cal P}_L=-{\cal P}\cdot N_L=-\frac{1}{{\cal N}(\omega,k\cdot p)}
\Big[\frac{p\cdot N_L}M {\cal N_{H_{\rm 1}}}(\omega,k\cdot p)
+\frac{q\cdot N_L}M{\cal N_{H_{\rm 2}}}(\omega,k\cdot p)\Big].\label{eq:PLdef}
\eea
where the appropriate  CM or LAB 
four-vectors should be used in each case.
As previously mentioned, $N_L^\mu$ corresponds to well defined helicity and, thus, ${\cal P}_L$ is related to
the helicity asymmetry via (see Eq.~\eqref{eq:GammafromO})
\bea
{\cal P}_L=-{\cal P}\cdot N_L=\frac{\overline{\sum}_{rr'}\,|{\cal M}(h=-1)|^2
-\overline{\sum}_{rr'}\,|{\cal M}(h=+1)|^2}{\overline{\sum}_{rr'}
\,|{\cal M}(h=-1)|^2
+\overline{\sum}_{rr'}\,|{\cal M}(h=+1)|^2},
\label{eq:plasi}
\eea
where here $h$ stands for the $\tau$ helicity measured in the CM or the LAB frames.
From Eq.~\eqref{eq:defsec}, it is then clear that ${\cal P}_L^{\rm CM\,, LAB }$ can be obtained from the experimental asymmetries 
\be
{\cal P}_L^{\rm CM}=\frac{\frac{d\Gamma(h_{\rm CM}=-1)}{d\omega
d\cos\theta_\tau}
-\frac{d\Gamma(h_{\rm CM}=+1)}{d\omega d\cos\theta_\tau}}{\frac{d\Gamma(h_{\rm CM}=-1)}
{d\omega d\cos\theta_\tau}
+\frac{d\Gamma(h_{\rm CM}=+1)}{d\omega d\cos\theta_\tau}}\, , \qquad
{\cal P}_L^{\rm LAB}=\frac{\frac{d\Gamma(h_{\rm LAB}=-1)}{d\omega dE_\tau}
-\frac{d\Gamma(h_{\rm LAB}=+1)}{d\omega dE_\tau}}{\frac{d\Gamma(h_{\rm LAB}=-1)}{d\omega dE_\tau}
+\frac{d\Gamma(h_{\rm LAB}=+1)}{d\omega dE_\tau}}\, , \qquad
 \label{eq:PLassy}
\ee
where, as already mentioned,  $\cos\theta_\tau$ is the cosine of the angle made by the CM
three-momenta of the final hadron and  $\tau$ lepton, and $E_\tau$ is the energy of 
the $\tau$ lepton in the LAB frame.
While the CM angle $\theta_\tau$ is not restricted,  the LAB energy $E_\tau$ is limited, 
for a given $\omega$ value, to the interval 
defined by  
\be 
E_\tau^{ \pm}(\omega)= 
\frac{M_\omega(q^2+m^2_\tau)  
  \pm M' \sqrt{\omega^2-1}(q^2-{m^2_\tau})}{2q^2}. \label{eq:Elimits}
\ee
%
Similarly, for the CM or LAB systems, one further has
\bea
{\cal P}_{T}&=&-{\cal P}\cdot N_{T}=\frac{-1}{{\cal N}(\omega,k\cdot p)}
\Big[\ \frac{p\cdot N_T}M {\cal N_{H_{\rm 1}}}(\omega,k\cdot p)+
\frac{q\cdot N_T}M {\cal N_{H_{\rm 2}}}(\omega,k\cdot p)\Big)\Big]\label{eq:PTdef},\\
{\cal P}_{TT}&=&-{\cal P}\cdot N_{TT}
=\frac{\epsilon^{k' q\,p\, N_{TT}}}{M^3}\frac{{\cal N_{H_{\rm 3}}}(\omega,k\cdot p)}
{{\cal N}(\omega,k\cdot p)}.\label{eq:ptt}
\eea
Note that both ${\cal P}_{T}$ and ${\cal P}_{TT}$ can also be  obtained from asymmetries of the decay distributions, as in Eq.~\eqref{eq:PLassy}, 
for polarizations along $N_T^\mu$ and $N_{TT}^\mu$ respectively. 

From the discussion above, a nonzero ${\cal P}_{TT}$ component  in the LAB 
or CM frames is a signal for time-reversal violation that originates from  
the presence of non-real
Wilson coefficients in the NP effective Hamiltonian. Since  $N_{TT}^\mu$ 
does not have a zero component,  ${\cal P}_{TT}$ comes from a non vanishing 
projection of the $\tau-$ polarization three-vector in the orthogonal direction to the plane  defined by the outgoing hadron and $\tau$ three-momenta.


Further details on the vector products 
appearing in the evaluation  of ${\cal P}_{L}$, ${\cal P}_{T}$ and ${\cal P}_{TT}$ 
are given in Appendix~\ref{app:cmlabkin}.

In Ref.~\cite{Ivanov:2017mrj},  the name polarization vector components is used for what  actually are averages. Here, we will denote those averages as $\langle{\cal P}_a\rangle(\omega),\,  a=L,T,TT$  and, within our scheme, they  are given by the expressions\footnote{Note that, apart from some differences in the notation, there is  a sign change in the definition we provide here. Besides we extend it to the LAB frame.} 
\bea
\langle{\cal P}^{\rm CM}_a\rangle(\omega)&=&
\frac{1}{{\cal N}_{\theta}(\omega)}\int_{-1}^{+1} d\cos\theta_\tau\, {\cal N}(\omega,k\cdot p)\,{\cal P}^{\rm CM}_a(\omega,k\cdot p), \nonumber \\
\langle{\cal P}^{\rm LAB}_a\rangle(\omega)&=&
\frac{1}{{\cal N}_E(\omega)} \int_{E_\tau^{-}(\omega)}^{E_\tau^{+}(\omega)} dE_\tau\, 
{\cal N}(\omega,k\cdot p)\,{\cal P}_a^{\rm LAB}(\omega,k\cdot p), \nonumber \\
 {\cal N}_{\theta}(\omega) &=&{\int_{-1}^{+1} d\cos\theta_\tau\, {\cal N}(\omega,k\cdot p)}, \quad {\cal N}_{E}(\omega)= \int_{E_\tau^{-}(\omega)}^{E_\tau^{+}(\omega)}
 dE_\tau\,  {\cal N}(\omega,k\cdot p) 
\label{eq:Pkorner1}
\eea
with the normalizations related by ${\cal N}_{E}= (E_\tau^{+}-E_\tau^{-}){\cal N}_{\theta}/2$, and ${\cal N}_{\theta}$ explicitly given in Eq.~\eqref{eq:ntheta}. These averages  correspond to the, easier to measure, experimental asymmetries 
\be
\langle{\cal P}^{\rm CM, \, LAB}_a\rangle(\omega)=\frac{\frac{d\Gamma(h^{\rm CM, \, LAB}_a=-1)}{d\omega
}
-\frac{d\Gamma(h^{\rm CM, \, LAB}_a=+1)}{d\omega }}{\frac{d\Gamma(h^{\rm CM, \, LAB}_a=-1)}
{d\omega }
+\frac{d\Gamma(h^{\rm CM, \, LAB}_a=+1)}{d\omega }},\label{eq:Pkorner2}
\ee
where $h^{\rm CM\,, LAB}_a=\pm1$ stand for positive/negative polarization along $N_a^\mu$ in the CM or LAB system, as appropriate. 
In particular, $\langle {\cal P}^{\rm CM\,, LAB}_{L}\rangle$ is nothing but the $\tau$
polarization asymmetry ${\cal A}^{\rm CM\,,LAB}_{\lambda_\tau}$ also used 
in the literature and evaluated for instance in
 Refs.~\cite{Harrison:2020nrv,Penalva:2020ftd}.

In Appendix~\ref{app:pltttavg} we give expressions for the $\langle{\cal P}^{\rm
CM}_a\rangle(\omega)$ and $\langle{\cal P}^{\rm \,LAB}_a\rangle(\omega)$ averages 
in terms of the scalar functions in
Eq.~(\ref{eq:pol2}). The equivalence between LAB and CM values present for 
the two-dimensional ${\cal P}_{L,T}$, and represented by Eq.~(\ref{eq:cmlabrel}), is now lost for the 
averages $\langle{\cal P}_{L,T}\rangle$, as can be easily be inferred from the expressions
in Appendix~\ref{app:pltttavg}. The reason is that the coefficients of the linear  combinations that relate CM and LAB ${\cal P}_{L,T}$ components depend on the  variable ($\cos\theta_\tau$ or $E_\tau$) which is integrated to obtain the averages.

Therefore, if only the averages $\langle{\cal P}_{L,T}\rangle(\omega)$ are measured, CM and LAB values give complementary information, as we already mentioned above for the case of tau helicity-polarized differential decay distributions.

One can also define the average
$\langle{\cal P}^2\rangle(\omega)$. In this case, it is the same in the CM and LAB frames 
as a consequence of both ${\cal P}^2(\omega,k\cdot p)$ and ${\cal N}(\omega,k\cdot p)$ being scalars.  Actually, $\langle{\cal P}^2\rangle(\omega)$  is a Lorentz invariant and in any reference system, for a given $\omega$, is given by
\bea
\langle{\cal P}^2\rangle(\omega) &=&\int_{-1}^{+1} \frac{d\cos\theta_\tau}{{\cal N}_\theta(\omega)}\, {\cal N}(\omega,k\cdot p)\,{\cal P}^2(\omega,k\cdot p) 
=\int_{E_\tau^{-}(\omega)}^{E_\tau^{+}(\omega)} \frac{dE_\tau}{{\cal N}_E(\omega)}\,{\cal N}(\omega,k\cdot p)\,{\cal P}^2(\omega,k\cdot p) \nonumber \\
&=&\int_{(k\cdot p)_{-}}^{(k\cdot p)_{+}} \frac{d(k\cdot p)}{{\cal N}(\omega)}\, {\cal N}(\omega,k\cdot p)\,{\cal P}^2(\omega,k\cdot p) \label{eq:defP2}
\eea
where $(k\cdot p)_{\rm \pm} = M\left(M_\omega- E_\tau^{\mp}(\omega)\right)$ and 
${\cal N}(\omega)$ is given by
\be
{\cal N}(\omega) =  \int_{(k\cdot p)_{\rm -}}^{(k\cdot p)_{\rm +}} d(k\cdot p)\, {\cal N}(\omega,k\cdot p) = M {\cal N}_E(\omega) = \frac{M (E_\tau^{+}-E_\tau^{-}){\cal N}_\theta(\omega)}{2}
\ee

We conclude the section with the trivial remark 
\be
\langle{\cal P}^2\rangle(\omega) \equiv -\langle{\cal P}^2_L+{\cal P}^2_T+{\cal P}^2_{TT}\rangle(\omega) \ne -\sum_{a=L,T,TT} \left[\langle{\cal P}_a\rangle(\omega)\right]^2 \equiv - |\vec{P}(\omega)|^2, \label{eq:P2noinv}
\ee
with $|\vec P|$ defined for instance in  Ref.~\cite{Ivanov:2017mrj} for the CM frame, and which is not even a Lorentz scalar. 

\section{Numerical Results }
\label{sec:results}
In this section we present the results for ${\cal P}_{L},\,{\cal P}_{T}$ and  
${\cal P}_{TT}$, evaluated for the
 $\Lambda_b\to\Lambda_c$ and $\bar B\to D^{(*)}$ 
 semileptonic decays. The averages introduced in Eq.~(\ref{eq:Pkorner1}) will be presented for 
 the above reactions as well as for the $\bar B_c\to\eta_c,J/\psi$ decays.  We  studied those decays in   Refs.~\cite{Penalva:2019rgt,Penalva:2020xup}
  ($\Lambda_b\to\Lambda_c$) and \cite{Penalva:2020ftd} ($\bar B_c\to\eta_c,J/\psi$ and 
  $\bar B\to D^{(*)}$), where we  analyzed  different observables related to the  unpolarized and helicity-polarized CM $d^2\Gamma/d\omega d\cos\theta_\tau$ and
  LAB $d^2\Gamma/d\omega dE_\tau$   
 distributions, and their possible role in distinguishing  between different
 NP scenarios. In this work, we shall show results for observables 
  mentioned above,  evaluated both in the CM  and LAB frames, and  within the 
 SM and with  the NP Wilson coefficients corresponding to  Fits 6 and 7 of 
 Ref.~\cite{Murgui:2019czp}.  
 
 For the particular case of the $\Lambda_b\to\Lambda_c$ decay, and in order to illustrate the 
 effect of complex Wilson coefficients, we will also show results 
 for one  more NP scenario  from Ref.~\cite{Shi:2019gxi}. It corresponds to 
 a $R_2$ leptoquark mediator model  that only gives contributions to the $C_{S_L}$ and $C_T$ Wilson coefficients
  and that was first analyzed for complex values of those coefficients in Ref.~\cite{
  Becirevic:2018afm}.

 For the $\Lambda_b\to \Lambda_c$ decay we use  form factors that are directly obtained (see Appendix E
 of  Ref.~\cite{Penalva:2020xup}) from those calculated in the  lattice quantum Chromodynamics (LQCD) simulations of Refs.~\cite{Detmold:2015aaa} (vector and axial ones) and \cite{Datta:2017aue} (tensor NP form factors) using $2+1$ flavors of
 dynamical domain-wall fermions.  The NP scalar and pseudoscalar
  form factors   are directly related to the vector and axial ones and we 
use  Eqs.~(2.12) and (2.13) of Ref.~\cite{Datta:2017aue} to evaluate them. We 
use errors and the statistical correlation-matrices, provided in the LQCD papers,
 to Monte Carlo transport the form-factor uncertainties to the different observables 
 shown in this work.  

For the case of $\bar B\to D^{(*)}$ decays, the form factors are  calculated
using a parameterization, based on heavy quark effective theory, that 
includes  corrections of order $\alpha_s$, $\Lambda_{\rm QCD}/m_{b,c}$ and 
partly $(\Lambda_{\rm QCD}/m_c)^2$~\cite{Bernlochner:2017jka}. In this case
 there exist also some experimental 
$q^2-$shape information~\cite{Lees:2013uzd, Huschle:2015rga}, which is used to further constrain some matrix elements.  Inputs from 
 LQCD~\cite{Bailey:2014tva, Lattice:2015rga,
Na:2015kha, Harrison:2017fmw}, light-cone \cite{Faller:2008tr} and QCD sum 
rules ~\cite{Neubert:1992wq,Neubert:1992pn, Ligeti:1993hw} are also available. 
 Here,  we  use the set of form factors 
and Wilson coefficients found in \cite{Murgui:2019czp}, since in that work, 
not only the Wilson coefficients, but also the $1/m_{b,c}$ and $1/m^2_{c}$ corrections
 to the form factors  were simultaneously fitted to experimental data. In this way for these decays, we can also consistently estimate theoretical uncertainties, since we shall use  statistical samples of Wilson coefficients and form
factors, selected such that the $\chi^2-$merit function computed in ~\cite{Murgui:2019czp} changes at most by one unit from its value at the fit
minimum.
 
 For the $\bar B_c\to\eta_c,\,J/\psi$ transitions, there exist no 
 systematic LQCD calculations, except 
 for the very recent work of the HPQCD collaboration~\cite{Harrison:2020gvo} where 
  the SM vector and axial form factors of the 
 $\bar B_c\to J/\psi$ decay have been determined. Here we use  the form factors
  obtained within the non-relativistic quark model  scheme 
 of Ref.~\cite{Hernandez:2006gt}. It has the  advantage of consistency, since all the
 form factors needed can be evaluated within the model. These
form factors  are  consistent with heavy quark spin symmetry  and its expected
  pattern of breaking corrections.  In addition, in Ref.~\cite{Hernandez:2006gt}, five different 
   inter-quark potentials were considered  allowing us  to provide an estimate of the 
  theoretical uncertainties.  We expect the systematic errors present in the NRQM
  evaluation of the form factors should
  largely cancel out in ratios\footnote{ In Ref.~\cite{Penalva:2020ftd} we found
  a  remarkable agreement for ${\cal R}_{J/\psi}=\Gamma(\bar B_c\to
  J/\psi\tau\bar\nu_\tau)/\Gamma(\bar B_c\to
  J/\psi\mu\bar\nu_\mu)$ between our SM results and the
  ones obtained in the lattice calculation of Ref.~\cite{Harrison:2020nrv}. }.

We should mention that in our previous works of Refs.~\cite{Penalva:2020xup, 
Penalva:2020ftd}, we discussed in great detail, for all these decays,  the helicity 
differential distributions obtained in the SM and NP Fits 6 and 7 of 
Ref.~\cite{Murgui:2019czp}. Thus, the analysis presented below for the longitudinal 
${\cal P}_L$ projection shows, using a different language, the same physical content, with the exception
of the results related to the NP tensor $R_2$ leptoquark model fit 
 of Ref.~\cite{Shi:2019gxi}, which were not  considered in \cite{Penalva:2020xup, Penalva:2020ftd}.

However, the study of the transverse component ${\cal P}_{TT}$ carried out here is 
novel, and it directly provides independent physics information (${\cal F_H}$ and 
${\cal G_H}$ SFs in Eqs.~\eqref{eq:FyG} and \eqref{eq:PNP}) to that inferred from our 
previous works. 
In what respects to ${\cal P}_T$, this projection is determined by the scalar functions ${\cal A}$,   ${\cal B}$, ${\cal C}$, ${\cal A_H}$, ${\cal B_H}$, ${\cal C_H}$, ${\cal D_H}$ and ${\cal E_H}$ (see Eq.~\eqref{eq:PNP}), as it occurs with ${\cal P}_L$. As described in detail in \cite{Penalva:2020xup}, all these eight functions can be extracted  from the combined study of the
CM $d^2\Gamma/(d\omega d\cos\theta_\tau)$ and LAB $d^2\Gamma/(d\omega dE_\tau)$ helicity-polarized distributions.
Therefore, though ${\cal P}_T (\omega, k\cdot p)$  can be indirectly obtained from 
the results shown in Refs.~\cite{Penalva:2020xup, Penalva:2020ftd}, this polarization 
projection was not explicitly discussed in any of these works.  

\subsection{CM and LAB two-dimensional distributions}
\begin{figure}[h!!!]
\begin{center}
\includegraphics[scale=.321]{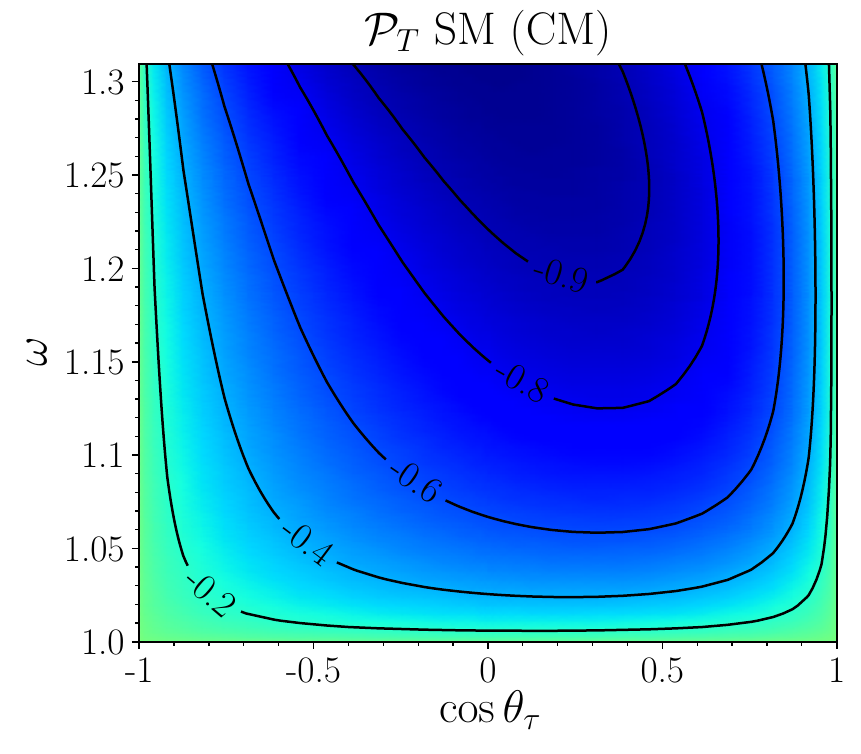}
\includegraphics[scale=.321]{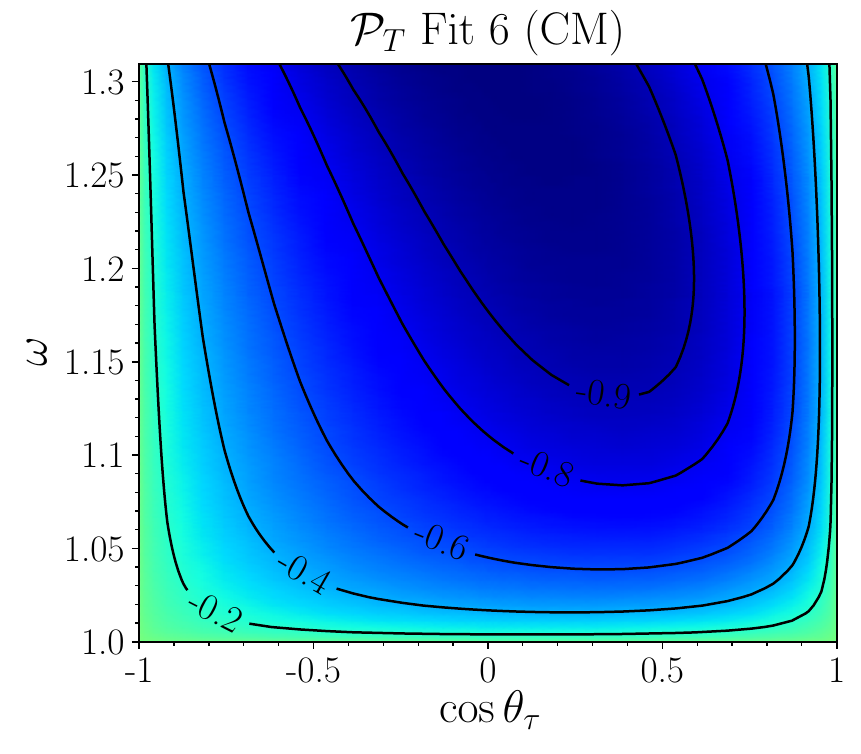} 
\includegraphics[scale=.321]{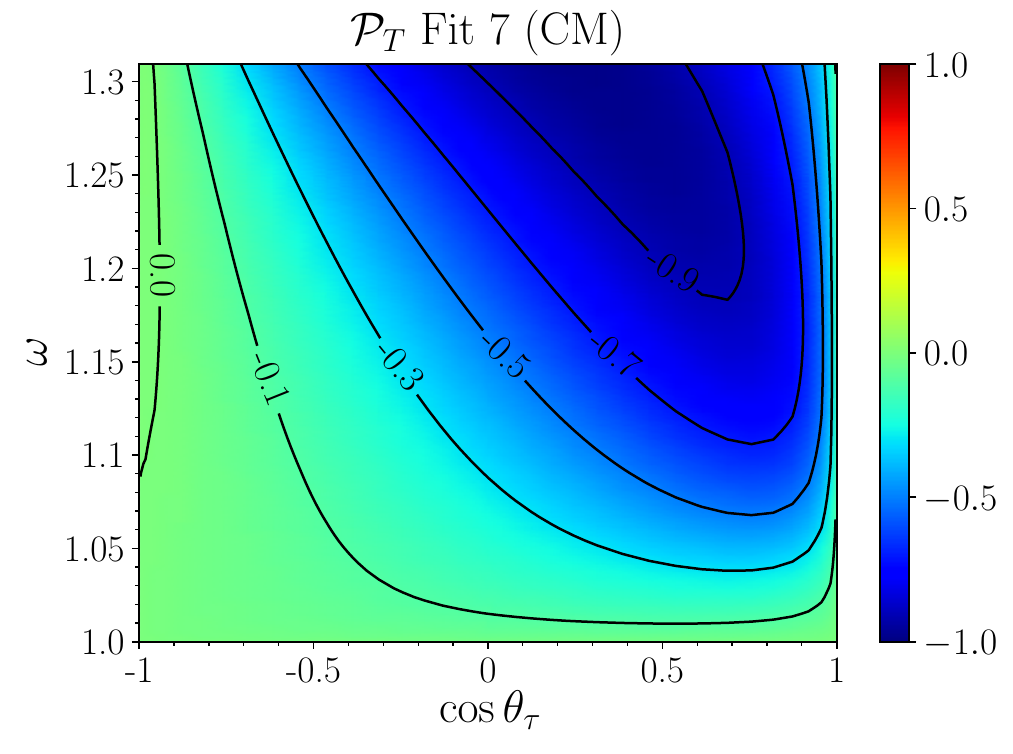}
\\
\includegraphics[scale=.321]{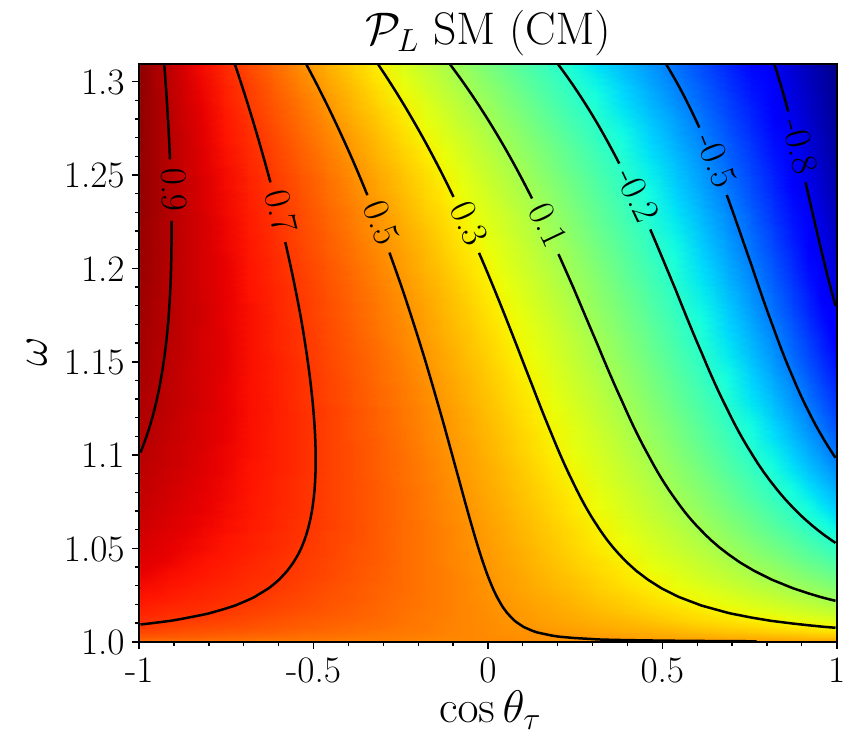}
\includegraphics[scale=.321]{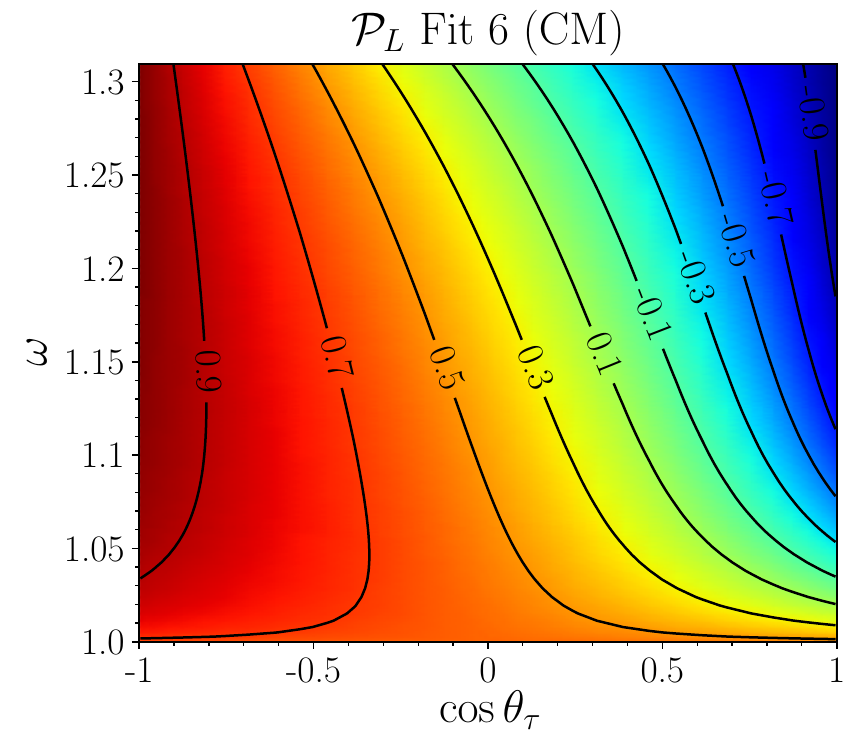} 
\includegraphics[scale=.321]{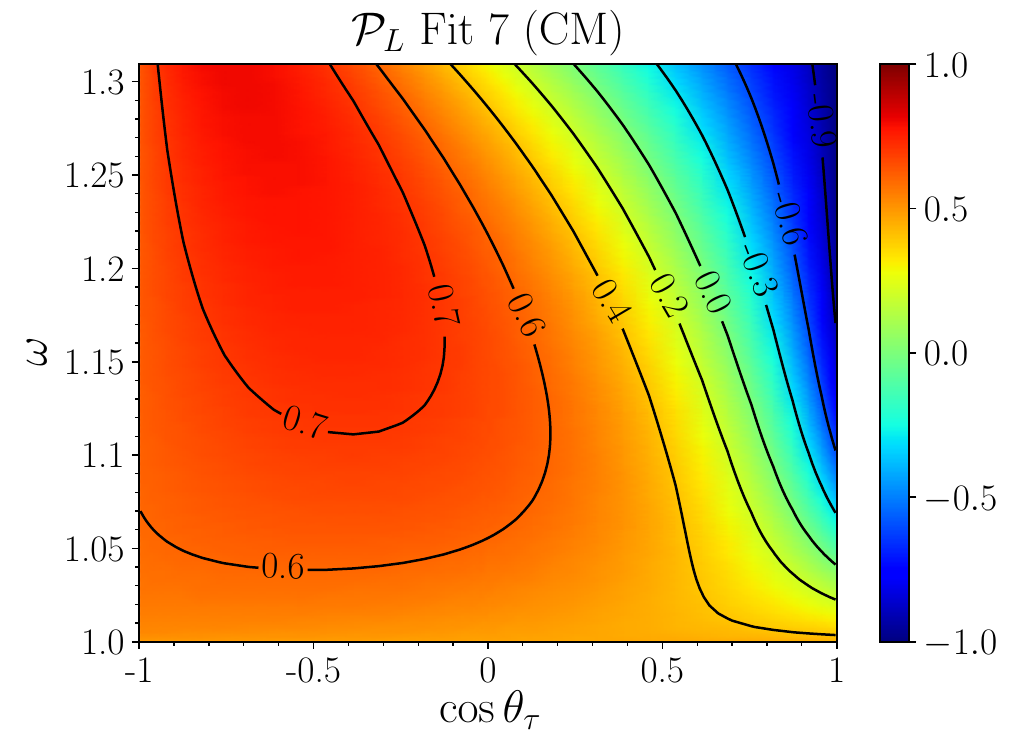}
\\
\includegraphics[scale=.321]{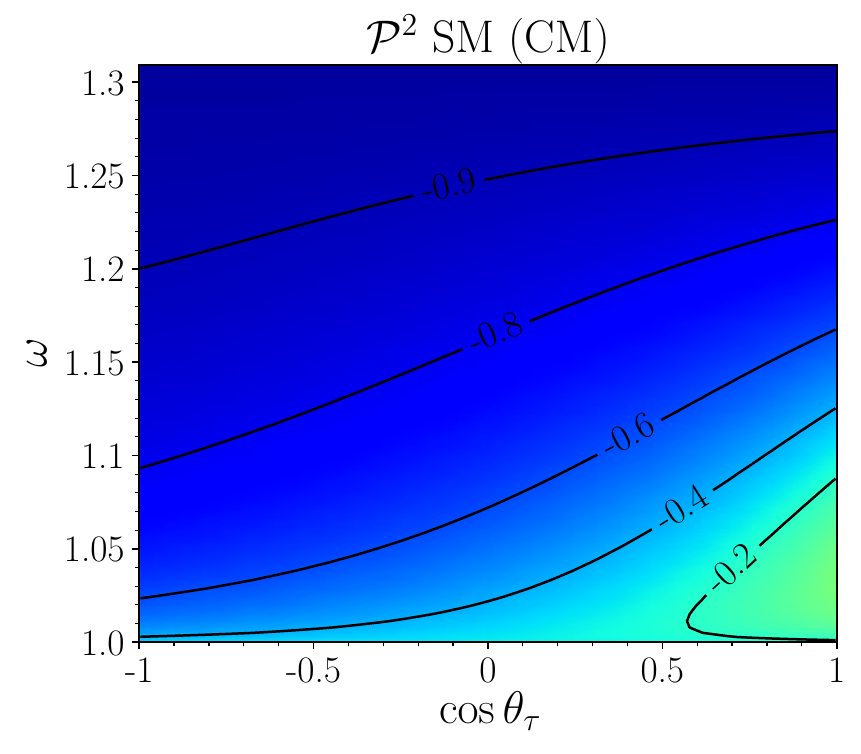}
\includegraphics[scale=.321]{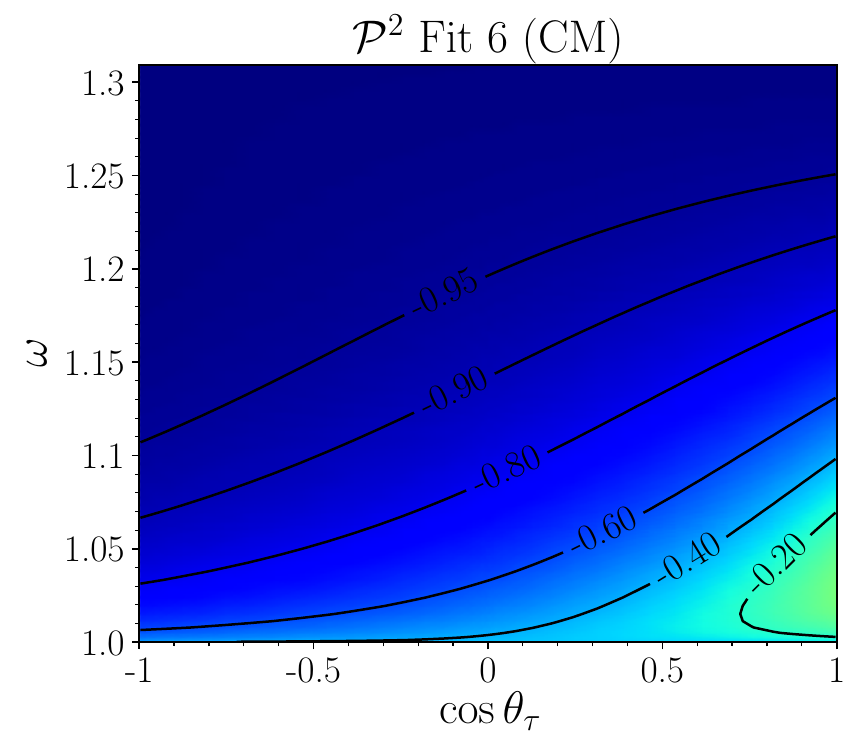} 
\includegraphics[scale=.321]{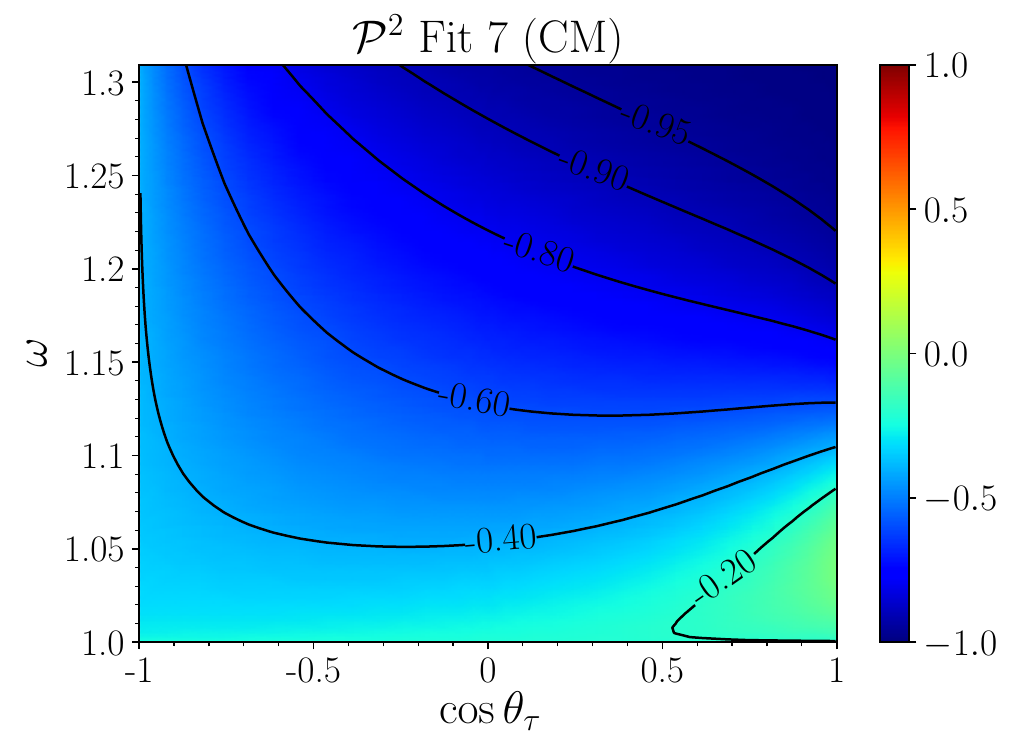}
\caption{ CM
${\cal P}_T$ (first row),  ${\cal P}_{L}$ (second row)  and ${\cal P}^2$ (third row) polarization observables  
 for the $\Lambda_b\to \Lambda_c\tau\bar\nu_\tau$
 decay evaluated within the SM (left column) and with the  NP   Wilson coefficients from 
Fits 6 (middle column) and  7 (right column) 
of Ref.~\cite{Murgui:2019czp}. We display the 2D distributions as a function of the $(\omega, \cos\theta_\tau)$ variables. In all cases,  central values for the form factors and
Wilson coefficients have been used. }  
\label{fig:LambdaPCM}
\end{center}
\end{figure}
\begin{figure}[t]
\begin{center}
\includegraphics[scale=.321]{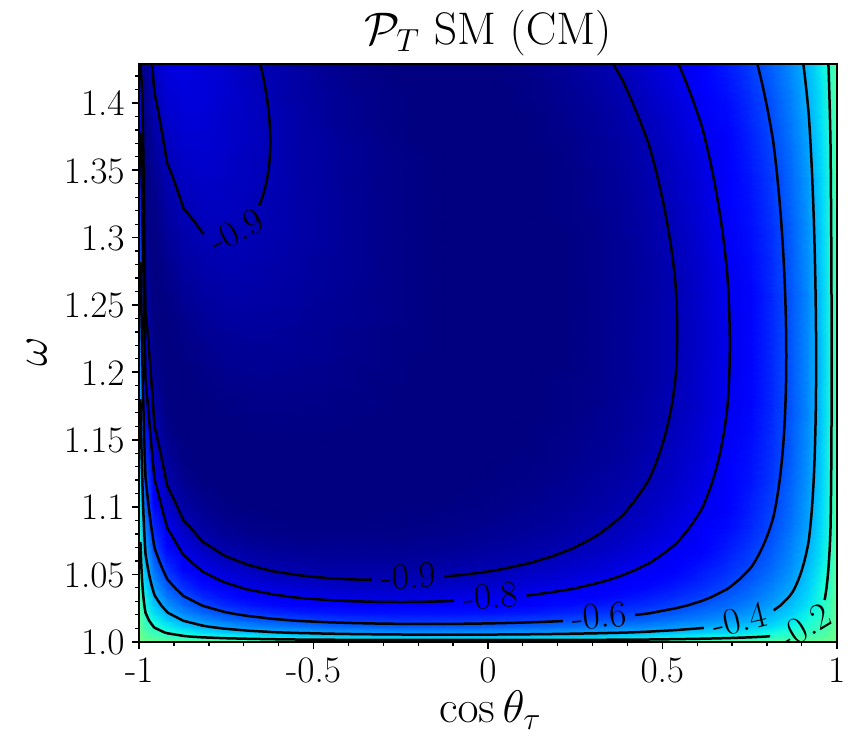}
\includegraphics[scale=.321]{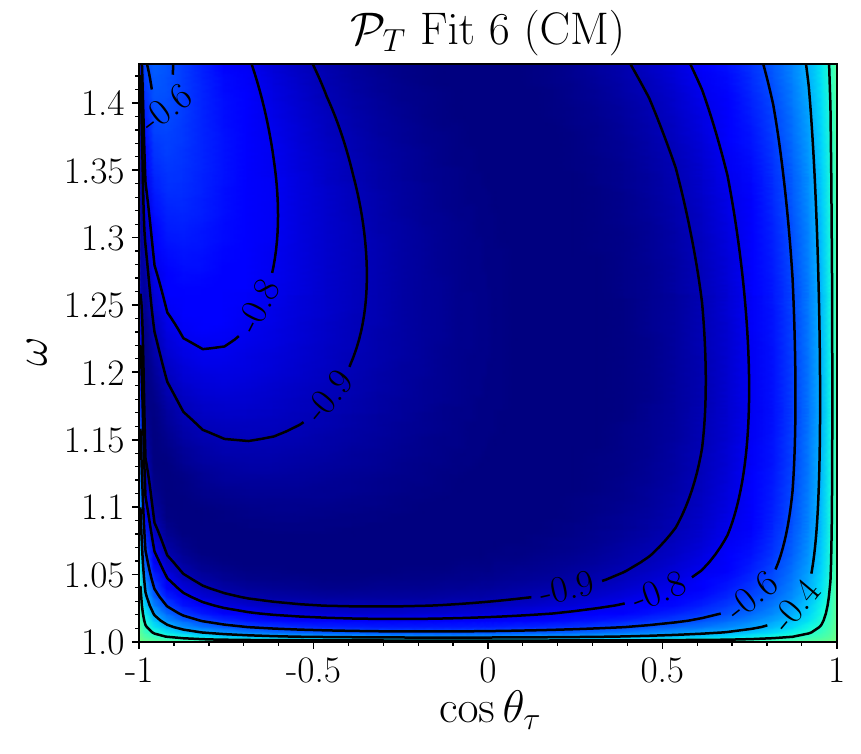} 
\includegraphics[scale=.321]{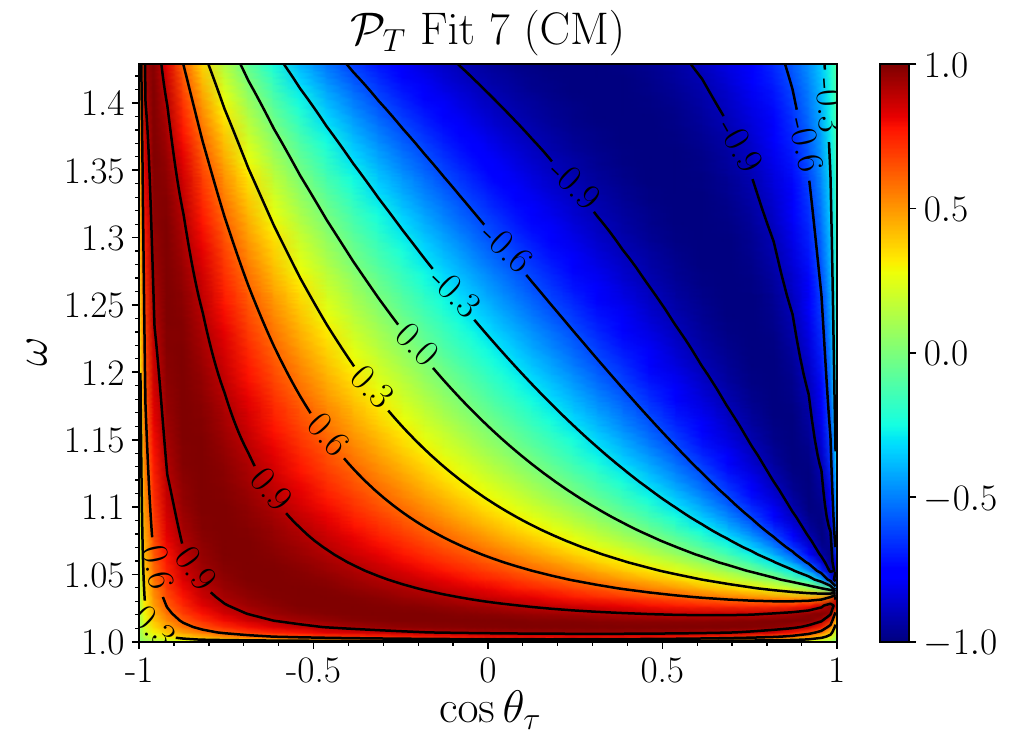}
\\
\includegraphics[scale=.321]{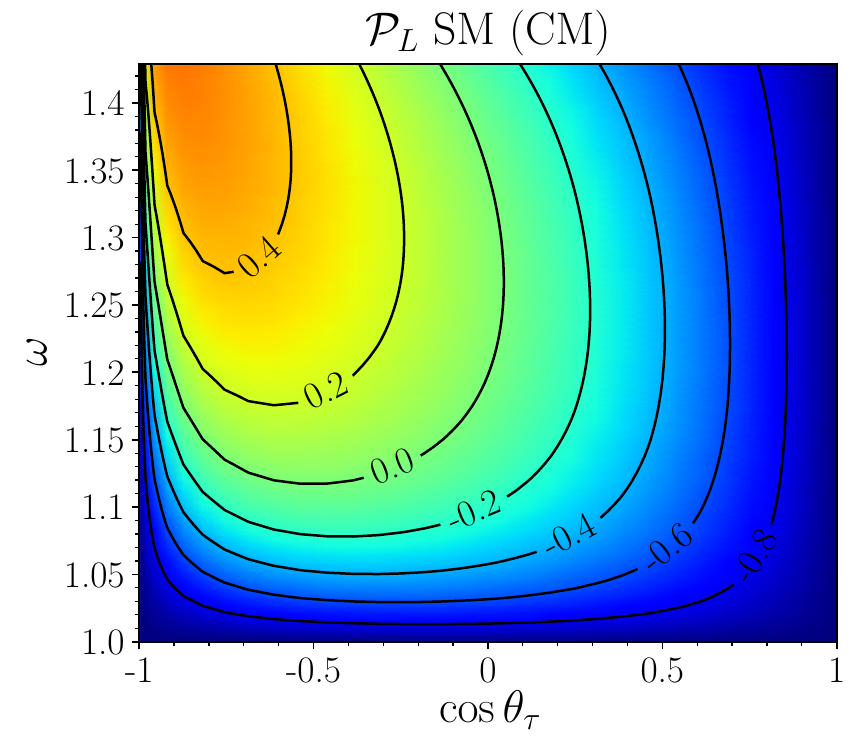}
\includegraphics[scale=.321]{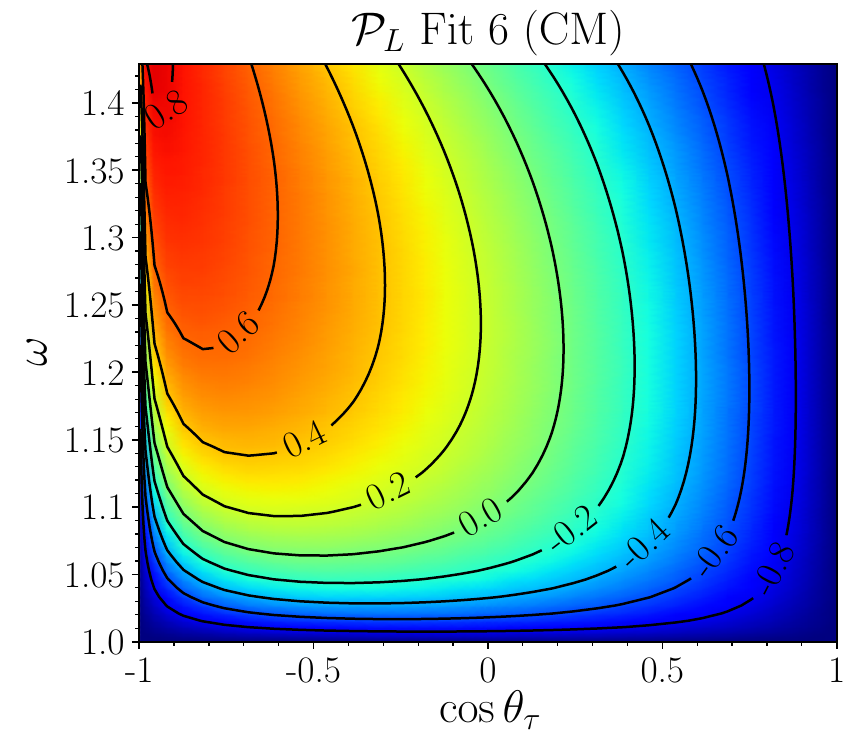} 
\includegraphics[scale=.321]{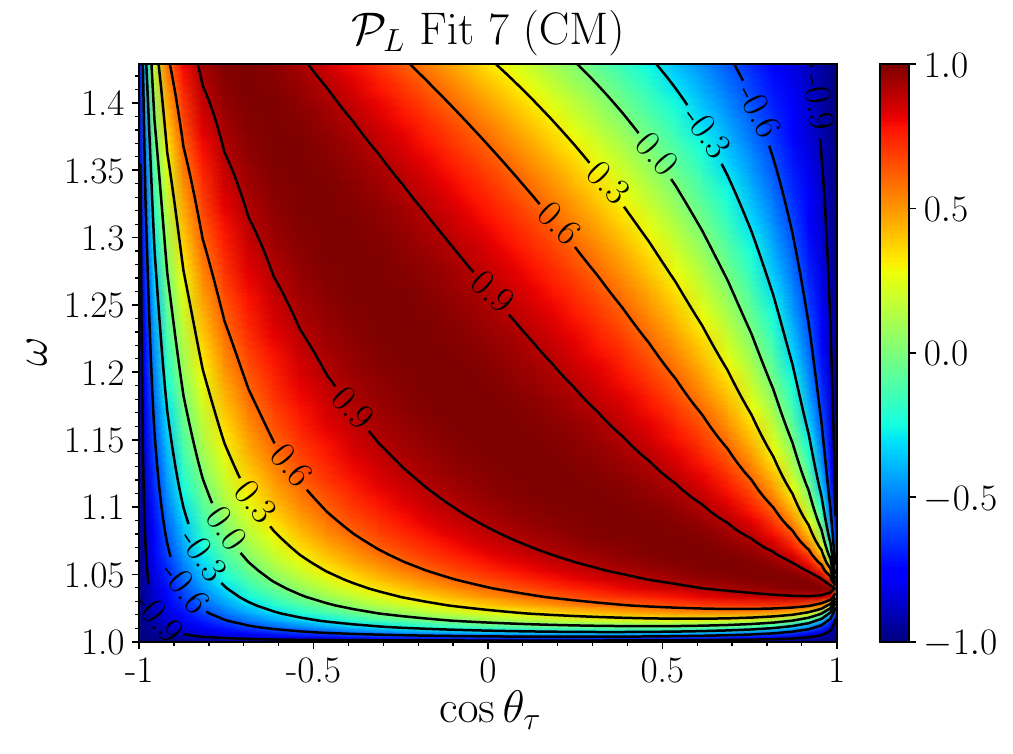}
\\
\includegraphics[scale=.321]{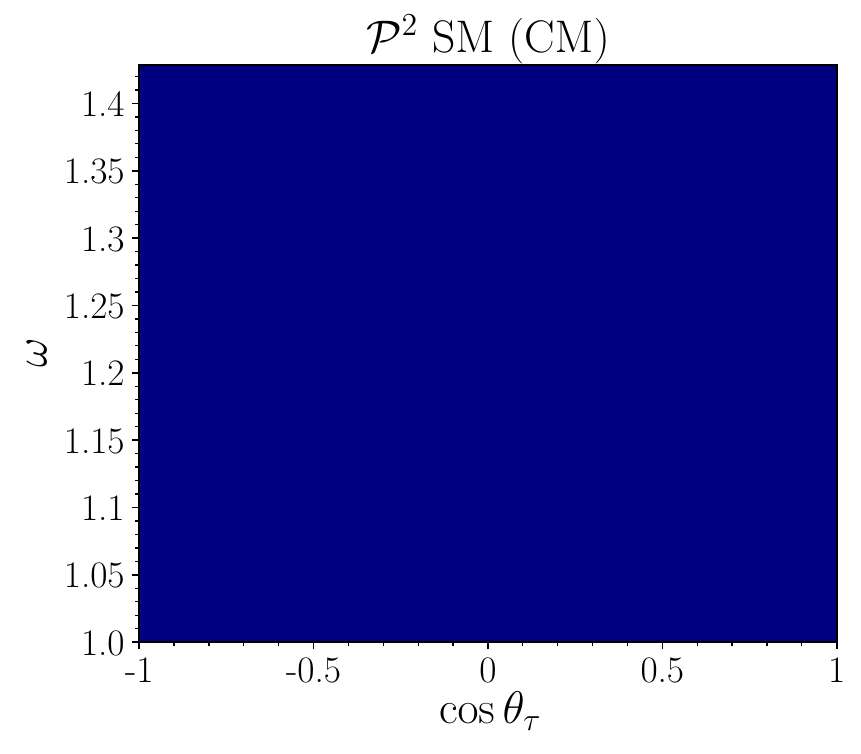}
\includegraphics[scale=.321]{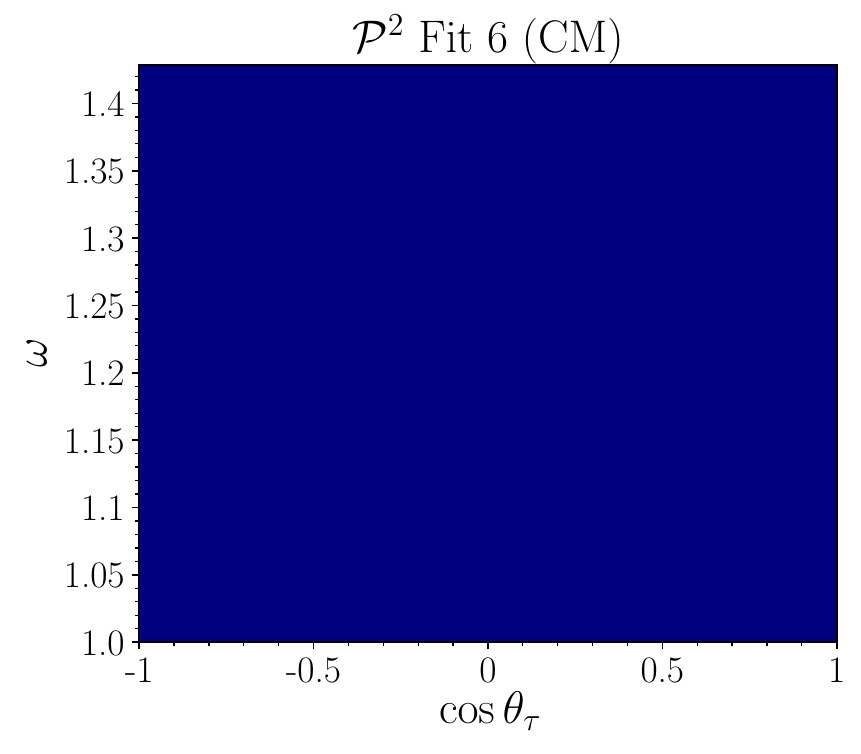} 
\includegraphics[scale=.321]{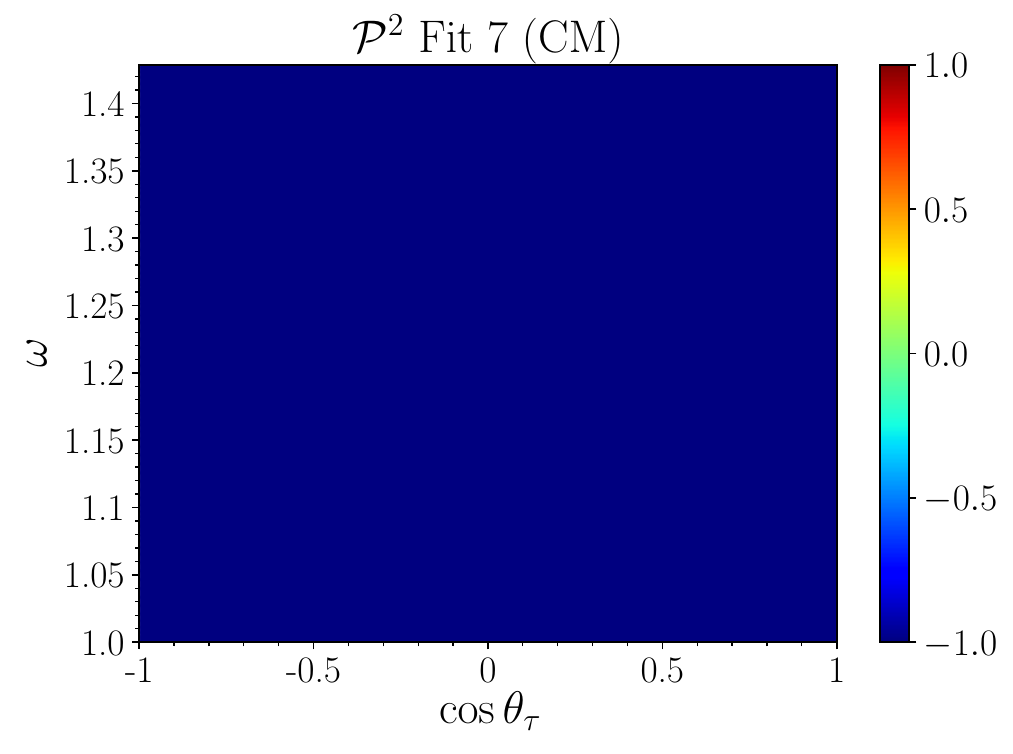}
\caption{ The same as in Fig.~\ref{fig:LambdaPCM}, but for the $\bar B\to D\tau\bar\nu_\tau$ decay.}  
\label{fig:DPCM}
\end{center}
\end{figure}
\begin{figure}[t]
\begin{center}
\includegraphics[scale=.321]{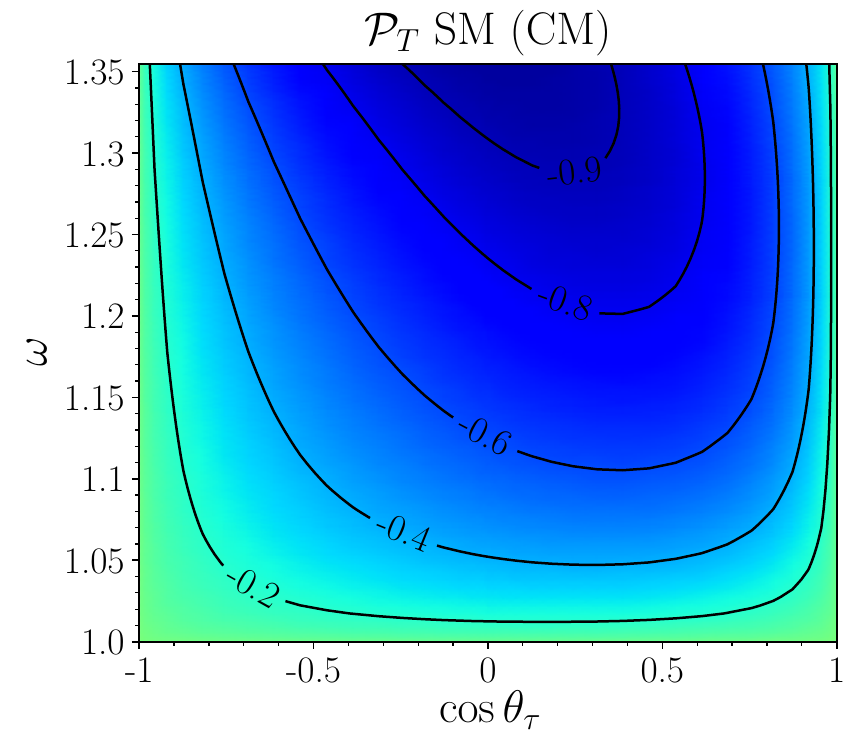}
\includegraphics[scale=.321]{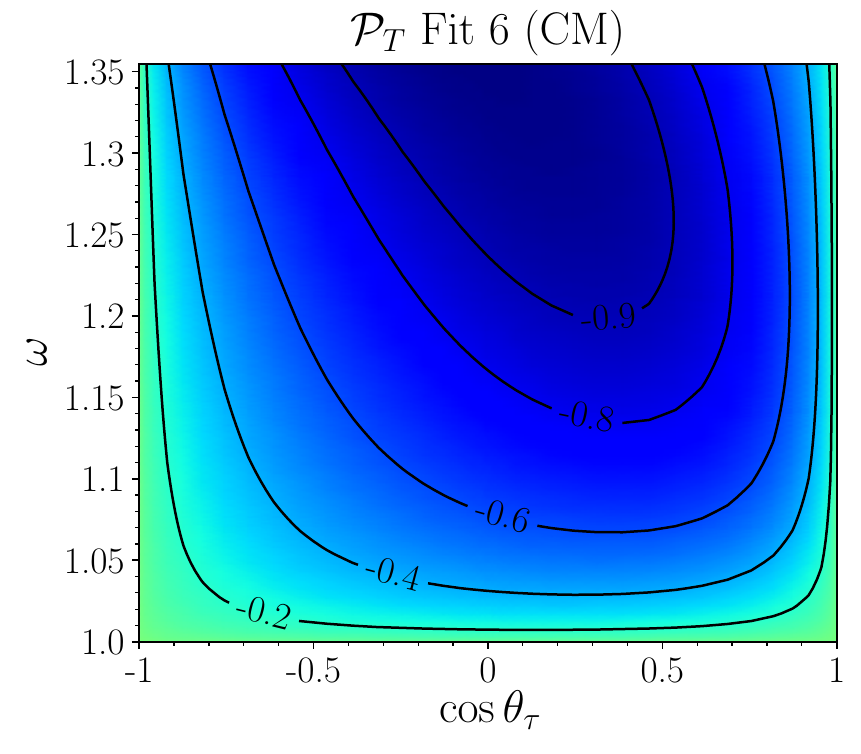} 
\includegraphics[scale=.321]{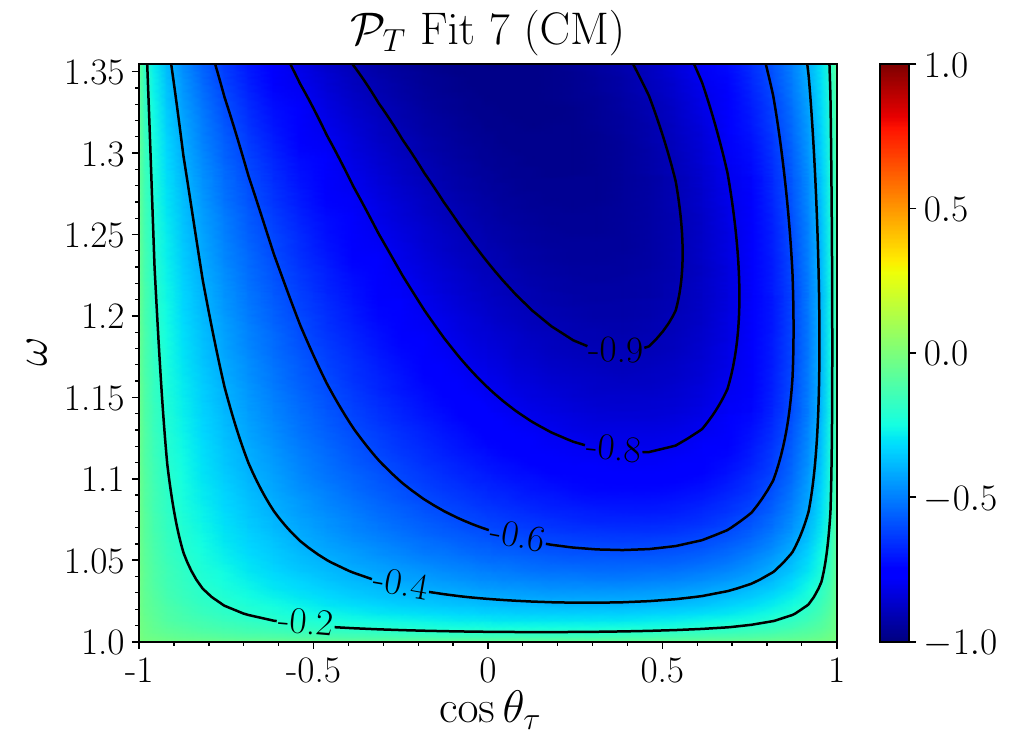}
\\
\includegraphics[scale=.321]{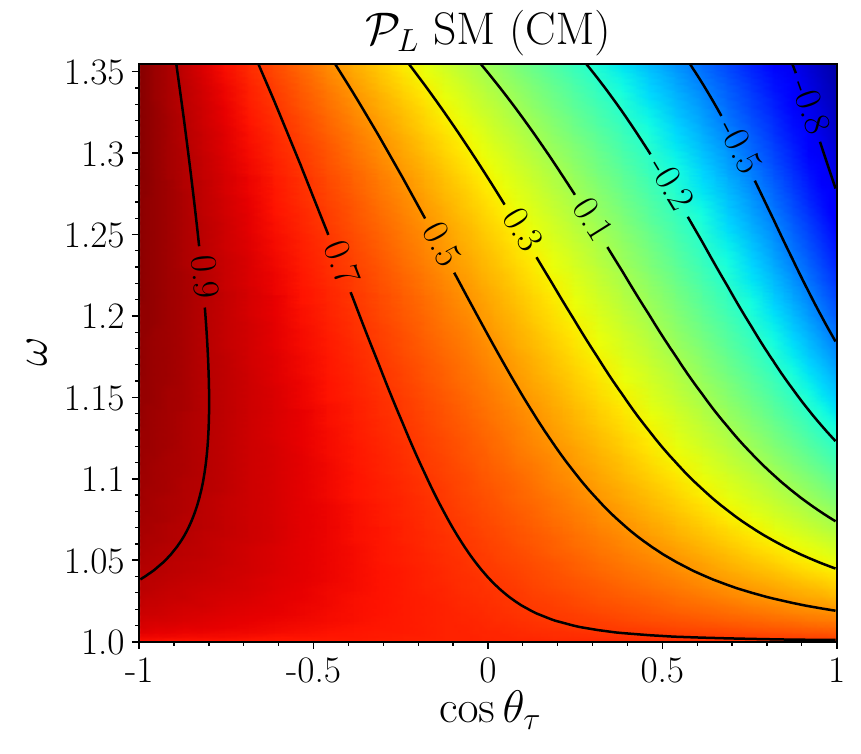}
\includegraphics[scale=.321]{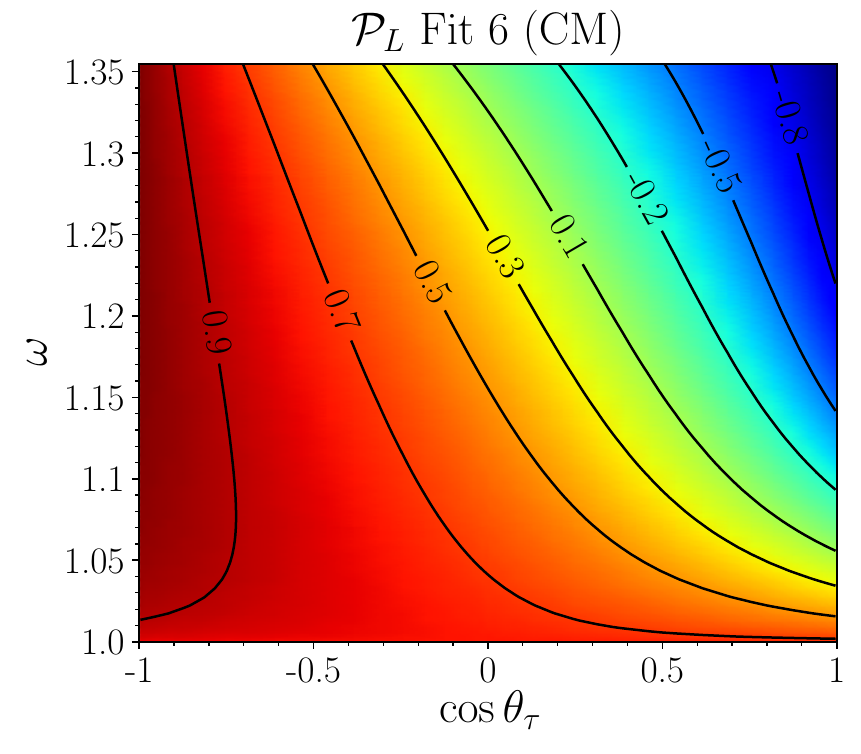} 
\includegraphics[scale=.321]{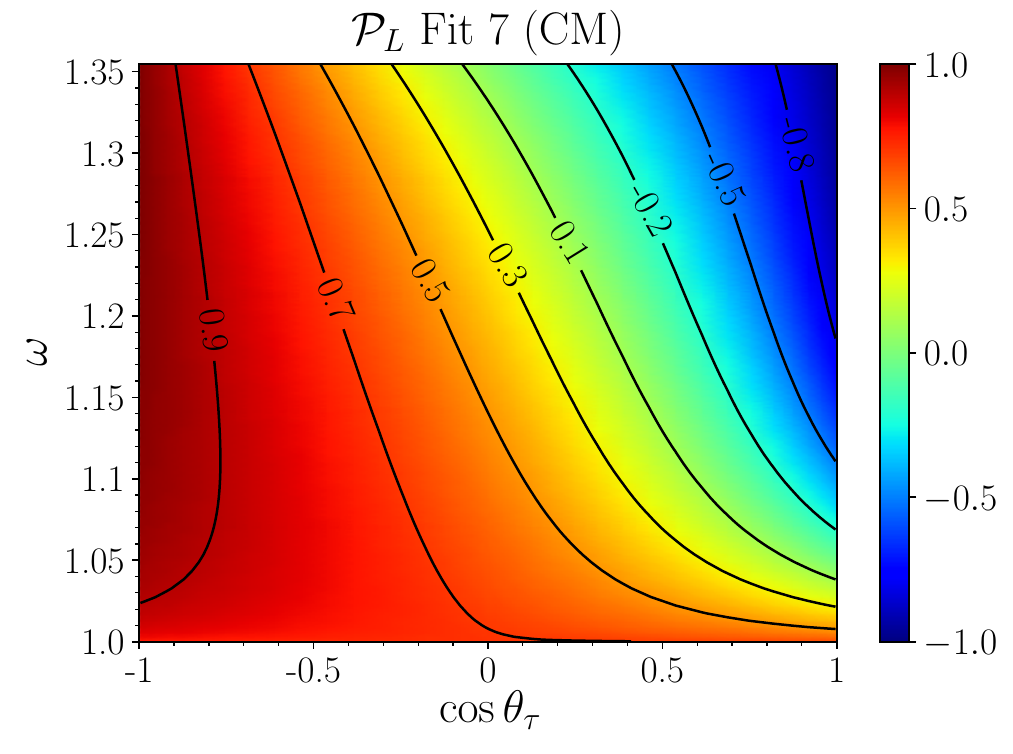}
\\
\includegraphics[scale=.321]{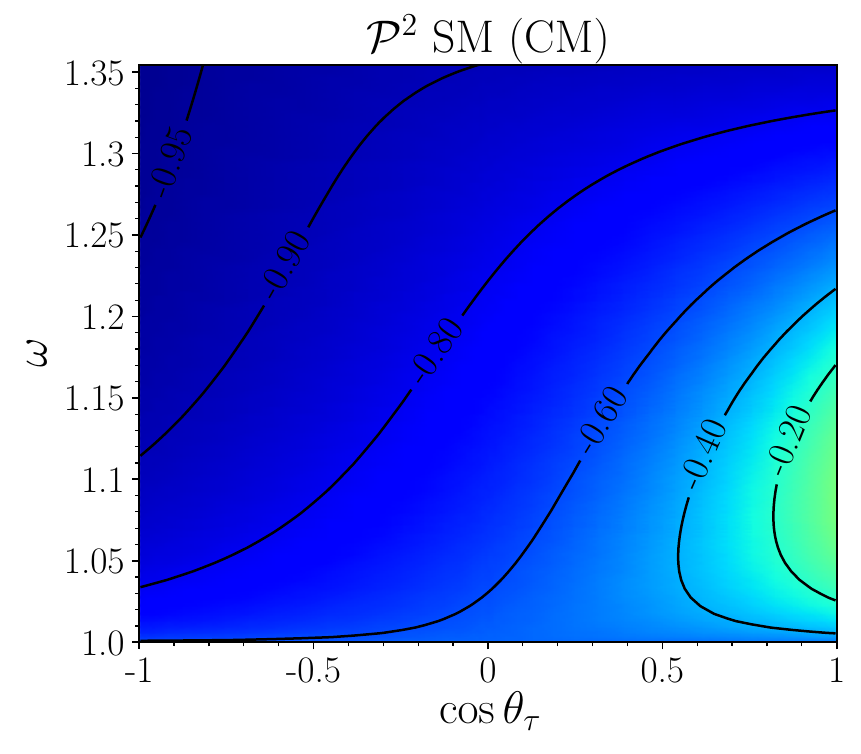}
\includegraphics[scale=.321]{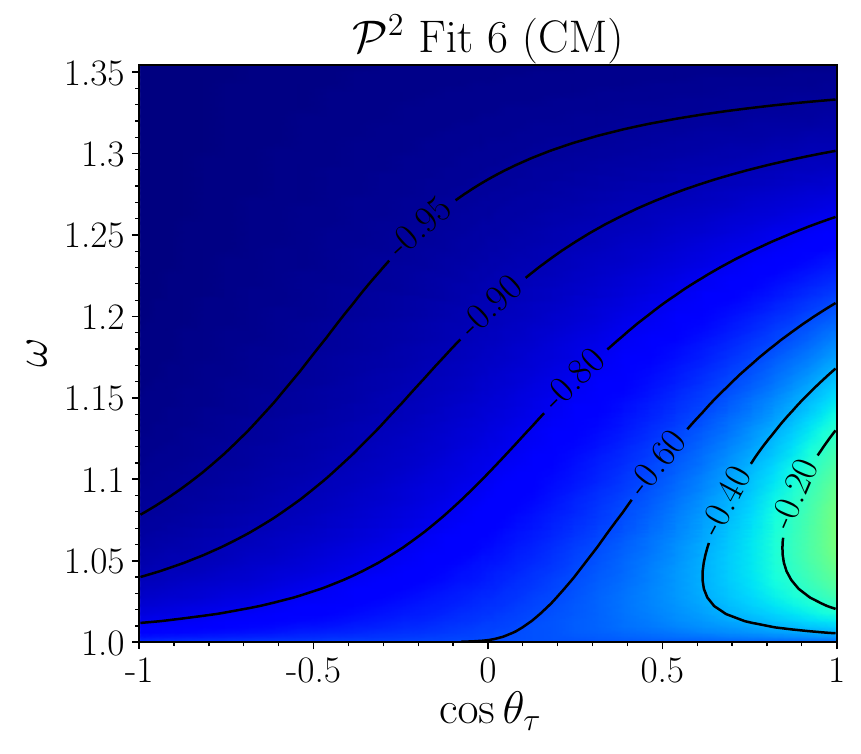} 
\includegraphics[scale=.321]{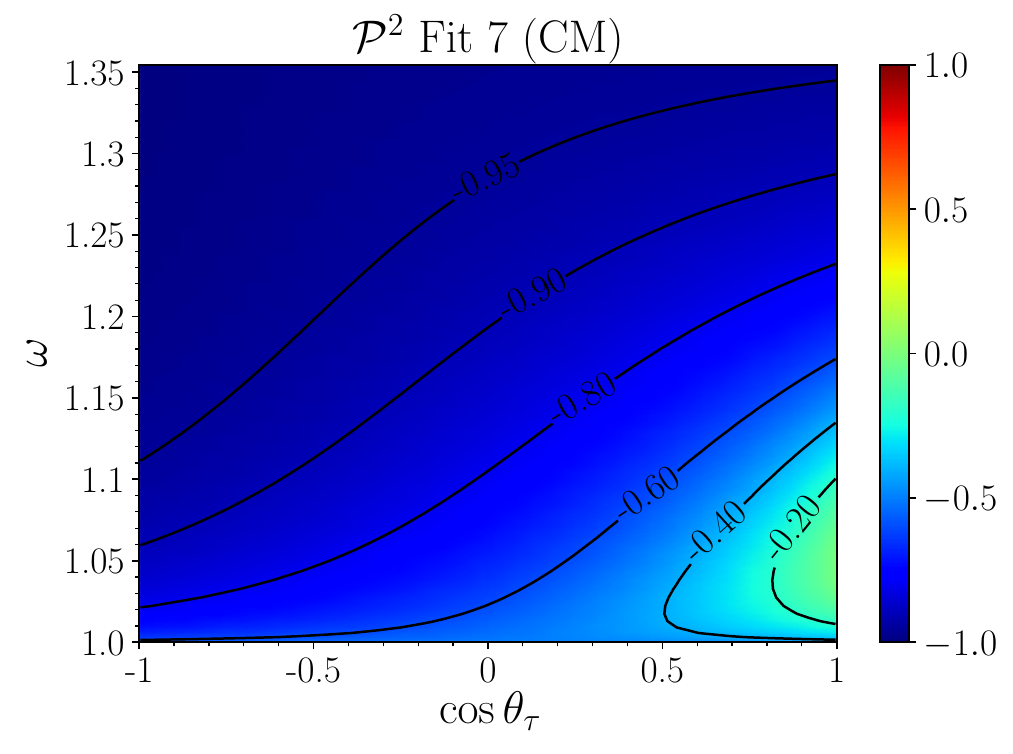}
\caption{ The same as in Fig.~\ref{fig:LambdaPCM}, but for the $\bar B\to D^*\tau\bar\nu_\tau$ decay.}  
\label{fig:DstarPCM}
\end{center}
\end{figure}
\begin{figure}[tbh]
\begin{center}
\includegraphics[scale=.321]{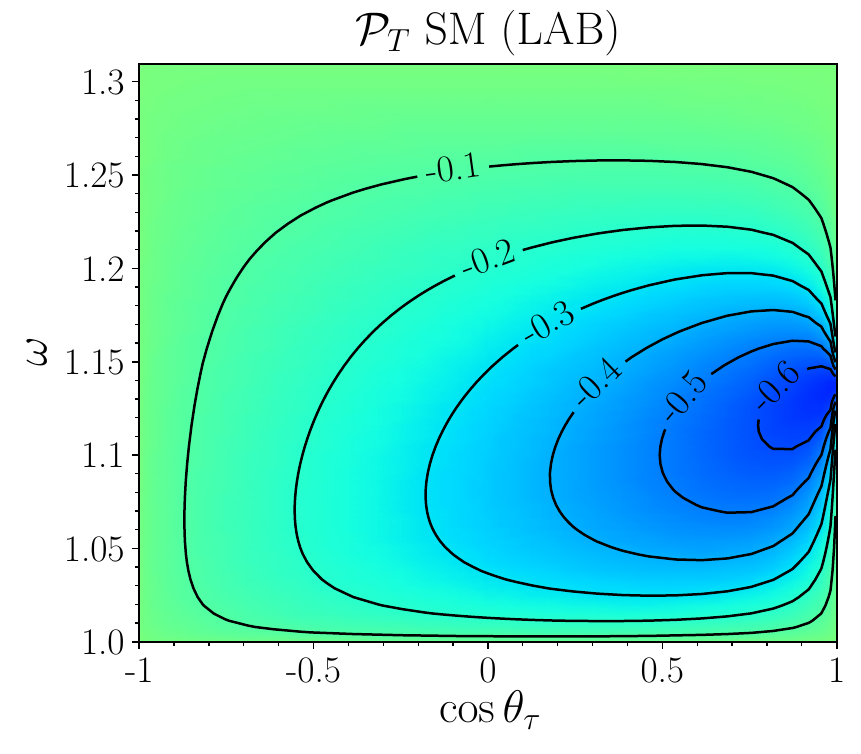}
\includegraphics[scale=.321]{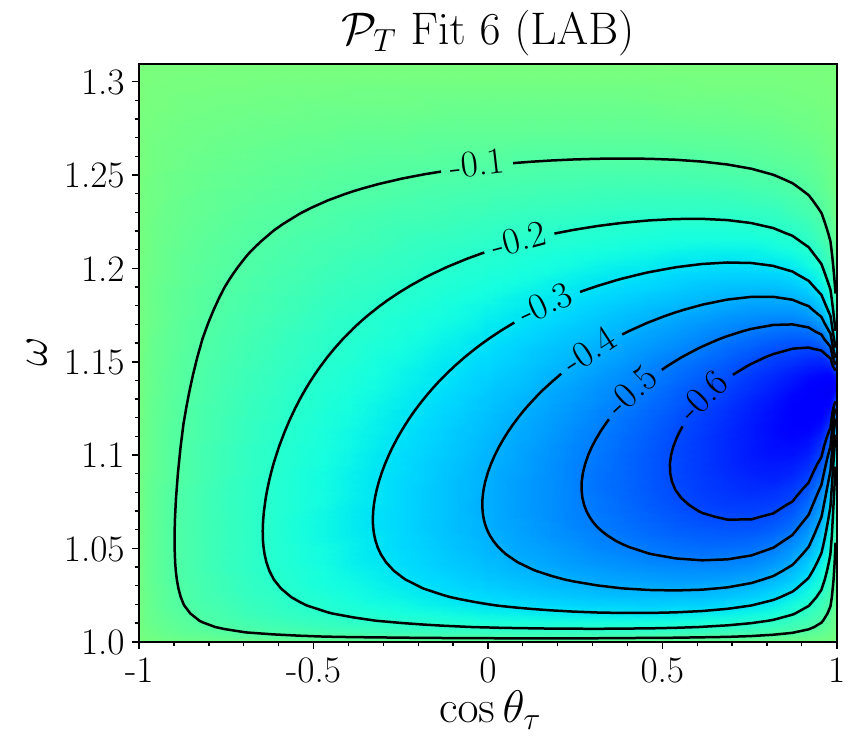} 
\includegraphics[scale=.321]{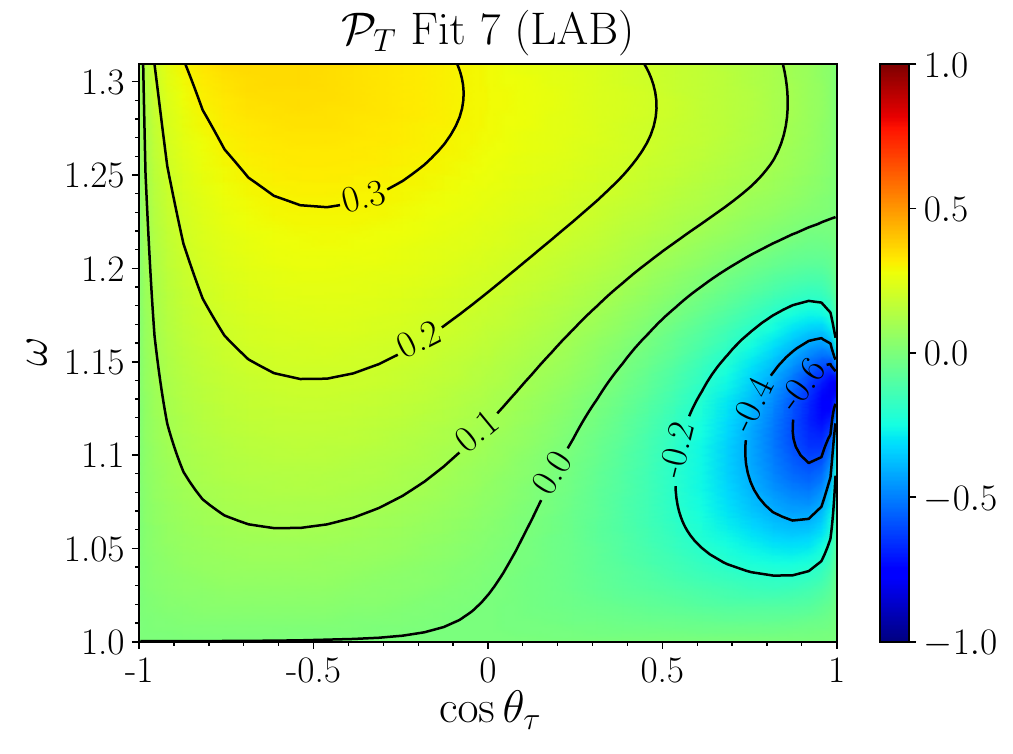}
\\
\includegraphics[scale=.321]{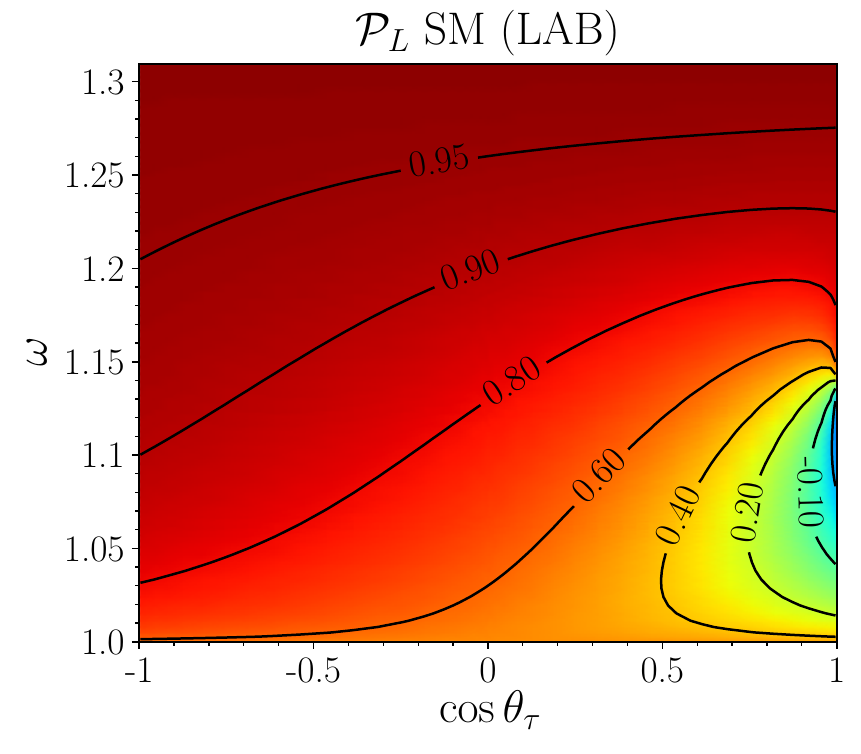}
\includegraphics[scale=.321]{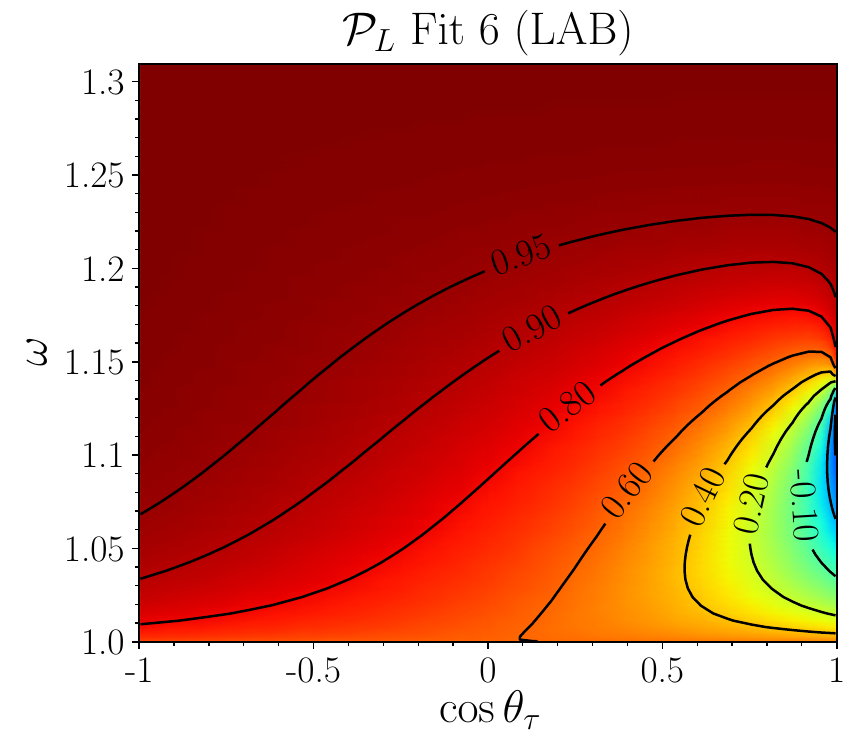} 
\includegraphics[scale=.321]{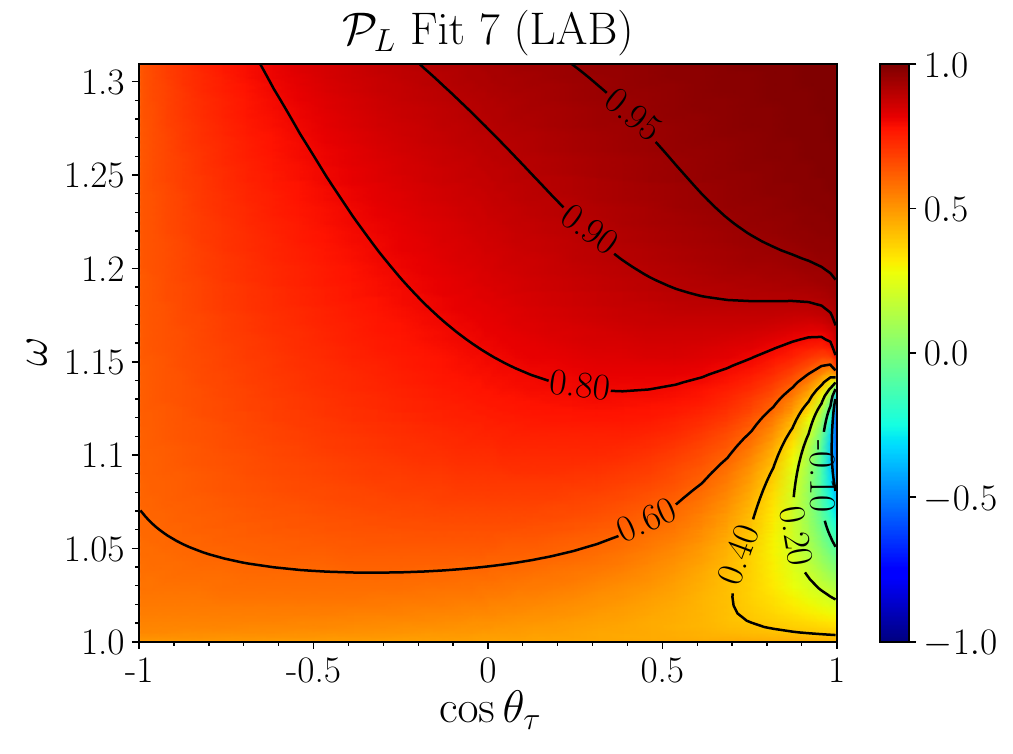}
\caption{ LAB ${\cal P}_T$ (first row) and  ${\cal P}_{L}$ (second row)  polarization observables
 for the $\Lambda_b\to \Lambda_c\tau\bar\nu_\tau$
 decay evaluated within the SM (left column) and with the  NP   Wilson coefficients from 
Fits 6 (middle column) and  7 (right column) 
of Ref.~\cite{Murgui:2019czp}. We display the 2D distributions as a function of the $(\omega, \cos\theta_\tau)$ variables, and use Eq.~\eqref{eq:relacEcso} to compute the $\cos\theta_\tau$ for fixed  $\omega$ and a given $E_\tau$ LAB energy. In all cases,  central values for the form factors and
Wilson coefficients have been used.}
\label{fig:LambdaPLAB}
\end{center}
\end{figure}
\begin{figure}[tbh]
\begin{center}
\includegraphics[scale=.321]{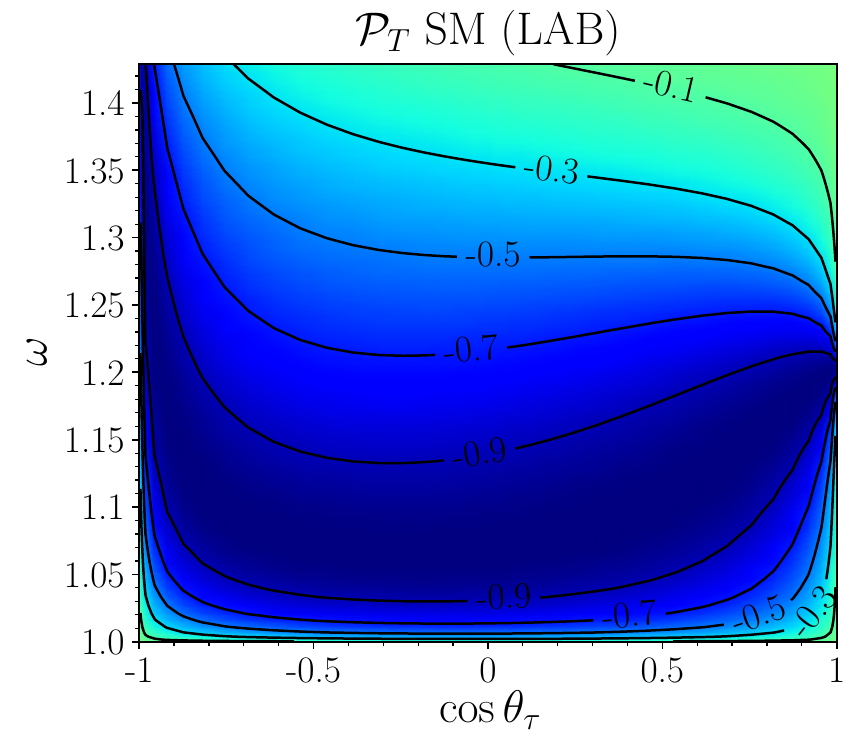}
\includegraphics[scale=.321]{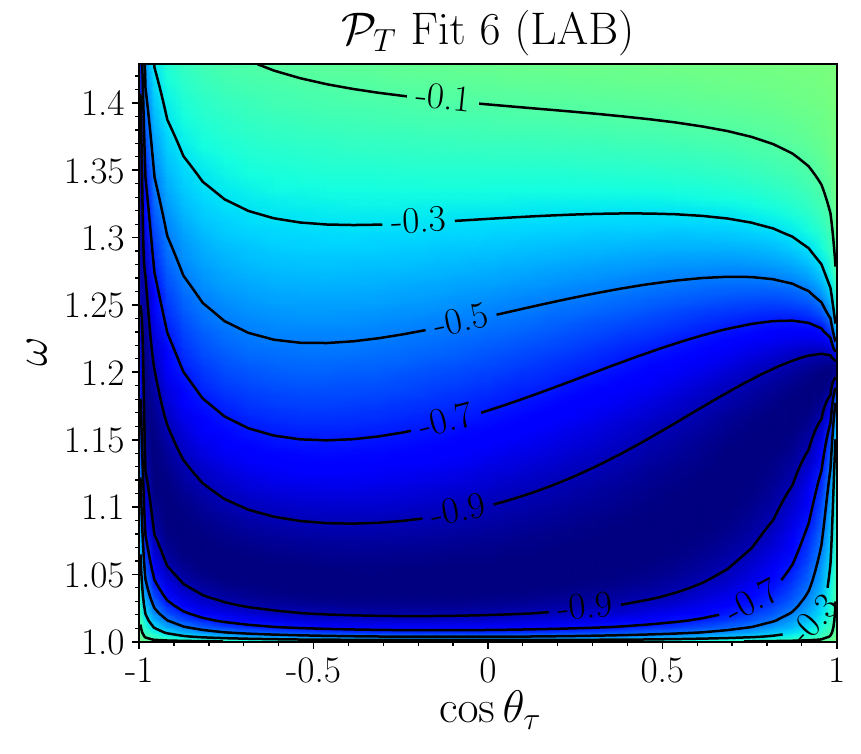} 
\includegraphics[scale=.321]{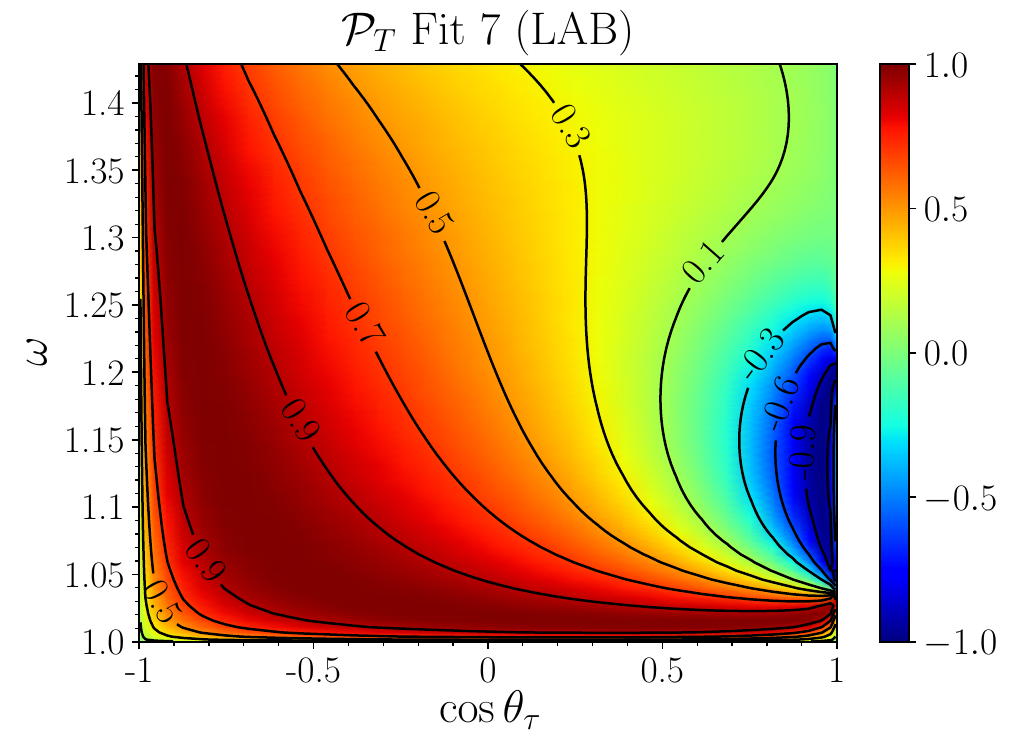}
\\
\includegraphics[scale=.321]{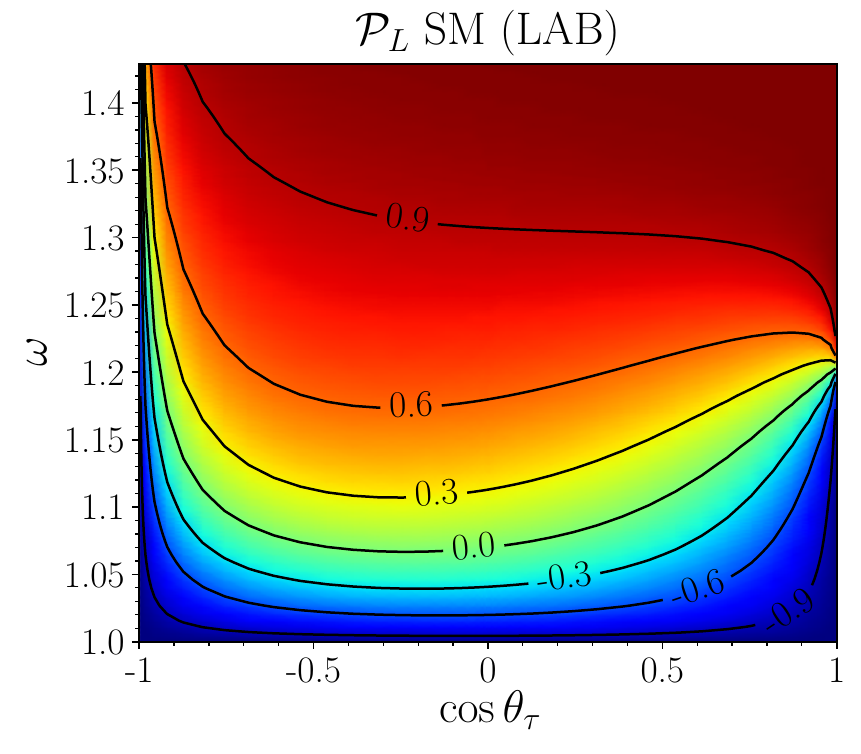}
\includegraphics[scale=.321]{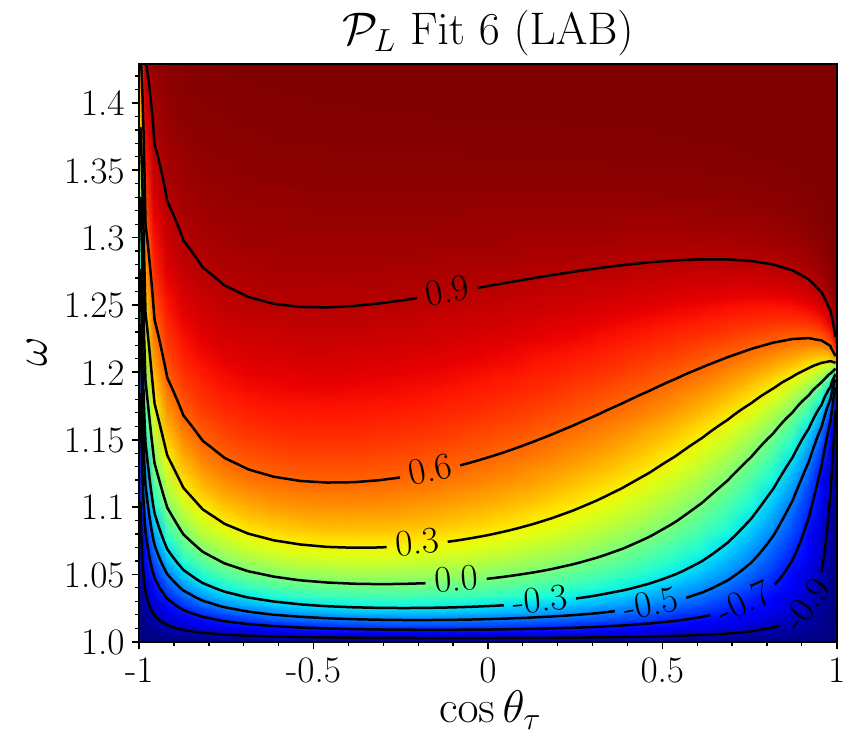} 
\includegraphics[scale=.321]{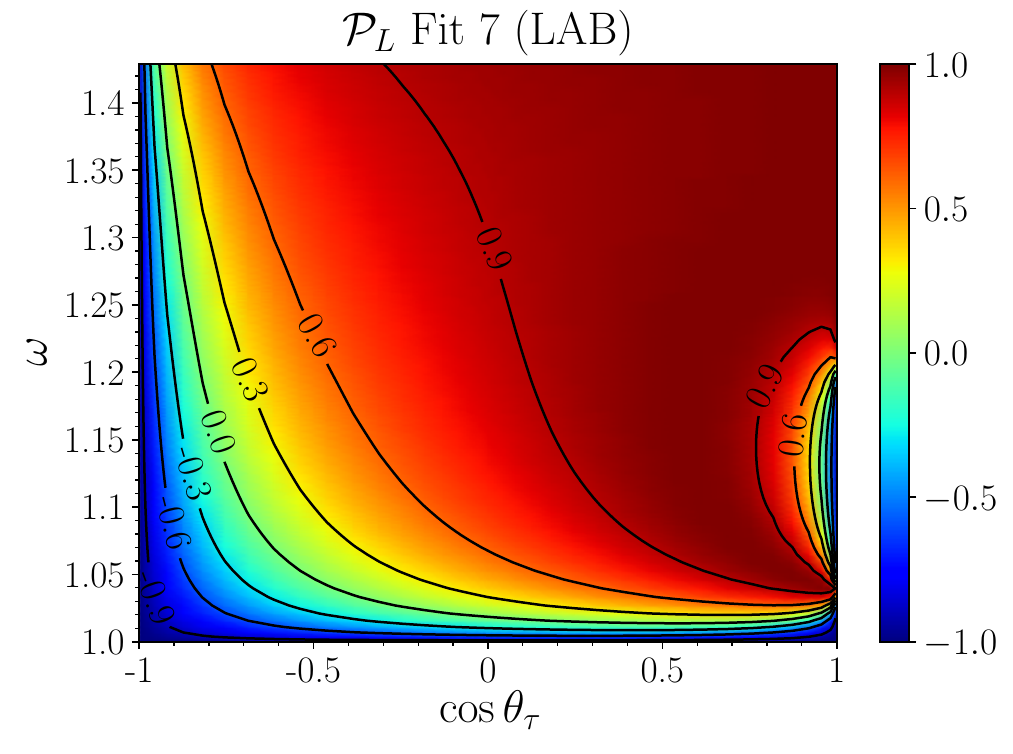}
\caption{The same as in Fig.~\ref{fig:LambdaPLAB}, but for the $\bar B\to D\tau\bar\nu_\tau$ decay. }  
\label{fig:DPLAB}
\end{center}
\end{figure}
\begin{figure}[h!!!]
\begin{center}
\includegraphics[scale=.321]{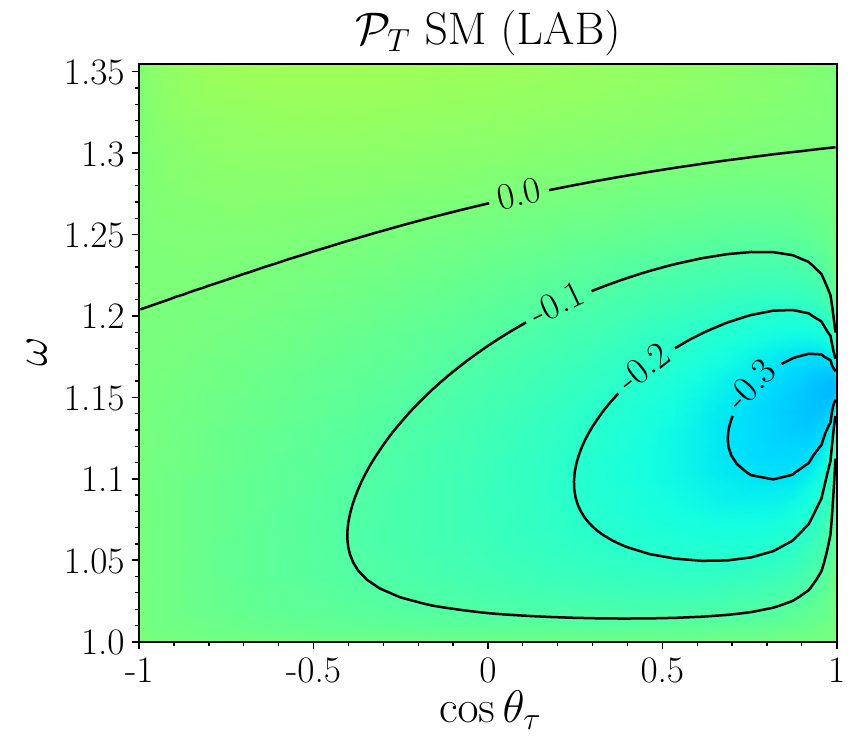}
\includegraphics[scale=.321]{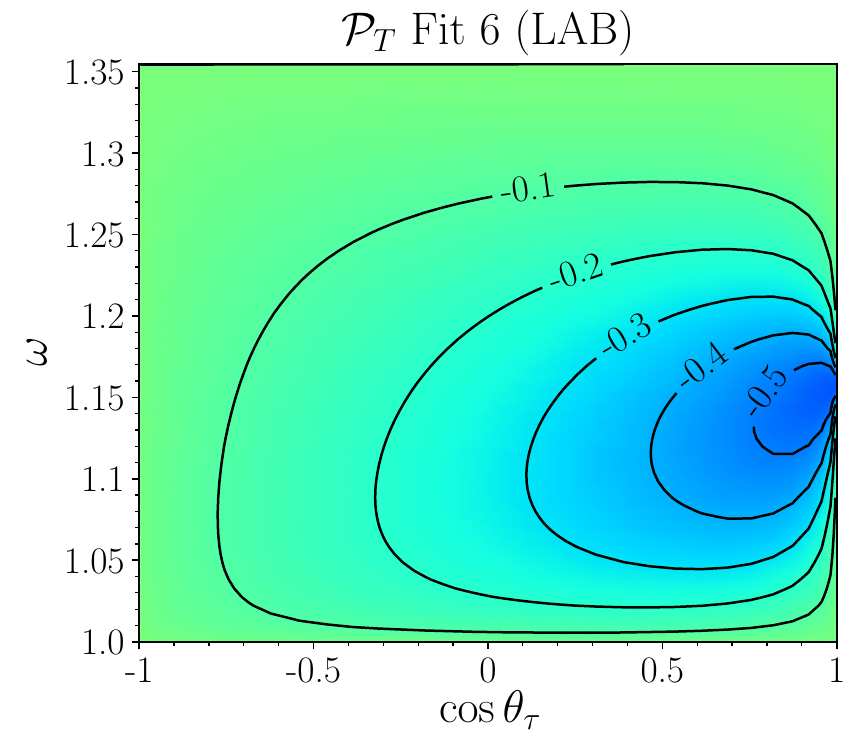} 
\includegraphics[scale=.321]{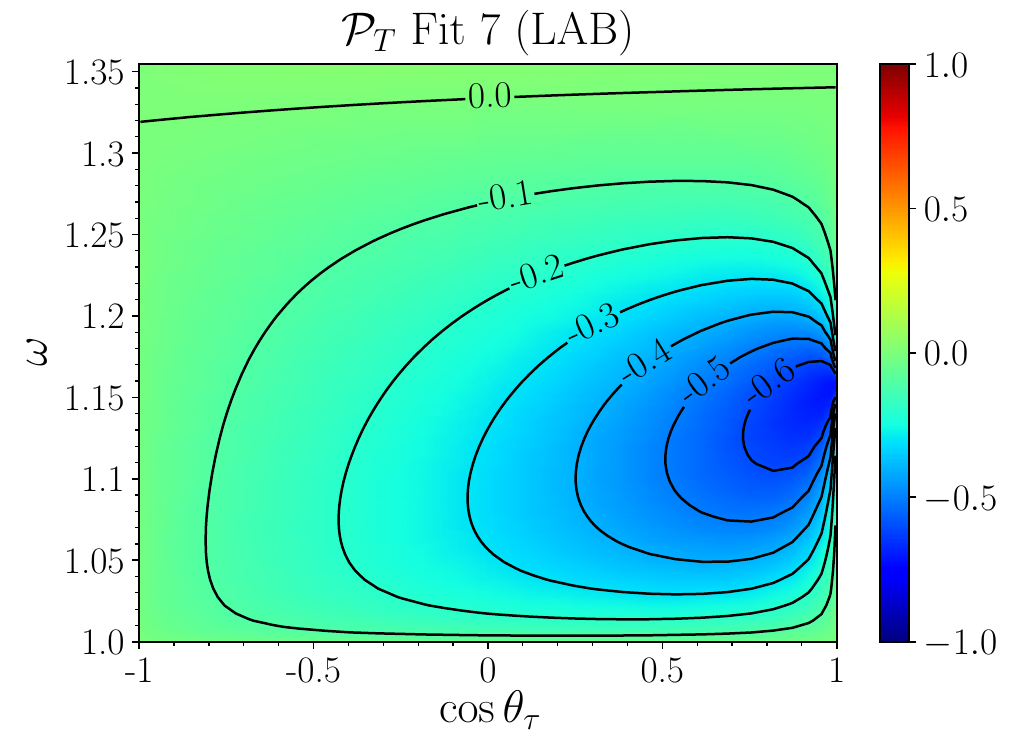}
\\
\includegraphics[scale=.321]{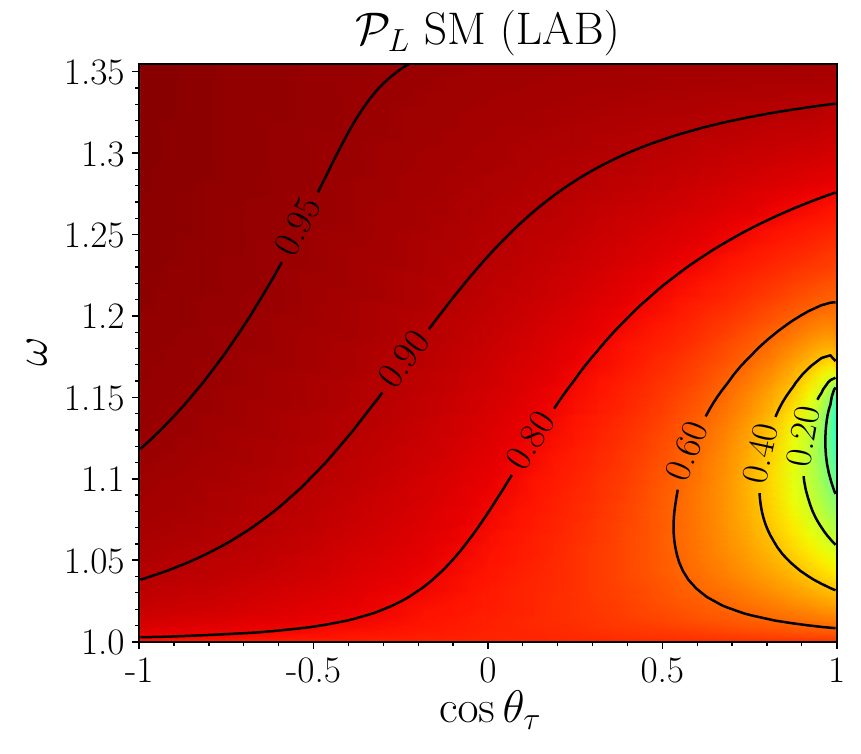}
\includegraphics[scale=.321]{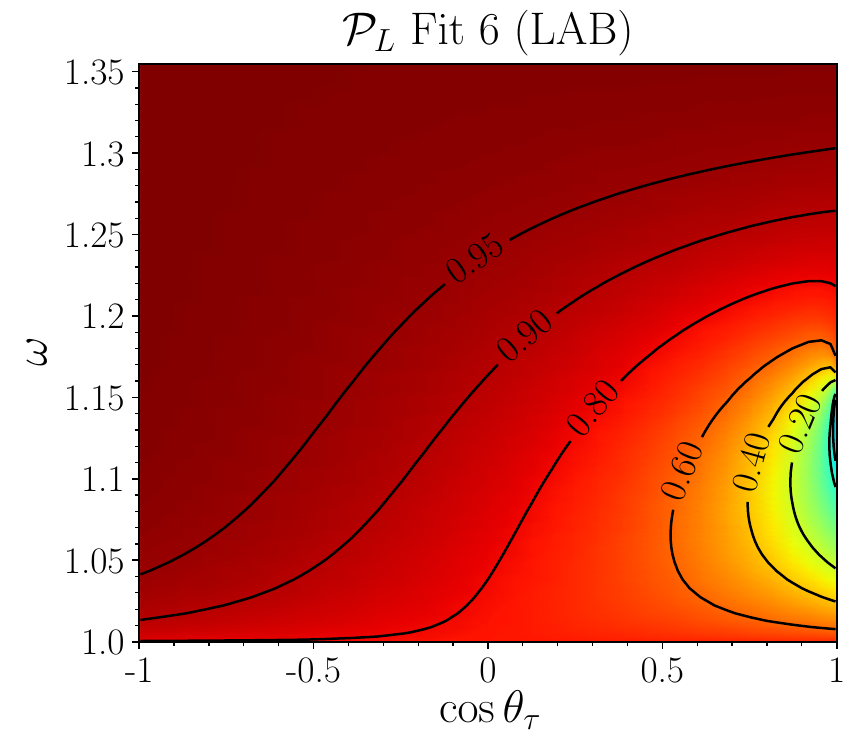} 
\includegraphics[scale=.321]{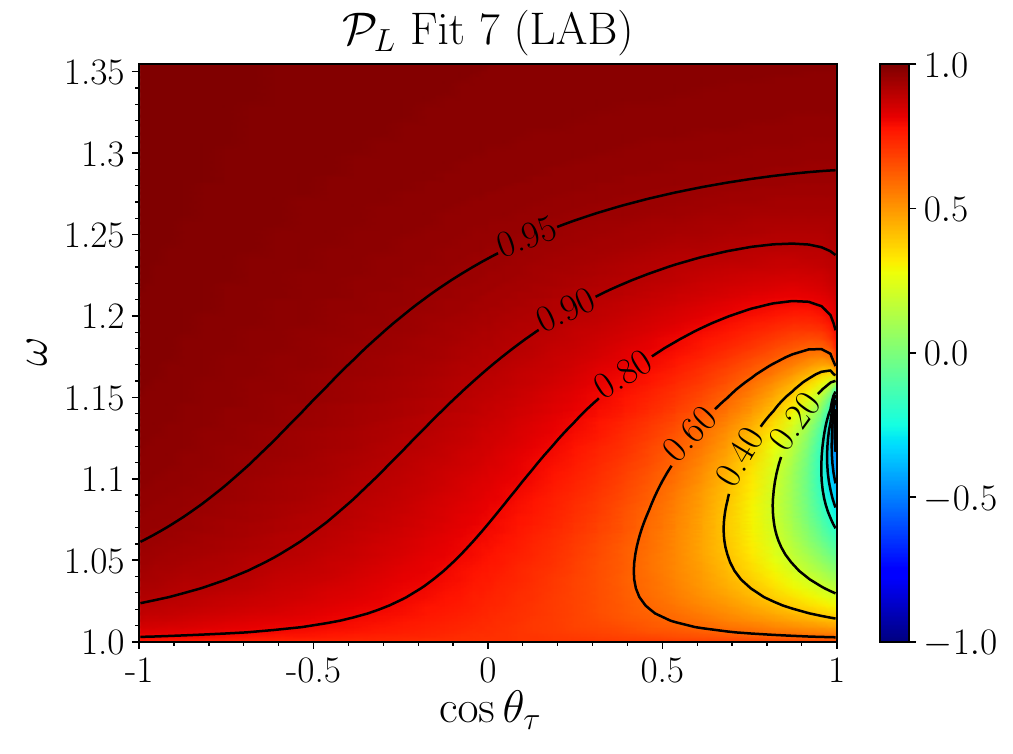}
\caption{The same as in Fig.~\ref{fig:LambdaPLAB}, but for the $\bar B\to D^*\tau\bar\nu_\tau$ decay. }  
\label{fig:DstarPLAB}
\end{center}
\end{figure}
 
Two-dimensional (2D) distributions of the ${\cal P}^\mu$ projections provide observables that can also be used 
to distinguish between different types of NP. In this subsection, we discuss results obtained within the SM and the NP scenarios corresponding to Fits 6 and 7 of Ref.~\cite{Murgui:2019czp}. Since in this latter work,  all Wilson coefficients are real,  
the ${\cal P}_{TT}$ component  comes out  identically zero. 

First in Figs.~\ref{fig:LambdaPCM}, \ref{fig:DPCM} and \ref{fig:DstarPCM}, we show CM 2D distributions for 
${\cal P}_L$,     ${\cal P}_{T}$  and ${\cal P}^2$ and the $\Lambda_b\to \Lambda_c$,  $\bar B\to D$ and $\bar B\to D^*$ decays, respectively, and obtained with the central values for the Wilson coefficients and form factors. In all cases, predictions from Fit 6 are closer to the SM results, and  we clearly observe,  except for the $\bar B\to D^*$ decay,  different 2D patterns for Fits 6 and 7, which would certainly allow  
to distinguish between both NP scenarios. 

The transverse component  ${\cal P}_{T}$ is always negative for $\Lambda_b\to \Lambda_c$ and   $\bar B\to D^*$ decays, with a $\cos\theta_\tau-$dependence that becomes flatter as $\omega$ decreases from $\omega_{\rm max}$ to the vicinity of zero recoil ($\omega=1$), where  ${\cal P}_{T}$ reaches, in modulus, its minimum  value. Large negative values of ${\cal P}_{T}$, which can reach $-0.9$, are found  for $\omega > 1.2$  and intermediate values of $\cos\theta_\tau$ far from the $\pm 1$ limits.  For these two decays, the longitudinal polarization shows a large variation, going from ${\cal P}_{L}\sim 0.9$ for angles close to $\pi$ to values in the $(-0.9,-0.8)$ range in the forward direction, where the dependence on $\omega$ is significantly more pronounced than at backward angles.  Moreover, we see regions close to zero recoil, and in the forward direction, where the $\tau-$lepton is produced largely unpolarized (${\cal P}^2\sim -0.2$), with  $|{\cal P}^2\,|$ growing as both $\theta_\tau$ and $\omega$ increases, reaching values in the interval $(-0.95,-0.9)$ for $\omega$ in the vicinity of $\omega_{\max}$ (see the 2D$-{\cal P}^2$  distributions in the bottom panels). The exception is found for  NP Fit 7 in the baryon decay, for which the $\tau$ is produced almost polarized, ${\cal P}^2 < -0.95$ at forward angles and close to $\omega_{\rm max}$ (right-top corner), with a large ${\cal P}_T$ polarization component, around $-0.9$. However, in this case for backward angles,  ${\cal P}^2$ does not become so close to $-1$ as $\omega$ approaches $\omega_{\rm max}$.

The discussion for the $\bar B\to D $ transition should take into account that for this decay ${\cal P}^2=-1$,  implying that the  $\tau$ emitted is always fully polarized. In Ref.~\cite{Penalva:2020ftd}, it was already pointed out that for $0^-\to 0^-$ transitions at zero recoil and $\theta_\tau=0$ or $\pi$, angular momentum conservation  forces the $\tau$ helicity to equal that of the antineutrino which is positive, thus ${\cal P}_L=-1$ (see Eq.~\eqref{eq:plasi} or \eqref{eq:PLassy}), 
which implies ${\cal P}^2=-{\cal P}_L^2=-1$ and ${\cal P}_T=0$. 

Indeed, we see in the bottom panels of Fig.~\ref{fig:DPCM} that ${\cal P}^2=-1$ in the whole $(\omega, \cos\theta_\tau)$ phase-space, and not only for $\theta_\tau=0$ or $\pi$ at zero recoil. Therefore,  longitudinal and transverse polarizations are not independent for non-CP violating physical scenarios, and in the full phase-space both components  satisfy the relation ${\cal P}_L^2+{\cal P}_T^2=1$. As in the other decays, Fit 7 predictions differ from SM ones significantly more than those obtained in the NP Fit 6, with  
${\cal P}_L$ exhibiting a pronounced dependence on $\cos\theta_\tau$, when $\omega$ departs from the zero recoil point. While ${\cal P}_L$ takes negative and positive values within the SM and both Fits 6 and 7 of Ref.~\cite{Murgui:2019czp}, we observe that ${\cal P}_T$ is negative for SM and Fit 6, while for the NP Fit 7, this transverse component also takes positive and negative values, and even it vanishes along a  $(\omega, \cos\theta_\tau)-$curve, for which ${\cal P}_L=+1$. As we will see below, for the kinematics encoded in this curve, the $\tau-$lepton is produced in a negative-helicity state.  

The reason why the $\tau$ is always fully polarized for a  general $0^-\to0^- $ transition  is the following. Since, 
in the massless limit, the $\bar\nu_\tau$ is fully polarized, we have 
that the invariant amplitude $ {\cal M}$, apart from momenta,  only depends on the $\tau$ spin degrees
 of freedom. If we have $ {\cal M}(h)$, where here $h=\pm1$ represents  the $\tau$ helicity, one 
 can always define two coefficients
\bea
a_{\pm 1}=\frac{\pm {\cal M}(h=\mp 1)}{\sum_{h'=\pm 1}|{\cal M}(h')|^2}
\eea
such that  $\sum_{h=\pm1}|a_h|^2=1$ and satisfy
\bea
\sum_{h=\pm1} a_h{\cal M}(h)=0,\
\eea
What this result tells us is that the probability to produce a $\tau$ in the state
$a^*_{+1}u^{\tilde s}_{+1}(k')+a^*_{-1}u^{\tilde s}_{-1}(k')$ is identically zero. Thus, 
the probability to produce a $\tau$ in the orthogonal state, $a_{-1}u^{\tilde s}_{+1}(k')
-a_{+1}u^{\tilde s}_{-1}(k')$, should be one. The $\tau$ is then fully polarized. 
Apart from irrelevant phases these two polarization states correspond to
\bea
a^*_{+1}u^{\tilde s}_{+1}(k')+a^*_{-1}u^{\tilde s}_{-1}(k')\equiv  u^{\cal P}_{+1}(k'),\quad 
a_{-1}u^{\tilde s}_{+1}(k')-a_{+1}u^{\tilde s}_{-1}(k')\equiv u^{\cal P}_{-1}(k'),
\eea
i.e., they are the two spin-covariant eigenstates associated to the four-vector\footnote{From Eq.~\eqref{eq:probS}, the probability of  measuring the $\tau$ in a state $u^{\cal P}_{h}(k')$, eigenstate of the operator $\gamma_5\slashed{\cal P}$ with eigenvalue $h$, is given by $(1-h)/2$ since ${\cal P}^2=-1$ for $\bar B \to D$ decays. Therefore, we assign the state $u^{\cal P}_{-1}(k')$ to the produced polarized tau.  This result
 is consistent with ${\cal P}^\mu [\omega=1,\cos\theta_\tau=\pm 1)] = -\tilde s^\mu [\omega=1,\cos\theta_\tau=\pm 1]$, since for these two CM kinematics ${\cal P}_L=-1$.} $N^\mu={\cal P}^\mu$. For a given $k'$, these states 
depend on the pair of variables $(\omega,\cos\theta_\tau)$ or 
$(\omega, E_\tau)$  that determine all ${\cal P}^\mu$ components (or equivalently ${\cal M}$) in the  CM or LAB frames  respectively. 
The above argumentation fails as soon as $\cal M$ depends on the spin variable of the hadrons involved in the decay. This is so since, in general, it is not possible to find $a_{\pm1}$ such that
\bea
\sum_{h=\pm1} a_h{\cal M}_{\lambda}(h)=0, \label{eq:demoP2one}
\eea
for all $\lambda\equiv(r,r')$ values, where  different $\lambda$  values represent 
different hadronic spin configurations. Note however that for a fixed $\lambda$ 
(corresponding to fixed $r/r'$ polarization of the initial/final hadron) 
Eq.~\eqref{eq:demoP2one} has always a solution. Thus, for fixed $\lambda$, the $\tau$ 
is also fully polarized but with a polarization state that depends on $\lambda$. This is 
in agreement with the results obtained in Ref.~\cite{Tanaka:1994ay} for $\bar B\to D^{(*)}$ 
decays. 

As noted above, for the rest of the transitions,  $ {\cal P}^2$  approaches $-1$  at maximum recoil  ($\omega_{\rm max}$), with the exception of Fit 7  for the $\Lambda_b\to\Lambda_c$ decay in the $\cos\theta_\tau<0$ region. This is better understood by looking at
the polarization projections in the laboratory frame. 

In Figs.~\ref{fig:LambdaPLAB}--\ref{fig:DstarPLAB}, we present the LAB  
${\cal P}_L$ and ${\cal P}_T$ 2D distributions for the same NP scenarios and decays  discussed previously in Figs.~\ref{fig:LambdaPCM}--\ref{fig:DstarPCM}. 
In the LAB  plots, we have made use of the relation in Eq.~(\ref{eq:relacEcso})
%
to represent the polarization observables as a function of $(\omega,\cos\theta_\tau)$ instead 
of $(\omega,E_\tau)$. 
On the other hand, since ${\cal P}^2$ is a scalar [and thus ${\cal P}^2_{\rm
LAB}(\omega, E_\tau(\omega,\cos\theta_\tau))={\cal P}^2_{\rm
CM}(\omega, \cos\theta_\tau)$ ] we will no show it again. 

Though the LAB ${\cal P}_L$ and ${\cal P}_T$ 2D distributions shown in Figs.~\ref{fig:LambdaPLAB}--\ref{fig:DstarPLAB} can be obtained from the CM ones depicted above in Figs.~\ref{fig:LambdaPCM}--\ref{fig:DstarPCM}, we stress that the coefficients of the linear combinations (Eq.~\eqref{eq:cmlabrel}) depend on $\omega$ and $\cos\theta_\tau$. Moreover, the longitudinal or transverse character is not preserved, which also makes interesting a short discussion of the main features of the LAB polarization components. In the LAB frame, the $\tau$'s are  mainly being emitted with negative helicity 
(${\cal P}_L^{\rm LAB}\approx 1$)  in the high $\omega$-region close to $\omega_{\rm max}$, as can be seen in the second row of plots in Figs.~\ref{fig:LambdaPLAB}--\ref{fig:DstarPLAB}.  The explanation for this behavior, at least in part, is that close to maximum recoil, the $\tau$ momentum in the LAB is  large and hence positive helicity is suppressed by the
 dominant contribution that selects negative chirality for the final
charged  lepton\footnote{At very large momentum helicity almost equals chirality}. As 
mentioned in Ref.~\cite{Penalva:2020ftd}, only the 
${\cal O}_{S_L,S_R}$ and ${\cal O}_T$ NP terms select positive chirality. Looking at 
the values for the corresponding Wilson coefficients (see Table 6 of Ref.~\cite{Murgui:2019czp})
 one  expects larger deviations from the above behavior for Fit 7.

\begin{figure}[tbh]
\begin{center}
\includegraphics[scale=1]{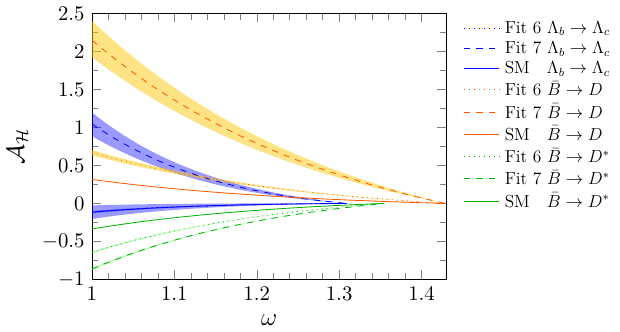}\\ 
\includegraphics[scale=1]{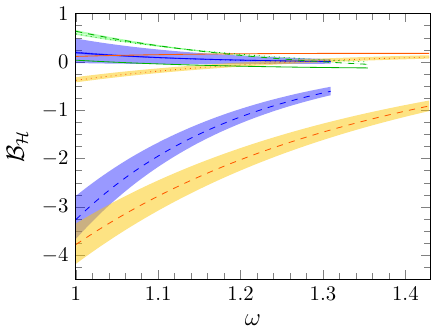}\hspace{.15cm}
\includegraphics[scale=1]{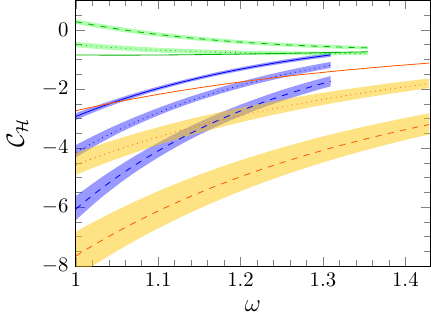}\\
\includegraphics[scale=1]{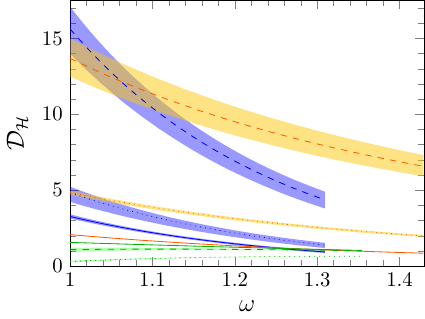}\hspace{.15cm}
\includegraphics[scale=1]{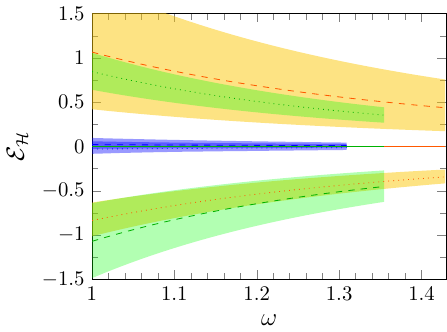}\\
\caption{ ${\cal A_H}(\omega)$,\,${\cal B_H}(\omega)$,\,${\cal C_H}(\omega)$,\,${\cal D_H}$ and
${\cal E_H}(\omega)$ functions (Eqs.~\eqref{eq:pol2} and \eqref{eq:PNP}) for the $\Lambda_b\to\Lambda_c$ (bluish), $\bar B\to D$ (reddish) and $\bar B\to D^*$ (greenish) decays evaluated
for the SM (solid), Fit 6 (dotted) and  Fit 7 (dashed) of Ref.~\cite{Murgui:2019czp}. Error bands take into account  the uncertainties associated to the Wilson coefficients and form factors, and they are calculated as explained in Refs.~\cite{Penalva:2020xup,Penalva:2020ftd}.}  
\label{fig:abcdeH}
\end{center}
\end{figure}
To finish this subsection, we recall here that for fixed $\omega$, the polarization components 
 turn out to be ratios of linear or quadratic functions of $(p\cdot k)$, as inferred  from Eqs.~\eqref{eq:PLdef} and  \eqref{eq:PTdef}. Restricting the discussion to CM observables, the denominator of these ratios, ${\cal N}(\omega, p\cdot k)$,  is proportional to $p_{\cal N}(\omega, \cos\theta_\tau)=a_0(\omega) + a_1(\omega) \cos\theta_\tau + a_2(\omega) \cos^2\theta_\tau$, with the coefficients $a_i(\omega)$ appearing in the angular decomposition of the tau-unpolarized  $d^2\Gamma/(d\omega d\cos\theta_\tau)$ differential decay width. We have already presented results for them in our previous works~\cite{Penalva:2020xup, Penalva:2020ftd}, and we will not make any further comment here. On the other hand, taking into account the dependence of $(p\cdot k), (p\cdot N_{L,T}^{\rm CM})$ and $(q\cdot N_{L,T}^{\rm CM})$ on $\cos \theta_\tau$, we find 
 \bea
  P_L^{\rm CM}(\omega,\cos\theta_\tau)&=& \frac{p_0(\omega)+ p_1(\omega) \cos\theta_\tau + p_2(\omega) \cos^2\theta_\tau}{a_0(\omega) + a_1(\omega) \cos\theta_\tau + a_2(\omega) \cos^2\theta_\tau}, \nonumber \\ P_T^{\rm CM}(\omega,\cos\theta_\tau)&=&\sin\theta_\tau \frac{p'_0(\omega)+ p'_1(\omega) \cos\theta_\tau }{a_0(\omega) + a_1(\omega) \cos\theta_\tau + a_2(\omega) \cos^2\theta_\tau}
 \eea
with the five  coefficients, $p_0, p_1,p_2, p'_0$ and $p'_1$, of the numerator polynomials being linear combination of the five ${\cal A_H}(\omega)$,\,${\cal B_H}(\omega)$,\,${\cal C_H}(\omega)$,\,${\cal D_H}$ and ${\cal E_H}(\omega)$ functions, introduced in Eqs.~\eqref{eq:pol2} and \eqref{eq:PNP} to generally describe the decay with polarized taus in the final state. We observe that  $P_L^{\rm CM}$ and $P_T^{\rm CM}$ are not just  polynomials in $\cos\theta_\tau$ and that the simultaneous knowledge/measure of both of them, in conjunction with the unpolarized  $d^2\Gamma/(d\omega d\cos\theta_\tau)$ distribution,  provide the maximum information which can be obtained from the decay with polarized taus\footnote{Non-conserving CP contributions, ${\cal F_H}$ and ${\cal G_H}(\omega)$, related to $P_{TT}$ are not considered in this discussion.}. In addition, the longitudinal component, or equivalently the CM tau-helicity  $d^2\Gamma/(d\omega d\cos\theta_\tau)$ double differential decay width, provides only three independent conditions ($p_0, p_1$ and $p_2$) and it is not enough to determine all undetermined  ${\cal A_H}, \cdots {\cal E_H}$ functions. This was already pointed out in Ref.~\cite{Penalva:2020xup}, where it is also shown that all these functions can be obtained using also input from the LAB tau-helicity  $d^2\Gamma/(d\omega dE_\tau)$ distribution (or equivalently $P_L^{\rm LAB}$), as expected from  the discussion in Eq.~\eqref{eq:cmlabrel} since this brings in some information of $P_T^{\rm CM}$. This is another way to point out that the CM and LAB tau-helicity differential distributions provide complementary results. 
  
  Thus, we  show results for ${\cal A_H}(\omega)$,\,${\cal B_H}(\omega)$,\,${\cal C_H}(\omega)$,\,${\cal D_H}$ and
${\cal E_H}(\omega)$, since  this is another, more simple, way of presenting the physical information contained in the above 2D polarization observables. 
This is done in Fig.\ref{fig:abcdeH}  for  the $\Lambda_b\to\Lambda_c\tau\bar\nu_\tau$ and $\bar B\to D^{(*)}
\tau\bar\nu_\tau$ decays.  We note that these functions could  also be reconstructed from the exhaustive results included in Refs.~\cite{Penalva:2020xup, Penalva:2020ftd} on the tau CM angular and LAB energy dependencies of the helicity $d^2\Gamma/(d\omega d\cos\theta_\tau)$ and $d^2\Gamma/(d\omega dE_\tau)$ distributions, but that they have never been directly shown.  In most cases we see the capability of these observables to distinguish the SM and Ref.~\cite{Murgui:2019czp} Fits 6 and 7 predictions, with the latter deviating more from the SM results. 
One can have direct access to these functions by measuring the $\tau$ polarization in the decay or, indirectly, through the measuring of the polarization vectors components ${\cal P}_{L,T}$. The latter can be obtained for instance from the 
analysis of the subsequent $\tau$ decay. Both methods require however to be able to measure the
$\tau$ momentum (in the first case also its polarization), something that it is
 extremely difficult, since the decay products of the tau include an
undetected neutrino. 
 
 In this sense, we should comment that the framework presented in Refs.~\cite{Asadi:2020fdo, Alonso:2017ktd} for $\bar B-$decays, where so-called visible  distributions of detectable particles from the $\tau$-decay are analyzed,  aims to determine the ${\cal A_H}, \cdots {\cal E_H}$ functions without having to  measure the $\tau$ momentum. Indeed, it is integrated out in these works, and the proposed (visible) kinematical variables are referred to the initial $\bar B$ and outgoing $D^{(*)}$ three-momenta. Further and complementary constrains, within this scheme of visible kinematics, can  also be   obtained from different angular asymmetries that can be constructed using the products of  the final hadron decay ($D^*\to D \pi$)~\cite{Bhattacharya:2020lfm}.

\subsection{One-dimensional polarization averages}
\begin{figure}[tbh]
\begin{center}
\includegraphics[scale=1]{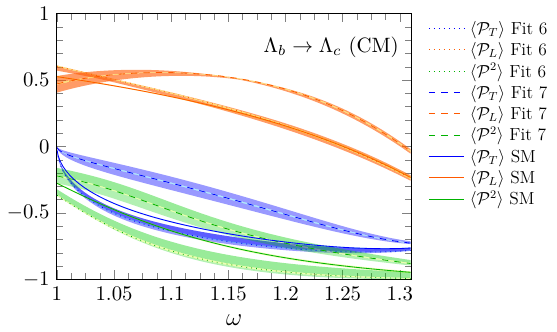}\\ 
\includegraphics[scale=1]{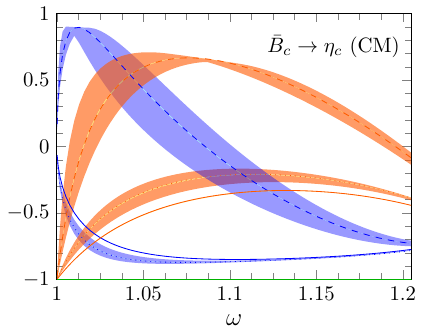}\hspace{.15cm}
\includegraphics[scale=1]{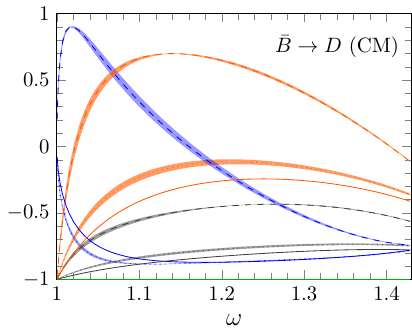}\\
\hspace{.35cm}\includegraphics[scale=1]{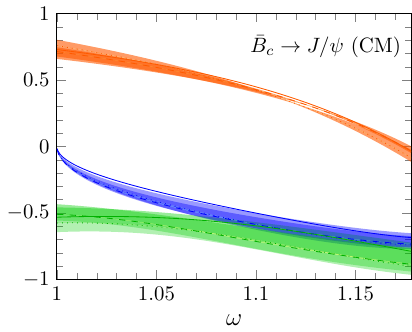}\hspace{.35cm}
\includegraphics[scale=1]{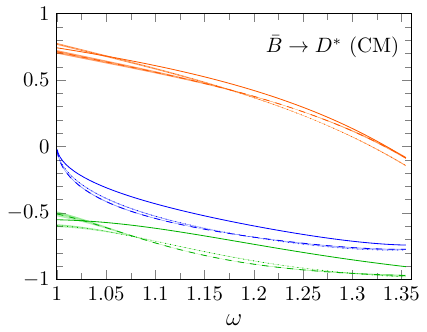}\\
\caption{ Polarization
$\langle {\cal P}_T^{\rm CM}\rangle$, $\langle {\cal P}_{L}^{\rm CM}\rangle$
and $\langle {\cal P}^{2}\rangle$ averages, defined in the CM system and calculated for the SM  and the  NP  Fits 6  and  7   of Ref.~\cite{Murgui:2019czp}, as a function of $\omega$. In addition, for the $\bar B \to D$ decay, the gray curves stand for the SM
(solid), Fit 6 (dotted) and  Fit 7 (dashed)  results obtained for $-|\vec P_{\rm CM}|^2$ (Eq.~\eqref{eq:P2noinv}). Error bands take into account  the uncertainties associated to the Wilson coefficients and form
factors, and they are calculated as explained in Refs.~\cite{Penalva:2020xup,Penalva:2020ftd}.}  
\label{fig:CM}
\end{center}
\end{figure}
\begin{figure}[tbh]
\begin{center}
\includegraphics[scale=1]{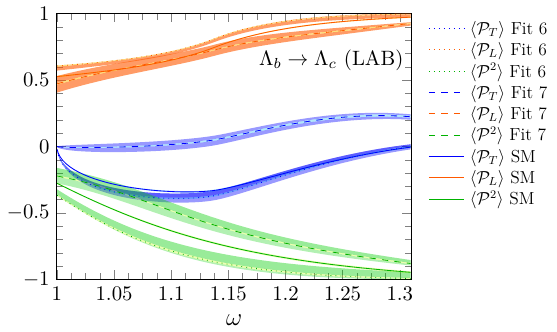}\\ 
\includegraphics[scale=1]{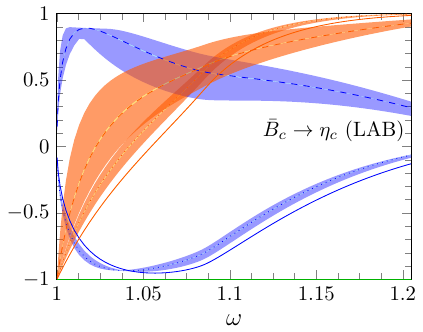}\hspace{.15cm}
\includegraphics[scale=1]{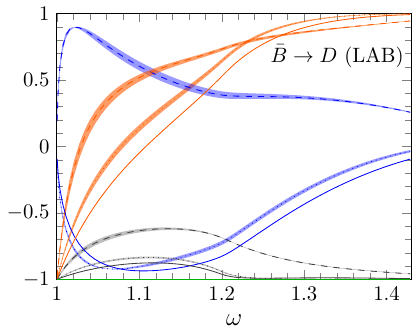}\\
\hspace{.35cm}\includegraphics[scale=1]{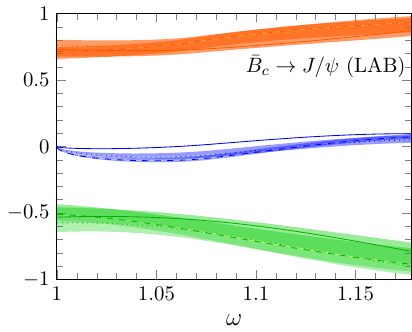}\hspace{.35cm}
\includegraphics[scale=1]{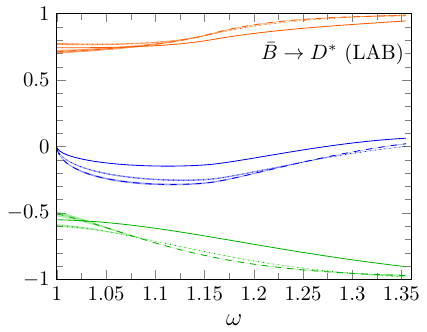}\\
\caption{ The same as in Fig.~\ref{fig:CM}, but for projections defined in the LAB system.}  
\label{fig:LAB}
\end{center}
\end{figure}
Some of the features discussed above in the presentation of the ${\cal P}_{L,T}$ results are  easier to observe in the one dimensional plots displayed in Figs.~\ref{fig:CM} and \ref{fig:LAB}. There,  we now show the CM and LAB $\langle{\cal
P}_L\rangle(\omega)$, $\langle{\cal P}_T\rangle(\omega)$ 
 and $\langle{\cal P}^2\rangle(\omega)$ averages, the latter given by 
 $\langle{\cal P}^2\rangle(\omega)=-\langle{\cal P}^2_L+{\cal P}^2_T+{\cal P}^2_{TT}\rangle(\omega)$, 
 as introduced in Eqs.~\eqref{eq:Pkorner1} and \eqref{eq:defP2}. As discussed in Subsec.~\ref{sec:cpv}, these averages are related to the CM/LAB tau polarization asymmetries obtained from $d\Gamma/d\omega$, whose measurement require the detection of the momentum and spin-state of the $\tau$. Equivalently, these averages can be obtained from  the analysis of the full angular distribution of the pion or rho mesons, originated in the subsequent hadron decay of the tau, measured in the $\tau$-rest frame~\cite{Ivanov:2017mrj}. Following the discussion at the end of the previous subsection, these observables seem more difficult to access experimentally than those proposed in Refs.~\cite{Asadi:2020fdo, Alonso:2017ktd}, which do not require the detection of the tau lepton and that we will study elsewhere.

In Figs.~\ref{fig:CM} and \ref{fig:LAB}, only NP Fits 6 and 7 of Ref.~\cite{Murgui:2019czp} are still considered, where all Wilson coefficients are real and therefore the ${\cal P}_{TT}$ component vanishes. Additionally,  we  also show results for the 
$\bar B_c\to \eta_c\tau\bar\nu_\tau$ and $\bar B_c\to J/\psi\tau\bar\nu_\tau$ decays, not presented for the 2D distributions and the ${\cal A_H}, \cdots {\cal E_H}$ functions discussed in the previous subsection.  We include, in all cases, 68\% confident-level (CL) error bands that take into account  the uncertainties associated to the Wilson coefficients and form
factors, as explained in Refs.~\cite{Penalva:2020xup,Penalva:2020ftd}. 

The $\omega-$shape patterns for the $\bar B  \to D$ and $\bar B_c  \to \eta_c$ or the $\bar B  \to D^*$ and $\bar B_c  \to J/\psi$ reactions are qualitatively similar, while those obtained from the $\Lambda_b \to \Lambda_c$ decay show some resemblances with the $0^- \to 1^-$ ones. A good number  of the distributions depicted in Figs.~\ref{fig:CM} and \ref{fig:LAB} can be used to disentangle between SM and the two NP cases considered there. In particular, Fit 7 leads to results clearly distinctive, even taking into account theoretical uncertainties bands, while  SM and Fit 6 predictions are more difficult to separate. Nevertheless,  from the results of Figs.~\ref{fig:CM} and ~\ref{fig:LAB} one can safely conclude that, with the exception of the  $\bar B_c\to J/\psi$, and to a lesser extent $\bar B\to D^*$, 
   the observables shown could theoretically tell apart Fit 6 from Fit 7.

We note that the averages of the LAB longitudinal and transverse projections 
$\langle {\cal P}_{L,T}^{\rm LAB}\rangle(\omega)$
can not be obtained as linear combinations of the 
$\langle {\cal P}_{L,T}^{\rm CM}\rangle(\omega)$ with known kinematical coefficients. They provide thus complementary information. 
 This is easily seen in the 
expressions collected in Appendix \ref{app:pltttavg}. In addition to ${\cal A}, {\cal B}$ 
and ${\cal C}$, which could be extracted from either the unpolarized CM 
$d^2\Gamma/(d\omega d\cos\theta_\tau)$ or the LAB $d^2\Gamma/(d\omega dE_\tau)$ 
differential decay widths~\cite{Penalva:2020xup}, we observe  that 
$\langle {\cal P}_T^{\rm CM}\rangle(\omega)$ depends on ${\cal A_H}$ and ${\cal C_H}$ while, 
in $\langle {\cal P}_L^{\rm CM}\rangle(\omega)$, the scalar $\omega$-functions ${\cal B_H}$, 
${\cal D_H}$ and ${\cal E_H}$ also appear. In turn, $\langle {\cal P}_T^{\rm LAB}\rangle$ 
provides an independent linear combination of ${\cal B_H}$, ${\cal D_H}$ and 
${\cal E_H}$, and the expression for $\langle {\cal P}_L^{\rm LAB}\rangle(\omega)$ 
involves all the  ${\cal A_H}$,  ${\cal B_H}$, ${\cal C_H}$,
 ${\cal D_H}$ and ${\cal E_H}$ functions. 
 Another consequence of this discussion is that for a given decay,  
all former five $\omega$-functions cannot be   determined  only from the 
four averages  $\langle{\cal P}_{L,T}^{\rm CM}\rangle(\omega)$ and 
$\langle{\cal P}_{L,T}^{\rm LAB}\rangle(\omega)$, and it would be necessary to have 
additional information, as for example the two-dimensional dependencies of the 
different polarization components discussed in the previous subsection. Alternatively, 
as noted above, all these scalars (${\cal A_H}, \cdots {\cal E_H}$)  can  also be obtained  
from the combined study of the CM $d^2\Gamma/(d\omega d\cos\theta_\tau)$ and LAB $d^2\Gamma/(d\omega dE_\tau)$ helicity-polarized distributions~\cite{Penalva:2020xup}. 

However, it is clear that the combined use of all averages, for the five decays, shown in Figs.~\ref{fig:CM} and \ref {fig:LAB} will greatly restrict the characteristics of possible extensions of the SM, and certainly in a more efficient way than if only one particular decay is considered.

Finally, for the $\bar B\to D$ decay,  we also show (gray curves and bands) the  frame 
dependent quantity  $-|\vec P|^2$ (see Eq.~\eqref{eq:P2noinv}), introduced in  
Ref.~\cite{Ivanov:2017mrj}.  Clearly, $-|\vec P|^2$  fails to convey the information 
on the degree of polarization of the $\tau$.  For the $\bar B\to D$ decay, and except 
at zero recoil, for which  $-|\vec P|^2=-\langle{\cal P}_L\rangle^2=\langle{\cal P}^2
\rangle=-1$,  its value is never exactly minus one. We also test that while 
$\langle{\cal P}^2\rangle$ is a scalar and leads to the same LAB and CM 
$\omega$-distributions, $-|\vec P|^2$ depends on the reference system where it has been
 defined. We observe that in the high $\omega$-region, $-|\vec P|^2$ in LAB is closer
  to $-1$ than when it is calculated in CM, being in the first frame almost 
  indistinguishable from $-1$ for the SM and Fit 6 cases. This follows from the 
  discussion above of the LAB 2D distributions, where we pointed out that ${\cal P}_L$ 
   approaches 1 at maximum recoil, as a consequence of an approximate negative-helicity 
   selection by  the dominant operators in that $\omega$ region (large $\tau$ LAB
    momentum).

\subsubsection{Complex Wilson coefficients}

\begin{figure}[tbh]
\begin{center}
\hspace{0cm}\includegraphics[scale=.8]{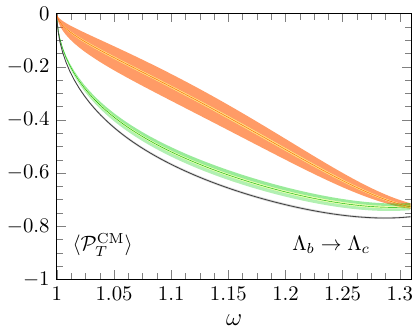}\hspace{2.6cm}
\includegraphics[scale=.8]{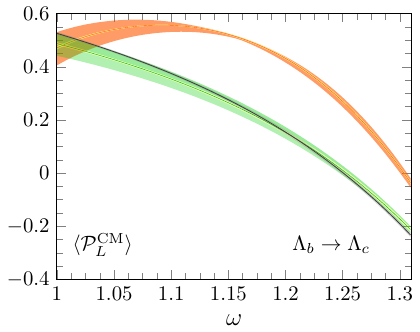}\\
\includegraphics[scale=.8]{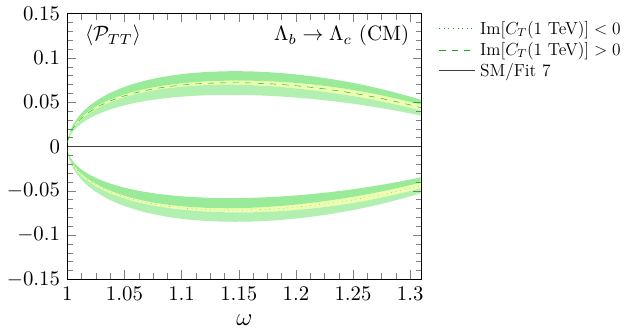}
\includegraphics[scale=.8]{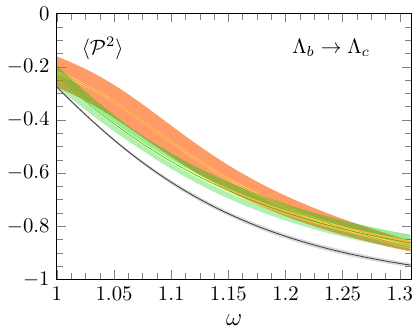}
\caption{ Polarization averages $\langle {\cal P}_T^{\rm CM}\rangle(\omega)$, 
$\langle {\cal P}_{L}^{\rm CM}\rangle(\omega)$,
 $\langle {\cal P}_{TT}^{\rm CM}\rangle(\omega)$
and $\langle {\cal P}^2\rangle(\omega)$ defined in the CM system and calculated
for the  SM  (black)  and the  NP   Wilson coefficients from 
Fit 7 (red)  of Ref.~\cite{Murgui:2019czp} and the $R_2$ leptoquark model fit (green)
of Ref.~\cite{Shi:2019gxi} (see text for details).  Error bands  take into account the uncertainties 
 associated to the Wilson coefficients and form factors, and they are calculated as 
 explained in the main text for the $R_2$ leptoquark model fit,  and in 
 Ref.~\cite{Penalva:2020xup}  for the SM and Fit 7. }  
\label{fig:CMLAMBDA}
\end{center}
\end{figure}
For the particular case of the $\Lambda_b\to\Lambda_c$ transition, we also show in 
Fig.~\ref{fig:CMLAMBDA} results for the $R_2$ leptoquark model fit of Ref.~\cite{Shi:2019gxi}. 
This fit is particularly interesting since the two nonzero Wilson coefficients, $C_{S_L}$ and $C_T$,
are complex giving rise to a nonzero $\langle {\cal P}_{TT}^{\rm CM}\rangle(\omega)$ value. For this particular model $C_{S_L}$ and $C_T$ at the bottom mass scale, appropriate for the present  calculation, are given in terms  of just 
 the value of  $\hat C_T$ at the scale of 1\,TeV, with   $\hat C_{S_L}(1\, \rm{TeV})=4\,\hat C_T(1\, \rm{TeV})$, and the corresponding
 evolution matrix (see Ref.~\cite{Shi:2019gxi}). The right panel of figure 4 of Ref.~\cite{Shi:2019gxi} shows the constraints on
the complex $C_{S_L}$ plane, with best fit point at $C_{S_L} = -0.08 \pm 0.30~ i$.   As we 
have done in Figs.~\ref{fig:CM} and \ref{fig:LAB}, the error on the observables inherited
 from the form-factor uncertainties is evaluated and propagated  via Monte Carlo, 
 taking into account statistical correlations between the different parameters. 
 It is shown as an inner error band that accounts for 68\% CL intervals. 
 The uncertainty  induced by  the fitted Wilson coefficients is determined 
  using  different $1\sigma$ Wilson coefficients configurations provided by the authors of
  Ref.~\cite{Shi:2019gxi}. The two sets of errors are then added in 
 quadrature  giving rise to the larger uncertainty band that can be seen in the figure. 

In Fig.~\ref{fig:CMLAMBDA}, and for the sake of comparison, we also include the polarization observables 
obtained with the SM and  Fit 7 of  Ref.~\cite{Murgui:2019czp}. We do not show in the 
figure any result for Fit 6 of \cite{Murgui:2019czp}, because this latter NP fit leads 
to predictions  close to the SM ones.
 
The results for $\langle {\cal P}_{L,T}\rangle$ obtained with the $R_2$ fit of Ref.~\cite{Shi:2019gxi} 
are closer to the SM ones than the ones
obtained from Fit 7 of  Ref.~\cite{Murgui:2019czp}. 
This is particularly true for $\langle {\cal P}_{L}\rangle$ where the SM result is contained in the error band of the
$R_2-$model prediction. Things change for $\langle {\cal P}_{TT}\rangle$. As mentioned above,
 the complex  $C_{S_L}$ and
$C_T$ Wilson coefficients of the $R_2$ fit of Ref.~\cite{Shi:2019gxi}
  generate a nonzero average-polarization $\langle {\cal P}_{TT}\rangle(\omega)$,
 which is shown  in the lower-left panel of Fig.~\ref{fig:CMLAMBDA}. The 
 nonzero-result for $\langle {\cal P}_{TT}\rangle $ comes from  the  interference of SM vector-axial
   with the NP  terms, as well as the interference between the NP terms themselves. While for the $R_2$ model most observables are 
 quadratic in the imaginary part of $\hat C_T(1\,  {\rm TeV})$, like $\langle {\cal P}^{\rm CM}_{L,T}\rangle$ here but also the 
  ${\cal R}_{D^{(*)}},\ {\cal R}_{\Lambda_c},{\cal R}_{J/\psi}$ ratios, and the $\tau$ ($A_{\lambda_\tau}$) and the longitudinal $D^*$ ($F_L^{D^*}$)  polarization asymmetries, $\langle {\cal P}_{TT}\rangle$ is indeed linear in the imaginary part of
 $\hat C_T(1\,  {\rm TeV})$. This allows to break the degeneracy present in the other observables with respect to
 the sign of  ${\rm Im\,}[\hat C_T(1\,  {\rm TeV})]$.

 As discussed above, the projection ${\cal P}_{TT}$ is invariant under co-linear boost transformations,
  and as a consequence  the LAB average $\langle {\cal P}_{TT}\rangle(\omega)$ would be identical to 
  that shown in  Fig.~\ref{fig:CMLAMBDA}, and evaluated in the CM frame. This average can be used to 
  determine the linear combination of the functions ${\cal F_H}$ and ${\cal G_H}$ given in 
  Eq.~\eqref{eq:pttexplicito} of the appendix.  However, additional information on the CM angular 
  dependence of the  ${\cal P}_{TT}$ projection would be required to separately extract   the 
  time-reversal odd functions ${\cal F_H}$ and ${\cal G_H}$. 
The experimental finding of  signatures of non-zero tau polarization in a direction perpendicular 
to the plane formed by the CM (or LAB) three momenta of the outgoing hadron and the $\tau$ would be 
a clear indication, not only of NP beyond the SM, but also of CP (or time reversal) violation.

The results for ${\cal R}_{\Lambda_c}$ are collected in Table~\ref{tab:ratios}. The result 
 obtained with the
 $R_2$ fit of Ref.~\cite{Shi:2019gxi} is not far from to the SM one. Part of the reason for this 
 behavior could be  in the use,  in Ref.~\cite{Shi:2019gxi}, of  
  $B\to D^{(*)}$  form factors  evaluated in the heavy quark limit. The use of the improved form factors obtained
  in Ref.~\cite{Murgui:2019czp}, which included sub-leading corrections, in the $R_2$ fit
 gives rise to  a larger  ${\rm Im\,}[\hat C_T(1\,  {\rm TeV})]$ value that  results in  
 ${\cal R}_{\Lambda_c}$ being larger around $\sim 0.385$.

\begin{table}
\begin{center}
\begin{tabular}{c|ccc}
                        &  SM  & Fit 7 \cite{Murgui:2019czp}& $R_2$ 
                         \cite{Shi:2019gxi}\\\hline\tstrut
 $\Gamma_{e(\mu)}/\left(10\times |V_{cb}|^2 {\rm ps}^{-1}\right)$ & ~$2.15\pm 0.08$ & $-$ & $-$\\ \tstrut 
 $\Gamma_\tau/ \left(10\times |V_{cb}|^2 {\rm ps}^{-1}\right)$ & ~$0.715\pm 0.015$ 
 &~  $0.89\pm 0.05$&~ $0.75\pm0.02$ \\ \tstrut
 ${\cal R}_{\Lambda_c}$ & ~$0.332 \pm 0.007$   & ~ $0.41\pm 0.02$ & ~ $0.350\pm0.010$\\ \hline
\end{tabular}
\end{center}
\caption{Total decay widths $\Gamma_\tau=\Gamma\left(\Lambda_b\to\Lambda_c\tau\bar\nu_\tau\right)$ 
and  $\Gamma_{e(\mu)}=\Gamma\left(\Lambda_b\to\Lambda_c\, e(\mu)\bar\nu_{e(\mu)}\right)$  
and  ratios ${\cal R}_{\Lambda_c} =\Gamma\left(\Lambda_b\to\Lambda_c\tau\bar\nu_\tau\right)
/\Gamma\left(\Lambda_b\to\Lambda_c\, e(\mu)\bar\nu_{e(\mu)}\right)$ obtained in the SM and 
in the NP scenarios corresponding to Fit 7 of Ref.~\cite{Murgui:2019czp} and the $R_2$ leptoquark model
 fit of Ref.~\cite{Shi:2019gxi} (see text for details).}
\label{tab:ratios}
\end{table}
\section{Summary} 
For a given configuration of the momenta of all particles involved, we have introduced 
the tau spin-density matrix $\bar \rho$ and the polarization vector ${\cal P}^\mu$ associated 
to a general $H_b\to H_c\tau\bar\nu_\tau$ decay. These two quantities contain all the 
information on the  spin state of the $\tau$  provided no other particle spin  is measured. 
For different semileptonic decays, we have evaluated ${\cal P}^\mu$ in the LAB and CM frames 
including the effects of NP.  We have seen that the independent components ${\cal P}_{L}$, 
${\cal P}_{T}$ and $ {\cal P}_{TT}$ provide useful information to  distinguish between different 
NP scenarios. This is specially true for the 
meson $0^-\to 0^-$ and also for the baryon $\Lambda_ b\to \Lambda_c$ reactions analyzed in this work. 
For this latter reaction, we have presented results for an extension of the SM that contains complex Wilson coefficients.

The LAB and CM helicity-polarized differential decay widths do not allow access to observables related 
to $ {\cal P}_{TT}$,  which is the 
 component of the polarization vector orthogonal to the plane defined by the final hadron and tau 
 three-momenta. Moreover,  ${\cal P}_{T}$, which is the projection of $\vec{{\cal P}}$ contained in the former 
 plane and perpendicular to the $\tau-$momentum,  can only be obtained indirectly from these 
 helicity-distributions, provided that results from both reference systems  are  analyzed 
 simultaneously.  The  transverse polarization ${\cal P}_{TT}$ is of special interest, since it 
 is only possible for complex Wilson coefficients.  Measuring a non-zero $ {\cal P}_{TT}$ value in
  any of the two frames will be a clear 
  indication of physics beyond the SM and of time reversal (or CP) violation. For the NP 
  scenarios  corresponding to Fits 6 and 7 of Ref.~\cite{Murgui:2019czp} the Wilson coefficients are 
  real and thus    ${\cal P}_{TT}$ is identically zero. In such a case, $\vec {\cal P}$ is contained 
  in the hadron-lepton plane. The $R_2$ fit of  Ref.~\cite{Shi:2019gxi}, which contains $C_{S_L}$ and
  $C_T$ complex Wilson coefficients generates, however, a 
  nonzero $ {\cal P}_{TT}$ value. 
 
The NP effective Hamiltonian in Eq.~\eqref{eq:hnp} contains five  Wilson coefficients, in 
general complex, although one
of them can always be taken to be real. Therefore, nine free parameters should be 
determined from data. Even assuming
that the form factors are known, and therefore the genuinely hadronic part ($W$) of 
the $\widetilde W$ SFs, it is difficult to determine all NP parameters  from a unique 
type of decay, since the experimental measurement of the required polarization 
observables is an extremely difficult task. It is therefore essential to simultaneously
analyze data from various types of semileptonic decays, as we have done in this work. 
We have used state of the art form-factors for all reactions, and the results presented 
in this work nicely complement those presented in our previous works of 
Refs.~\cite{Penalva:2020xup, Penalva:2020ftd}, and all together can be efficiently 
employed to disentangle among different NP scenarios. 

Finally, we would like to draw the attention to the hadron-tensor method used in 
this work, previously derived in \cite{Penalva:2020xup}, which has shown to be a 
particularly suited tool to study processes where all final/initial hadron polarizations 
have been summed up. The scheme leads to compact expressions, valid for any baryon/meson 
semileptonic decay for unpolarized hadrons in the presence of NP and it clearly is an 
alternative to the helicity amplitude framework commonly used in the literature. 
Subsequent decays of the produced $\tau$, after the $b\to c \tau \bar \nu_\tau$ 
transition, 
\bea
H_b \to H_c &\tau^-& \bar \nu_\tau \nonumber \\
&\,\drsh & \nu_\tau\mu^-\bar\nu_\mu,\, \nu_\tau \pi^-,\, \nu_\tau \rho^-\,\cdots  
\eea
can be straightforwardly studied within this hadron-tensor scheme and they will be  presented elsewhere. 

\section*{Acknowledgements}
We warmly thank Jorge Martin Camalich by providing us with the statistical uncertainties and correlations of the Wilson coefficients for $R_2$ leptoquark model fit. This research has been supported  by the Spanish Ministerio de
Econom\'ia y Competitividad (MINECO) and the European Regional
Development Fund (ERDF) under contracts FIS2017-84038-C2-1-P and PID2019-105439G-C22, 
the EU STRONG-2020 project under the program H2020-INFRAIA-2018-1, 
grant agreement no. 824093 and by  Generalitat Valenciana under contract PROMETEO/2020/023. 
 
\appendix
\section{CM and LAB kinematics}
\label{app:cmlabkin}
In this appendix we collect the different  vector products needed to evaluate
the ${\cal P}_{L},\,{\cal P}_{T}$ and ${\cal P}_{TT}$
polarization vector components  in the CM and LAB reference frames. We have
\begin{equation}
 p^2=M^2,\,\, k^2=0,\,\, k^{\prime 2}= m^2_\tau,\,\, p\cdot q= MM_\omega, 
 \,\, k\cdot k'= q\cdot k=\frac{q^2-m^2_\tau}{2}, 
 \,\, q\cdot k'=\frac{q^2+m^2_\tau}{2},
\end{equation}
with $M_\omega=M-M'\omega$. In addition, the scalar products that depend
 explicitly on the charged lepton variables used in the differential decay
 widths  read \\

\underline{LAB}: In this case, $p^\mu=(M,\vec 0\,)$, $q^\mu=\left(M_\omega, \,M'\sqrt{\omega^2-1}\, \hat q_{\rm LAB}\right)$
and 
\begin{eqnarray}
 &&k\cdot p= M(M_\omega-E_\tau), \quad p\cdot N_{L}=\frac{M\sqrt{E^2_\tau-m^2_\tau}}{m_\tau}, \quad 
p\cdot N_{T}=0,\nonumber\\
&& 
 q\cdot N_L= \frac{M_\omega\sqrt{E_\tau^2-m_\tau^2}}{m_\tau}
 +\frac{E_\tau \,M'\sqrt{\omega^2-1}}{m_\tau}\,
 \cos\theta_\tau^{\rm LAB},
\quad q\cdot N_{T}=M'\sqrt{\omega^2-1}\,\sin\theta_{\tau}^{\rm LAB},\nonumber\\
&& \epsilon^{k'q\,p\,N_{TT}}=-MM'\sqrt{\omega^2-1}\,\sqrt{E^2_\tau-m^2_\tau}
\,\sin\theta_{\tau}^{\rm LAB},
\end{eqnarray}
with $\theta_{\tau}^{\rm LAB}$ the angle made by the final hadron and $\tau$ lepton
 LAB three-momenta, which is fixed, once $E_\tau$ and $\omega$ are known, by the relation
\be
\cos\theta_{\tau}^{\rm LAB}=\frac{q^2+m^2_\tau-2M_\omega E_\tau }
{2M'\sqrt{\omega^2-1}\,\sqrt{E^2_\tau-m^2_\tau}}.
\ee
\noindent
For a given $\omega$, the fact that $|\cos\theta_{\tau}^{\rm LAB}|\le1$ limits the possible $E_\tau$ energies to the interval
\bea
E_\tau\in[E_\tau^{-}(\omega),\,E_\tau^{+}(\omega)],
\eea
with $E_\tau^{-}$  and $E_\tau^+$ given in Eq.~\eqref{eq:Elimits}. In terms of $E_\tau^{\pm}(\omega)$ one also can write
\bea
\sin\theta_{\tau}^{\rm LAB}=\frac{\sqrt{q^2}\sqrt{\big[E_\tau-E_\tau^{-}(\omega)\big]\big[E_\tau^{+}(\omega)-E_\tau\big]}}{M'\sqrt{\omega^2-1}\sqrt{E_\tau^2-m_\tau^2}}.
\eea

\noindent
\underline{CM}: Now $q^\mu=(\sqrt{q^2},\vec 0\,)$ and $p^\mu=\frac{1}{\sqrt{q^2}}\left(M
M_\omega,\,-MM'\sqrt{\omega^2-1}\, \hat q_{\rm LAB} \right)$, and in addition
\begin{eqnarray}
&& k\cdot p= \frac{M}{2} \left(1-\frac{m^2_\tau}{q^2}\right)\left(M_\omega + M'\sqrt{\omega^2-1}
 \cos\theta_\tau\right),\ \ q\cdot N_L= \frac{q^2-m^2_\tau}{2m_\tau}, \ \ q\cdot N_{T}=0, \nonumber\\
&& p\cdot N_{L}=\frac{MM_\omega(q^2-m^2_\tau)-MM'\sqrt{\omega^2-1}\,(q^2+m^2_\tau)
\cos\theta_\tau}{2m_\tau q^2}, \nonumber \\
&&p\cdot N_{T}=-\frac{MM'\sqrt{\omega^2-1}}{\sqrt{q^2}}\,\sin\theta_\tau, 
\nonumber \\ 
&&\epsilon^{k'q\,p\,N_{TT}}=-MM'\sqrt{\omega^2-1}\,\frac{q^2-m^2_\tau}{2\sqrt{q^2}}
\sin\theta_\tau.
\end{eqnarray}

Note that, since the three-vector components transverse  to the velocity defining a 
boost do not change, we obtain
\bea
\sqrt{E^2_\tau-m^2_\tau}
\,\sin\theta_{\tau}^{\rm LAB}=|\vec k^{\,\prime\,}|^{\rm LAB}
\sin\theta_\tau^{\rm LAB}=|\vec k^{\,\prime\,}|^{\rm CM}\sin\theta_\tau=\frac{q^2-m^2_\tau}{2\sqrt{q^2}}
\sin\theta_\tau 
\eea
and then $\epsilon^{k'q\,p\,N_{TT}}\big|_{\rm LAB}=
\epsilon^{k'q\,p\,N_{TT}}\big|_{\rm CM}$, which shows that ${\cal P}_{TT}^{\rm LAB}={\cal P}_{TT}^{\rm CM}$, since  the other factor 
${\cal N_{H_{\rm 3}}}(\omega,\,k\cdot p)/{\cal N}(\omega,\,k\cdot p)$ in Eq.~\eqref{eq:ptt} is a Lorentz scalar.

\section{Expressions for $\langle{\cal P}_{L,T,TT}^{\rm CM}\rangle(\omega)$ and $\langle{\cal P}_{L,T,TT}^{\rm LAB}\rangle(\omega)$}
\label{app:pltttavg}
In this appendix we give expressions for
 $\langle{\cal P}_{L,T,TT}^{\rm CM,\,LAB}\rangle(\omega)$ in terms of the 
 ten scalar functions,  ${\cal A}, {\cal B}$, ${\cal C}$, ${\cal A_H}, {\cal B_H}, 
{\cal C_H}, {\cal D_H}$, ${\cal E_H}$, ${\cal F_H}$ and  ${\cal G_H}$, 
introduced in Eq.~(\ref{eq:pol2}). One has

\bea
\langle {\cal P}_{L}^{\rm CM} \rangle (\omega) &=& -\frac{1}{m_\tau }\frac{1}
{M{\cal N}_\theta(\omega)}\left(1-\frac{m^2_\tau}{q^2}\right) 
\left[ MM_\omega{\cal A_H}(\omega)+ q^2{\cal B_H}(\omega) +
 \frac{M_\omega\left(q^2-m^2_\tau\right)}{2M}{\cal D_H}(\omega)\right. \nonumber  \\
&+& \left. \frac{{\cal C_H}(\omega)}{6}\left(q^2+m^2_\tau+2M_ \omega^2-
\frac{4m^2_\tau M^2_\omega}{q^2}\right) -\frac{\left(q^2-m^2_\tau\right)^2}{12M^2q^2}
\left(q^2-4M_\omega^2\right){\cal E_H}(\omega)\right],\nonumber \\
\langle {\cal P}_{T}^{\rm CM} \rangle (\omega) &=& \frac{\pi M'}{{\cal N}_\theta(\omega)}
\frac{\sqrt{\omega^2-1}}{4\sqrt{q^2}} \left[2{\cal A_H}(\omega)+ \frac{M_\omega}{M}
\left(1-\frac{m^2_\tau}{q^2}\right){\cal C_H}(\omega)\right],\nonumber \\
\nonumber\\
\langle {\cal P}_{TT}^{\rm CM} \rangle (\omega) &=& -\frac{\pi M'}{{\cal N}_\theta(\omega)}
\frac{\sqrt{\omega^2-1}}{8\sqrt{q^2}} \left(1-\frac{m^2_\tau}{q^2} \right)
\left[\frac{2q^2}{M^2} {\cal F_H}(\omega)+ \frac{M_\omega(q^2-m^2_\tau)}{M^3}{\cal G_H}(\omega)
\right], \label{eq:pttexplicito}
\eea
with
\be
{\cal N}_\theta(\omega) = {\cal A}(\omega)+ \left(1-\frac{m^2_\tau}{q^2} \right)
\left[ \frac{M_\omega}{2M}{\cal B}(\omega) + \left(1-\frac{m^2_\tau}{q^2} \right)
\left(\frac{4M_\omega^2-q^2}{12M^2} \right){\cal C}(\omega)\right]. \label{eq:ntheta}
\ee
Note that from Eq.~\eqref{eq:Pkorner2}, $\langle {\cal P}_{L}^{\rm CM} \rangle (\omega)$ can be written as
\bea
\langle {\cal P}_{L}^{\rm CM} \rangle (\omega)=\frac{[a_0(\omega,h=-1)-a_0(\omega,h=1)]+
\frac13[a_2(\omega,h=-1)-a_2(\omega,h=1)]}{a_0(\omega)+\frac13a_2(\omega)}
\eea 
where the $a_{0,2}(\omega)$ and $a_{0,2}(\omega,h=\pm1)$ functions are given in
Eqs.~(18) and (25) of Ref.~\cite{Penalva:2020xup}  in terms of the eight ${\cal A}(\omega)$,\,
${\cal B}(\omega)$,\,${\cal C}(\omega)$,\,${\cal A_H}(\omega)$,\,${\cal B_H}(\omega)$,\,${\cal C_H}(\omega)$,\,${\cal D_H}(\omega)$ and 
${\cal C_H}(\omega)$ ones.\\

In the LAB frame one has that, 
\bea
\langle {\cal P}_{L}^{\rm LAB} \rangle (\omega)&=&\frac{1}
{\left(E_\tau^+-E_\tau^-\right)\,{\cal N}_\theta(\omega)}\nonumber \\
&\times&\bigg\{M \hat c_0(\omega)\,\ln\left(\frac{E_\tau^+ +p_\tau^+}
{E_\tau^-+p_\tau^-}\right)+ \left(c_0(\omega)+\hat c_1(\omega)\right)\left(p_\tau^+-p_\tau^-\right)\nonumber\\
&+&
\frac{c_1(\omega)+\hat c_2(\omega)}{2M}\Big[E_\tau^+p_\tau^+-
E_\tau^-p_\tau^-+m_\tau^2\,\ln\left(\frac{E_\tau^++p_\tau^+}
{E_\tau^-+p_\tau^-}\right)
\Big]\nonumber\\
&+&
\frac{c_2(\omega)+\hat c_3(\omega)}{3M^2}\Big[\left(E_\tau^{+\,2}+2 m_\tau^2\right)p_\tau^+-
\left(E_\tau^{-\, 2}+2m_\tau^2\right)p_\tau^-\Big]\bigg\}, 
\eea
with $p_\tau^\pm(\omega)=\sqrt{\left[E_\tau^\pm(\omega)\right]^2-m_\tau^2}$ and where $\hat c_{0}(\omega)$, 
$\left(c_0(\omega)+\hat c_{1}(\omega)\right)$, $\left( c_1(\omega)+\hat c_{2}(\omega)\right)$ and
$\left(c_2(\omega)+\hat c_{3}(\omega)\right)$ are given in Eq.~(27) of Ref.~\cite{Penalva:2020xup} in terms of 
the ${\cal A_H}(\omega)$,\,
${\cal B_H}(\omega)$,\,${\cal C_H}(\omega)$,\,${\cal D_H}(\omega)$ and 
${\cal E_H}(\omega)$ functions.
Besides,
\bea
\langle {\cal P}_{T}^{\rm LAB} \rangle (\omega) &=& -\frac{\pi \sqrt{q^2}M'
\sqrt{\omega^2-1}}{4{\cal N}_\theta(\omega)M^2\left(1-\frac{m^2_\tau}{q^2}\right)} 
\bigg[ \Big(\frac{4I_0(\omega)M_\omega}{M_\omega+\sqrt{q^2}} -I_1(\omega)\Big)
{\cal D_H}(\omega) +
\nonumber \\ 
&&  \frac{4 I_0(\omega)M}{M_\omega+\sqrt{q^2}}{\cal B_H}(\omega) + 
\Big(\frac{8I_0(\omega)M_\omega}{M_\omega+\sqrt{q^2}}+I_2(\omega)-4I_1(\omega)\Big)
\frac{M_\omega}{2M}{\cal E_H}(\omega) \bigg]\nonumber \\
\nonumber\\
\langle {\cal P}_{TT}^{\rm LAB} \rangle (\omega) &=& \langle {\cal P}_{TT}^{\rm CM} 
\rangle (\omega),
\eea
where we have introduced the (kinematical) functions $I_{0,1,2}(\omega)$, 
\bea
I_n(\omega) &=& \frac{1}{ K_n} \int_{E_\tau^-(\omega)}^{E_\tau^+(\omega)}\frac{dE_\tau\, E_\tau^n}{\sqrt{E_\tau^2-m^2_\tau}} \sqrt{\left(E_\tau^{+}(\omega)-E_\tau\right)\left(E_\tau-E_\tau^{-}(\omega)\right)}, \quad n=0,1,2 \nonumber \\
K_0&=& \frac{\pi}{2} \left(M_\omega-\sqrt{q^2}\right), \quad K_1= \frac{\pi}{8} \left(M_\omega^2-q^2\right), \quad K_2= \frac{\pi M_\omega}{16} \left(M_\omega^2-q^2\right)
\eea
which are normalized such that $I_0=I_1=I_2=1$ in the $m_\tau \to 0$ limit.

The formulae are general and they can be also used for muon or electron decay modes, taking appropriate values for the NP Wilson coefficients.

\bibliographystyle{jhep}

\bibliography{B2Dbib}

\end{document}